\documentclass[fleqn,usenatbib]{mnras}

\usepackage[T1]{fontenc}

\DeclareRobustCommand{\VAN}[3]{#2}
\let\VANthebibliography\thebibliography
\def\thebibliography{\DeclareRobustCommand{\VAN}[3]{##3}\VANthebibliography}

\usepackage{graphicx}	
\usepackage{amsmath}	
\usepackage{amssymb}	
\usepackage{pdflscape}
\usepackage[section]{placeins} 

\usepackage{newtxtext,newtxmath}

\title[Mapping the core of the Tarantula Nebula with VLT-MUSE. III]{Mapping the core of the Tarantula Nebula with VLT-MUSE. III. A template for metal-poor starburst regions in the visual and far-ultraviolet}

\author[Crowther \& Castro]{
Paul A. Crowther$^{1}$\thanks{paul.crowther@sheffield.ac.uk}, 
N. Castro$^{2}$\\
$^{1}$ Department of Physics and Astronomy, University of Sheffield, Sheffield, S3 7RH, UK\\
$^{2}$  Leibniz-Institut f\"{u}r Astrophysik Potsdam, An der Sternwarte 16, 14482 Potsdam, Germany}

\date{Accepted 2023 November 24. Received 2023 November 24; in original form 2023 September 19}

\pubyear{2023}

\begin{document}
\label{firstpage}
\pagerange{\pageref{firstpage}--\pageref{lastpage}}
\maketitle

\begin{abstract}
We present the integrated VLT-MUSE spectrum of the central $2'\times 2'$ (30$\times$30 pc$^{2}$) of NGC~2070, the dominant giant H\,{\sc ii} region of the Tarantula Nebula in the Large Magellanic Cloud, together with
an empirical far-ultraviolet spectrum constructed via LMC template stars from the ULLYSES survey and Hubble Tarantula Treasury Project UV photometry. NGC~2070 provides a unique opportunity to compare results from
 individual stellar populations (e.g. VLT FLAMES Tarantula Survey) in a metal-poor starburst region to the integrated results from the population synthesis tools Starburst99, Charlot \& Bruzual and BPASS. 
 The metallicity of NGC~2070 inferred from standard  nebular strong line calibrations is $\sim 0.4\pm 0.1$ dex lower than obtained from direct methods. The  H$\alpha$ inferred age of 4.2 Myr from Starburst99 is close to the median age of OB stars within the region, although individual stars span a broad range of 1--7 Myr. The inferred stellar mass is close to that obtained for the rich star cluster R136 within NGC~2070, although this contributes only 21\% to the integrated far-UV continuum. He\,{\sc ii} $\lambda$1640 emission is dominated by classical WR stars and main sequence WNh+Of/WN stars. 18\% of the NGC~2070 far UV continuum flux arises from very massive stars with $\geq$100 $M_{\odot}$, including several very luminous Of supergiants. None of the predicted population synthesis models at low metallicities are able to successfully reproduce the far-UV spectrum of NGC~2070. We attribute issues to the treatment of mass-loss in very massive stars, the lack of contemporary empirical metal-poor templates, plus WR stars produced via binary evolution.
\end{abstract}




\section{Introduction}

The Tarantula Nebula (30 Doradus) in the Large Magellanic Cloud (LMC) is intrinsically the brightest star-forming region within the Local Group of galaxies \citep{2019Galax...7...88C}. It has been the subject of numerous studies across the electromagnetic spectrum \citep{1995ApJ...444..647V, 2013AJ....146...53S,  2022ApJ...932...47W, 2022MNRAS.515.4130C, 2023arXiv231106336F}. Its low (half-solar) metallicity and high star-formation intensity are more typical of knots star-forming galaxies at $z\sim$2--3 \citep{2016ApJ...826..159S, 2017ApJ...843L..21J}  than local systems, owing to its very rich stellar content \citep{2018Sci...359...69S}. Indeed, 30 Doradus has nebular conditions which are reminiscent of Green Pea galaxies \citep{2009MNRAS.399.1191C}, which are local extreme emission-line galaxies, some of which are known to be Lyman continuum leakers \citep{2017ApJ...845..165M}. 

The Tarantula Nebula is host to hundreds of massive stars, including very massive stars (VMS) located in the central, dense star cluster R136 \citep{1998ApJ...493..180M, 2010MNRAS.408..731C} and the extended giant H\,{\sc ii} region NGC~2070, the central ionized nebula within the Tarantula \citep{2014A&A...570A..38B}. Its proximity permits observation and analysis of individual massive OB and Wolf-Rayet stars \citep{1985A&A...153..235M, 1999A&A...341...98S, 2011A&A...530A.108E}. Star formation in the Tarantula Nebula began at least 15--30 Myr ago, as witnessed by the Hodge~301 cluster, with an upturn in its rate of star formation in the last 5--10 Myr \citep{2018A&A...618A..73S}. Star formation is still ongoing, as witnessed by clumps of molecular gas observed with the Atacama Large Millimeter Array \citep[ALMA,][]{2022ApJ...932...47W}.

The proximity of the LMC provides a unique opportunity to study a rich, intensively star-forming region individually, via its resolved stellar content \citep{2013A&A...558A.134D}, and collectively, via its integrated light via application of population synthesis models. If the LMC were located at a distance of 10 Mpc, R136, NGC~2070 and the Tarantula would subtend diameters of 0.04 arcsec, 0.8 arcsec and 6 arcsec, respectively \citep{2019Galax...7...88C}. Population synthesis models are widely employed to interpret far-UV spectroscopy of unresolved star clusters at Mpc distances \citep{2004ApJ...604..153C, 2014ApJ...795..109J, 2022AJ....164..208S}, local star-forming galaxies \citep{2013ApJ...765..118W, 2022ApJS..261...31B} plus those at $z>2$ observed with large ground-based telescopes \citep{2016ApJ...826..159S, 2020A&A...636A..47S} or {\it James Webb Space Telescope}  \citep{2023MNRAS.518L..45C, 2023NatAs...7..622C}.

\begin{figure*}
\centering
\includegraphics[width=\columnwidth]{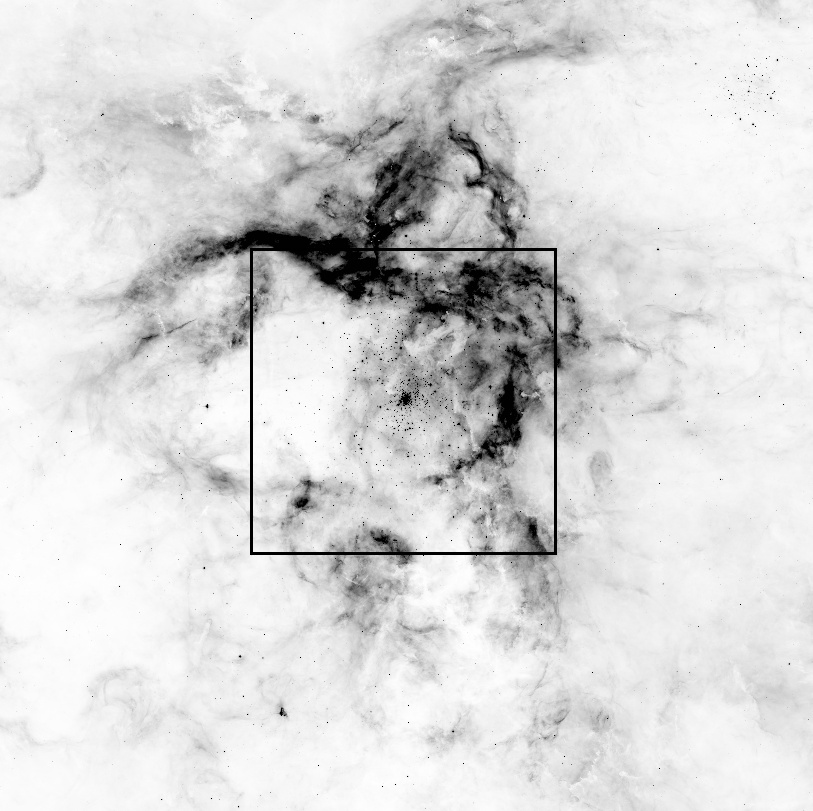}
\includegraphics[width=\columnwidth]{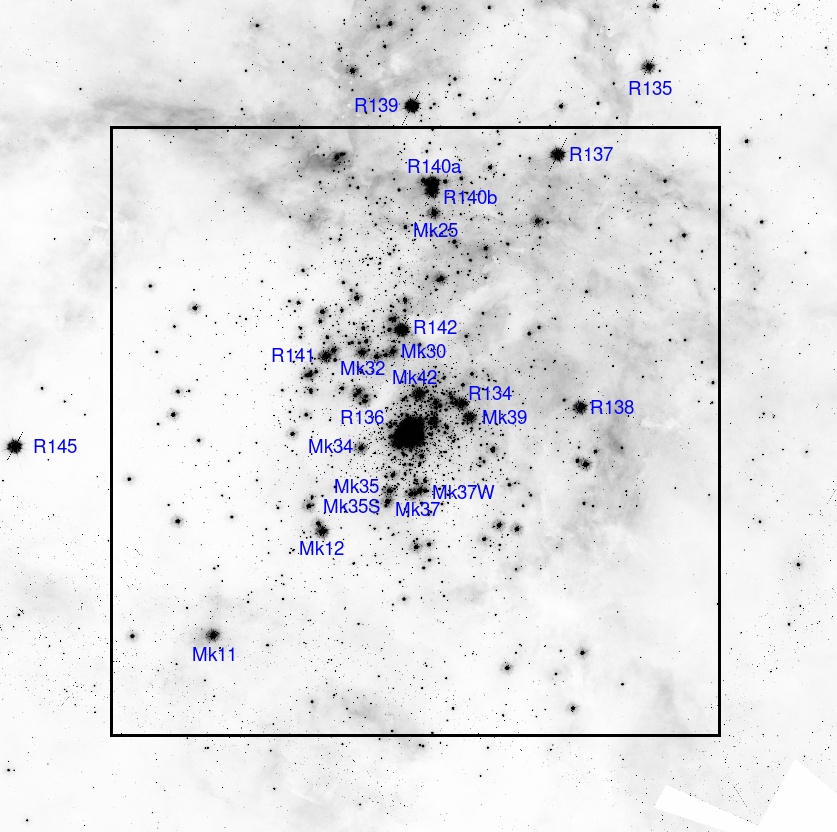}
\caption{Left panel: HST ACS/F658N image of the central region of 30 Doradus (320$\times$320 arcsec$^{2}$) from HTTP \citep{2013AJ....146...53S} including MUSE field of view (black box). North is up, East to the left. Hodge~301 can be seen to the upper right. Right panel: HST WFC3/F336W image of central region of NGC~2070 (165$\times$165 arcsec$^{2}$ from HTTP, with selected bright sources labelled.  The white region to the lower right is not included in the F336W footprint}
\label{30dor-F658N-F336W}
\end{figure*}


Although the nebular properties of the entire 30 Doradus region has previously been studied \citep{1995AJ....109..594K, 2010ApJS..191..160P}, here we focus on the central region of NGC~2070
observed with the Multi Unit Spectroscopic Explorer (MUSE) mounted at the Very Large Telescope (VLT), as part of its original Science Verification programme. \citet{2018A&A...614A.147C} introduce the dataset, and provide a stellar census and nebular kinematic properties, while \citet{2021A&A...648A..65C} present a spectroscopic analysis of OB stars. This region is host to the central R136 star cluster, several WR stars including the R140 complex, plus several cool supergiants such as Melnick~9.

Age estimates of OB stars within NGC~2070 (external to R136) range from 1--7 Myr, with a median age of 3.6 Myr \citep{2018A&A...618A..73S}. To date, only the central
cluster R136 has been observed in the far-ultraviolet, both  collectively \citep{1992ESOC...44..347H} and individually \citep{2016MNRAS.458..624C}, the latter obtaining a cluster age of $\sim$1.5 Myr  \citep[see also][]{2022A&A...663A..36B}. Several other luminous early-type stars within NGC~2070 have been observed in the far-UV with COS or STIS instruments aboard {\it Hubble Space Telescope (HST)}, plus a large sample of far-UV template spectra of LMC OB stars have been obtained via the {\it HST} initiative ULLYSES \citep{2020RNAAS...4..205R, 2022arXiv220708690C}. Consequently we are able to construct an empirical integrated spectrum of the MUSE field-of-view in the far-UV, for comparison with predictions from (theoretical) population synthesis models Starburst99 \citep{1999ApJS..123....3L, 2014ApJS..212...14L}, Charlot \& Bruzual \citep{2003MNRAS.344.1000B, 2019MNRAS.490..978P}, and BPASS \citep{2017PASA...34...58E, 2018MNRAS.479...75S}.

The present study completes the analysis of NGC~2070 MUSE Wide Field Mode (WFM) observations and is structured as follows. We provide a brief summary of visual MUSE observations of NGC~2070 and describe how the far-UV spectrum of NGC~2070 is constructed in Section~\ref{obs}. The integrated MUSE dataset is analysed in Section~\ref{muse}, with an emphasis on nebular properties and optical Wolf-Rayet bumps, with the far-UV spectrum compared to predictions from various population synthesis models in Section~\ref{pop-syn}. A comparison between individual and cumulative results is provided in Section~\ref{discussion}, together with brief conclusions. Initial results for nebular, stellar and integrated properties of the MUSE WFM datasets were presented in \citet{2017Msngr.170...40C}.

\section{NGC~2070 spectroscopic datasets}\label{obs}

\subsection{Visual observations}\label{visual}

MUSE is a wide-field, integral-field spectrograph providing intermediate resolution ($R\sim$2000) spectroscopy from 4600-9350\AA\ \citep{2010SPIE.7735E..08B}. Four overlapping MUSE Wide Field Mode pointings were obtained at the Very Large Telescope (VLT) in August 2014 via a Science Verification programme (PI: J. Melnick), providing a 2$\times2$ arcmin$^{2}$ (30$\times$30 pc$^{2}$) mosaic which encompasses both the R136 star cluster and R140, an aggregate of WR stars to the north \citep{2018A&A...614A.147C}. The MUSE field of view is indicated on Hubble Tarantula Treasury Project\footnote{https://archive.stsci.edu/hlsp/http} \citep[HTTP,][]{2013AJ....146...53S}. 
320$\times$320 arcsec$^{2}$ ACS/F658N and 165$\times$165 arcsec$^{2}$ WFC3/F336W images  in Fig.~\ref{30dor-F658N-F336W} in which selected bright sources are identified in the latter. A larger footprint, especially to the north, would be required to fully sample the bright nebulosity of NGC~2070, but MUSE field of view includes $\sim$50\% of the far-UV continuum of the entire Tarantula\footnote{We have followed the approach set out in Section~\ref{fuv} to estimate $F_{1500} \sim 1.2 \times 10^{-11}$ erg\,s$^{-1}$\,cm$^{-2}$\AA$^{-1}$ for the region of the Tarantula exterior to the MUSE footprint}. 

\begin{figure}
\includegraphics[angle=-90,width=\columnwidth, bb = 15 15 515 780]{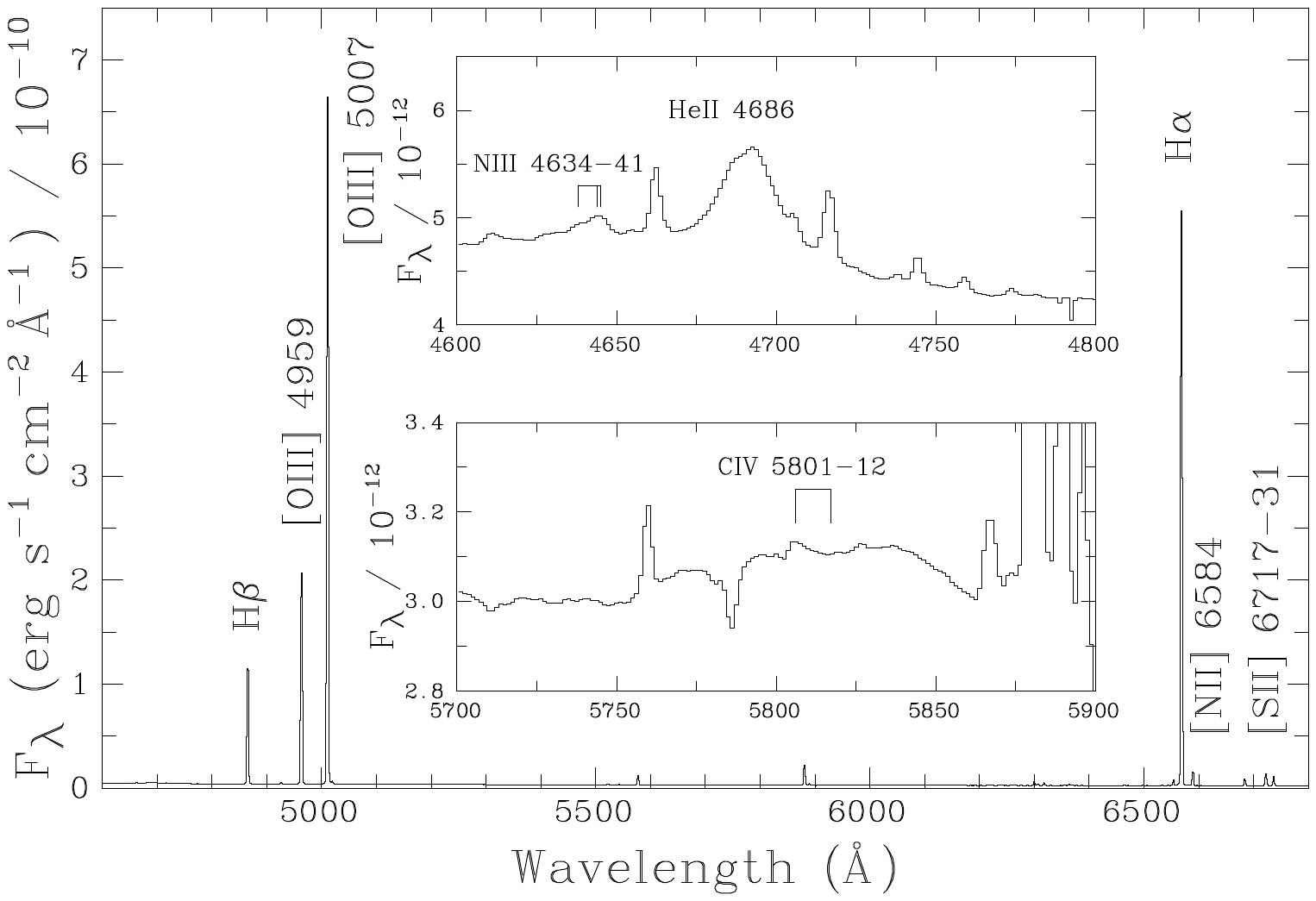}
\caption{Integrated MUSE spectrum of NGC~2070, updated from \citet{2017Msngr.170...40C}, revealing a striking emission line spectrum, with nebular
properties (e.g. high [O\,{\sc iii}] $\lambda$5007/H$\beta$, low [N\,{\sc ii}] $\lambda$6584/H$\alpha$) reminiscent of Green Pea galaxies \citep{2009MNRAS.399.1191C}. WR bumps are observed in the blue (upper inset, He\,{\sc ii} 
$\lambda$4686 arising from primarily WN stars) and yellow (lower inset, C\,{\sc iv} $\lambda\lambda$5801-12 due to WC stars). Nebular lines in the insets include [Fe\,{\sc iii}] $\lambda$4658, [Ar\,{\sc iv}] $\lambda$4711, [N\,{\sc ii}] $\lambda$5755 and He\,{\sc i} $\lambda$5876.}  
\label{muse-sp}
\end{figure}

The MUSE spatial resolution spanned 0.7 to 1.1 arcsec, corresponding to a mean spatial resolution of 0.22$\pm$0.04 pc. Four exposures of 600s for each pointing provided a continuum S/N exceeding 50 for 600 sources in the yellow.  The integrated MUSE green--red spectrum of NGC~2070 is presented in Fig.~\ref{muse-sp}, with nebular properties reminiscent of Green Pea galaxies \citep{2009MNRAS.399.1191C}, plus WR bumps in the blue (upper inset) and red (lower inset). 
Recent MUSE Narrow Field Mode (NFM) observations of the R136 cluster are presented in \citet{2021Msngr.182...50C}.

\begin{table}
\centering
\caption{Summary of sources brighter than $F_{1500} = 10^{-14}$ erg\,s$^{-1}$\,cm$^{-2}$\AA$^{-1}$ contributing to the far-UV continuum of the MUSE pointing, broken down by
subtype (primary in binaries) and whether empirical UV spectroscopy or templates were utilised. The R136a GHRS spectrum comprises 3 WNh stars and 26 O stars (assuming HSH95-17 is an O star, \citet{2022ApJ...935..162K}), each contributing $\sim$40\% and 60\% of the far-UV flux of R136a. 20 sources are confirmed spectroscopic binaries, comprising 15 systems with an O-type primary, 
3 with an Of/WN or WN5h primary, plus 1 each with a Wolf-Rayet or B-type primary. 180 additional faint sources with $5 \times 10^{-16} \leq F_{1500} < 1.0 \times 10^{-14}$ erg\,s$^{-1}$\,cm$^{-2}$\AA$^{-1}$
are incorporated via a B0\,V template scaled to the sum of their far-UV fluxes. }
\begin{tabular}{l @{\hspace{0mm}} r @{\hspace{2mm}} r @{\hspace{4mm}} r @{\hspace{2mm}} r @{\hspace{4mm}} r @{\hspace{2mm}} r}
\hline
Subtype & \multicolumn{2}{c}{Empirical} & \multicolumn{2}{c}{Template} & \multicolumn{2}{c}{Total} \\
              & N & $F_{1500}$                     & N & $F_{1500}$                          & N & $F_{1500}$  \\                    
\hline
O                      &   34 &  1.92$\times 10^{-12}$    &  159 &  7.44$\times 10^{-12}$     & 193 &  9.36$\times 10^{-12}$           \\ 
Of/WN+WNh &    7 &  1.31$\times 10^{-12}$ &   5  & 0.46$\times 10^{-12}$      &  12  & 1.77$\times 10^{-12}$          \\ 
B                      &    0 & $\cdots$                           &   16 & 1.40$\times 10^{-12}$     &  16  & 1.40$\times 10^{-12}$           \\
Wolf-Rayet                  &    1 & 0.39$\times 10^{-12}$       &   5 & 0.67$\times 10^{-12}$        & 6    &  1.06$\times 10^{-12}$          \\
\hline
Sum             & 42 & 3.62$\times 10^{-12}$       & 185 & 9.97$\times 10^{-12}$       & 227 & 13.59$\times 10^{-12}$       \\
Sum + Faint & $\cdots$ & $\cdots$                  & 180 & 0.83$\times 10^{-12}$        & 407 & 14.42$\times 10^{-12}$      \\
\hline
\end{tabular}\par
\label{tab:fuv-census}
\end{table}

\subsection{Far-ultraviolet observations}\label{fuv}

The combination of VLT/MUSE \citep{2018A&A...614A.147C}, VLT/FLAMES \citep{2011A&A...530A.108E} and {\it HST}/STIS \citep{2016MNRAS.458..624C} spectroscopy provide near complete spectral type census of bright early-type stars in the Tarantula \citep{2018Sci...359...69S, 2019Galax...7...88C}. This census permits us to construct an integrated far-UV spectrum of the region within the MUSE mosaic discussed in Section~\ref{visual},
by combining empirical datasets for a subset of stars (those with far-UV spectroscopy) with suitable templates for the remainder (those lacking far-UV spectroscopy at $R\geq$ 2000). Empirical
{\it HST} COS or STIS spectroscopy is available for a subset of individual UV-bright sources, which are rebinned to $R\sim$2000. These are supplemented by Goddard High Resolution Spectroscopy (GHRS) G140L observations of the central 2$\times$2 arcsec$^{2}$ R136a cluster \citep{1992ESOC...44..347H}, also obtained at $R\sim$2000\footnote{The spectral resolution of STIS/G140L observations from \citet{2016MNRAS.458..624C} achieved $R\sim$1000, too low for our purposes here}. The cumulative far-UV spectrum was constructed used the Starlink spectroscopic package {\sc dipso} \citep{dipso}.

\begin{figure}
\includegraphics[width=0.5\columnwidth, bb = 20 205 515 667]{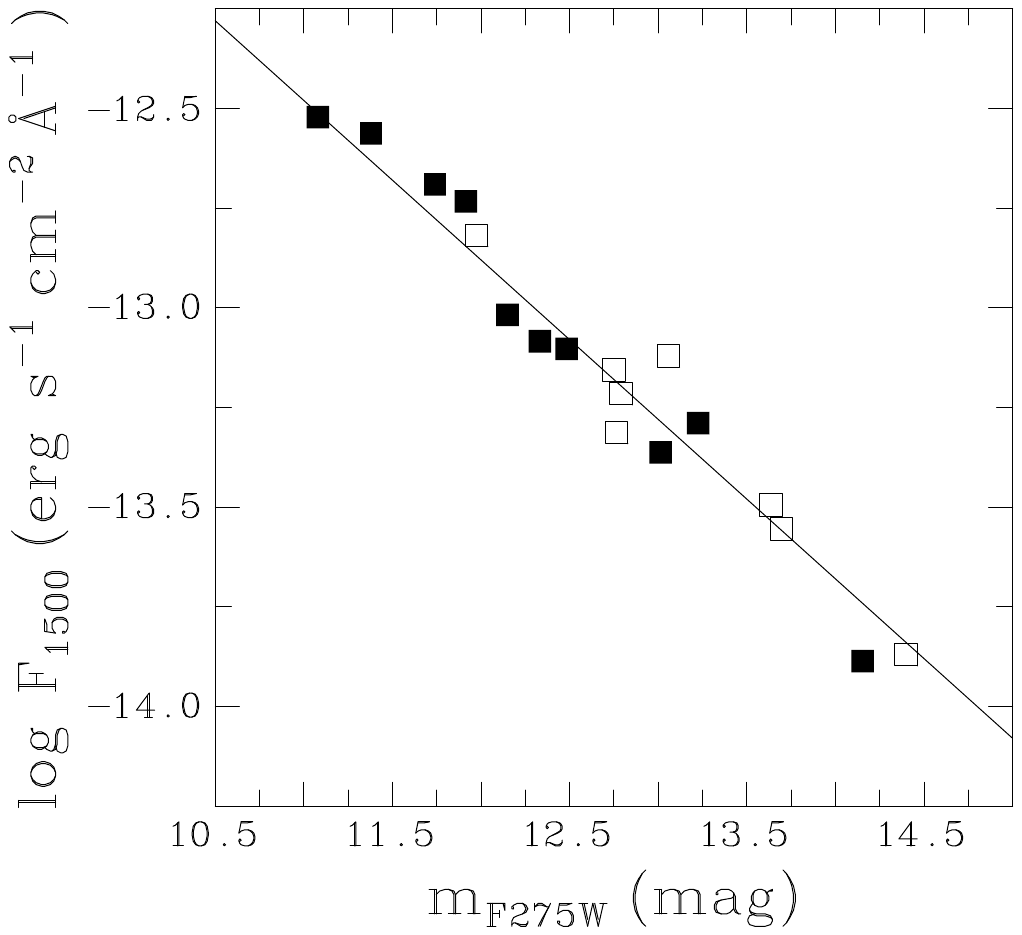}
\includegraphics[width=0.5\columnwidth, bb = 20 205 515 667]{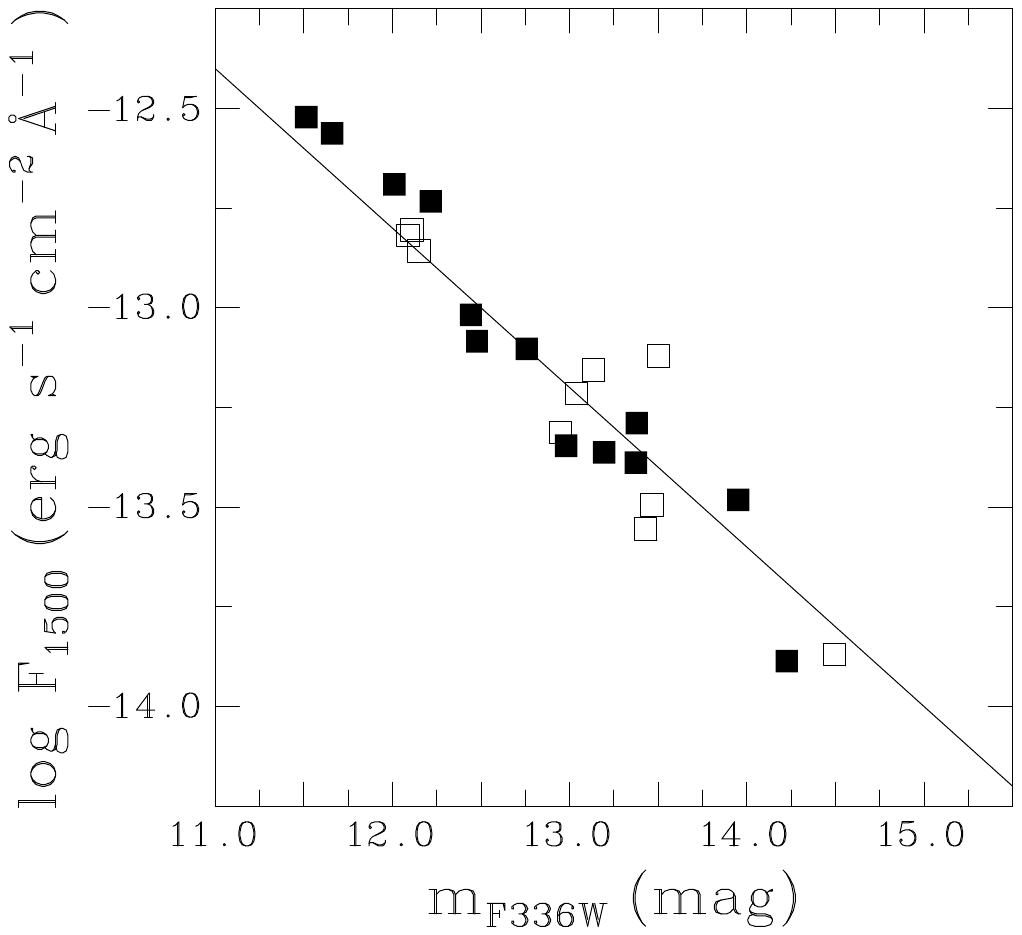}
\caption{Relationship between $m_{\rm F275W}$ (left) or $m_{\rm F336W}$ (right) and far-UV flux, $F_{\rm 1500}$, for O stars in 30 Doradus (solid within MUSE field) with far-UV spectroscopy and HST/WFC3 photometry \citep{2013AJ....146...53S}. The solid line is a linear fit to all observations.}
\label{F275W-F336W}
\end{figure}

For the majority of sources, far-UV spectroscopy of LMC OB and WR templates from ULLYSES\footnote{https://ullyses.stsci.edu/} \citep{2020RNAAS...4..205R} are utilised (up to DR6), supplemented by COS and STIS datasets from GO programmes (GO~15629, Mahy; GO~16272, Shenar).
Templates are anchored to estimates of 1500\AA\ fluxes determined from F275W or F336W photometry from HTTP \citep{2016ApJS..222...11S}. Since HTTP photometry of 30 Doradus is incomplete, we also utilise WFC3/F336W photometry from \citet{2011ApJ...739...27D} or WFPC2/F336W photometry from \citet{1995ApJ...448..179H}. We provide details of templates in Tables~\ref{tab:templates1}-\ref{tab:templates2}. Gaps in spectra arise from incomplete spectral coverage of templates (e.g. COS G130M+G160M).

We estimate $\lambda$1500 fluxes of O and WR stars lacking far-UV spectroscopy from a calibration of F275W photometry anchored by O stars within 30 Doradus for which far-UV spectroscopy is available, 
as shown in the left panel of Fig~\ref{F275W-F336W}, whose linear fit is 
\begin{equation} 
\log F_{1500} = -0.4 (m_{\rm F275W} + 20.20\pm0.08) 
\label{F275W}
\end{equation}
F275W photometry is not available throughout the MUSE field of view, whereas F336W is available for all sources, for which a  calibration is presented in the right panel of Fig.~\ref{F275W-F336W}, and a linear fit is
\begin{equation} 
\log F_{1500} = -0.4 (m_{\rm F336W} + 20.00\pm0.15).
\label{F336W}
\end{equation}
Fluxes  for individual stars in 30 Doradus used in the calibration are provided in Appendix~C (Table~\ref{calib}). 
For B supergiants, $\lambda$1500 fluxes are reduced by a scale factor of 0.85 (B0.5$\pm$0.5), 0.65 (B1--3) or 0.5 (B4--9) from a comparison between spherical non-LTE  model atmospheres of far- to near-UV fluxes of O and B stars \citep{1998ApJ...496..407H}. For B dwarfs, $\lambda$1500 fluxes are reduced by a factor of 0.77 (B0.5$\pm$0.5) or 0.65 (B1.5$\pm$0.5) from a comparison between TLUSTY plane parallel non-LTE model atmospheres of O and B stars \citep{2007ApJS..169...83L}. We also adjust the $\lambda\lambda$1160--1700 slopes of templates according to their $m_{\rm F336W} - m_{\rm F555W}$ colours as discussed in Appendix~\ref{FUV}

We incorporate a total of 227 sources with spectral types and estimated fluxes  $F_{1500} \geq 10^{-14}$ erg\,s$^{-1}$\,cm$^{-2}$\AA$^{-1}$ into our cumulative far-UV spectrum, which are listed in Appendix B (Table~\ref{tab:fuv-list}). Of these, 29 lie within the R136a 2$\times$2 arcsec$^{2}$ GHRS footprint,  13 possess HST COS (G130M+G160M or G140L) or STIS (E140M) spectroscopy, such that the remaining 185 require ULLYSES templates, several of which have known far-UV fluxes courtesy of HST/STIS G140L spectroscopy \citep{2005ApJ...627..477M}. Although R136a stars exterior to the GHRS footprint have been observed in the far-UV with HST/STIS G140L \citep{2016MNRAS.458..624C}, their slit loss corrections are uncertain, such that $F_{1500}$ from GHRS is $\sim$20\% higher than the sum of individual fluxes. Consequently we adopt $F_{1500}$ from calibrations for stars exterior to the GHRS aperture in common with \citep{2016MNRAS.458..624C}. 

HSH95 17, alias \#9 from \citet{2022ApJ...935..162K}, is included in Table~\ref{tab:fuv-list} despite an uncertain spectral type since it lies within the GHRS 2$\times$2 arcsec$^{2}$ footprint such that we do not require a bespoke UV template. Several other sources which would qualify on the basis of their far-UV fluxes are excluded due to unknown spectral types (e.g. HSH95 76, HSH95 87, SMB 136) which is necessary to incorporate suitable UV templates. Their contribution to the cumulative far-UV flux is negligible ($\sim$1\%).


20 sources in Table~\ref{tab:fuv-list} are confirmed spectroscopic binaries, although not all stars have been subject to spectroscopic monitoring, so the true binary frequency will be significantly higher.  Confirmed
multiple systems amongst the UV-brightest sources ($F_{1500} \geq 5 \times 10^{-14}$ erg\,s$^{-1}$\,cm$^{-2}$\AA$^{-1}$) include O-type binaries HSH95 39, HSH95 42 within R136a \citep{2002ApJ...565..982M}, 
Wolf-Rayet and O-type binaries R140b, c and d \citep{2019A&A...627A.151S, 2014A&A...564A..40W}, plus colliding wind binaries R136c, Mk~33Na, Mk 34 and Mk 39 \citep{2022MNRAS.515.4130C}.

The cumulative far-UV flux of the individual 227 sources is 1.36$\times 10^{-11}$ erg\,s$^{-1}$\,cm$^{-2}$\AA$^{-1}$. An additional 180 sources with known spectral types possess far-UV fluxes in the range $5 \times 10^{-16} \leq F_{1500} < 1.0 \times 10^{-14}$ erg\,s$^{-1}$\,cm$^{-2}$\AA$^{-1}$. We account for these collectively via a B0\,V template scaled to their cumulative far-UV flux ($8.35 \times 10^{-13}$
erg\,s$^{-1}$\,cm$^{-2}$\AA$^{-1}$), such that they contributes an additional $\sim$6\% to the total far-UV continuum. Table~\ref{tab:fuv-census} provides a spectral subtype of the sources contributing to the far-UV continuum. Stars possessing empirical far-UV spectroscopy contribute 25\% of the total. The dominant contribution arises from large numbers of O stars (65\%), although modest populations of very massive main-sequence WN (WNh) and transition Of/WN stars (12\%), B stars, primarily supergiants (10\%) and classical WR stars (7\%) also make non-negligible contributions. 

\begin{figure}
\includegraphics[angle=-90,width=\columnwidth,bb=13 46 553 757]{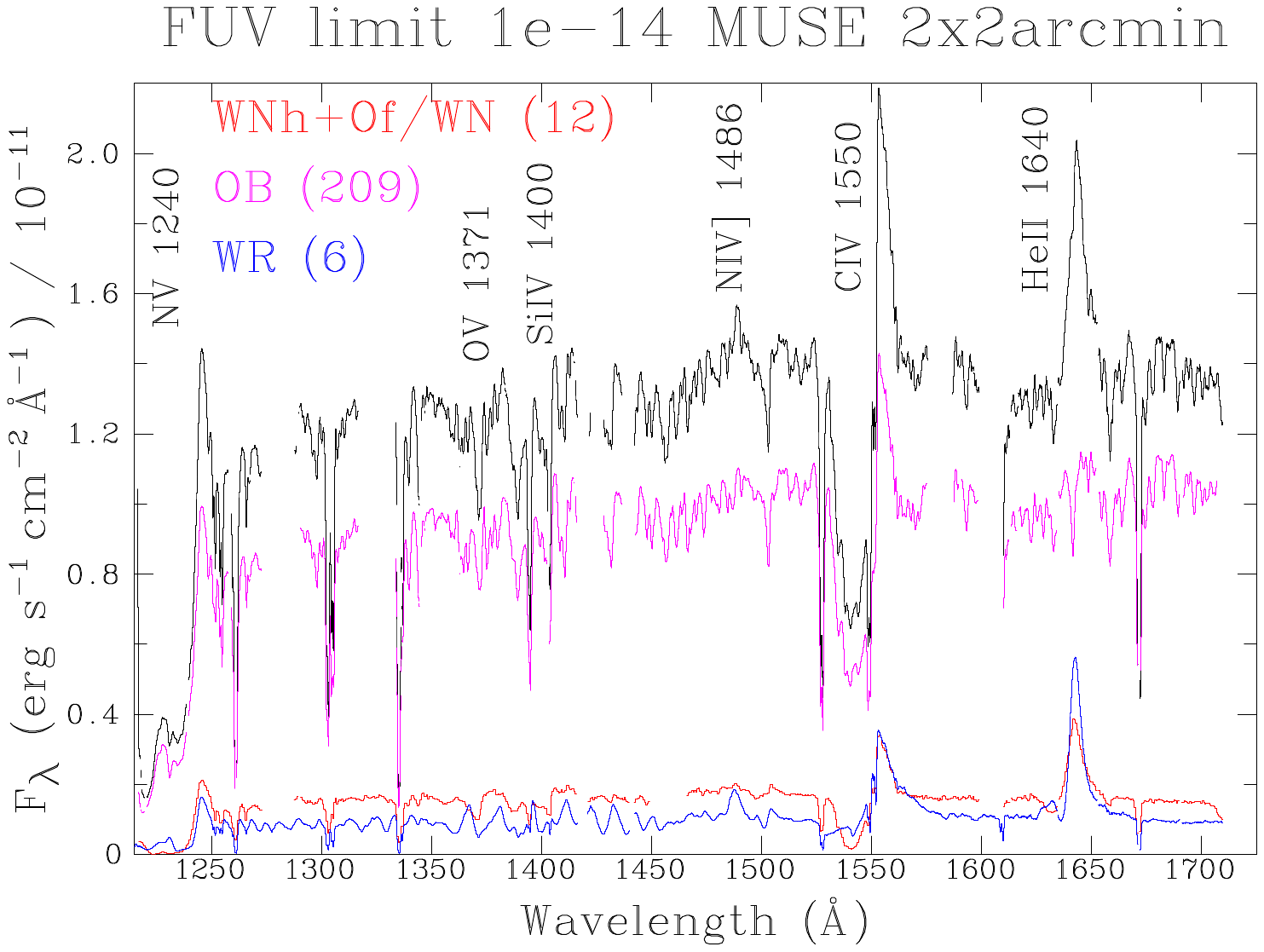}
\caption{Cumulative far-UV spectrum of the central region of NGC~2070 (black) inferred from a combination of empirical  (42 stars, 25\% of total), LMC templates (185 stars, 69\% of total) plus faint OB stars (180 stars, 6\% of total), highlighting contributions from OB stars (pink), Of/WN and WNh stars (red), and classical WR stars (blue). He\,{\sc ii} $\lambda$1640 emission is dominated by classical WR stars (56\%) and very massive main sequence stars (31\%), with the remainder arising from Of supergiants (e.g. Mk~42, R136a5). Gaps in spectra arise from incomplete spectral coverage of templates (e.g. COS G130M+G160M).}
\label{fig:far-UV}
\end{figure}

\begin{figure}
\includegraphics[angle=-90,width=\columnwidth,bb=13 46 553 757]{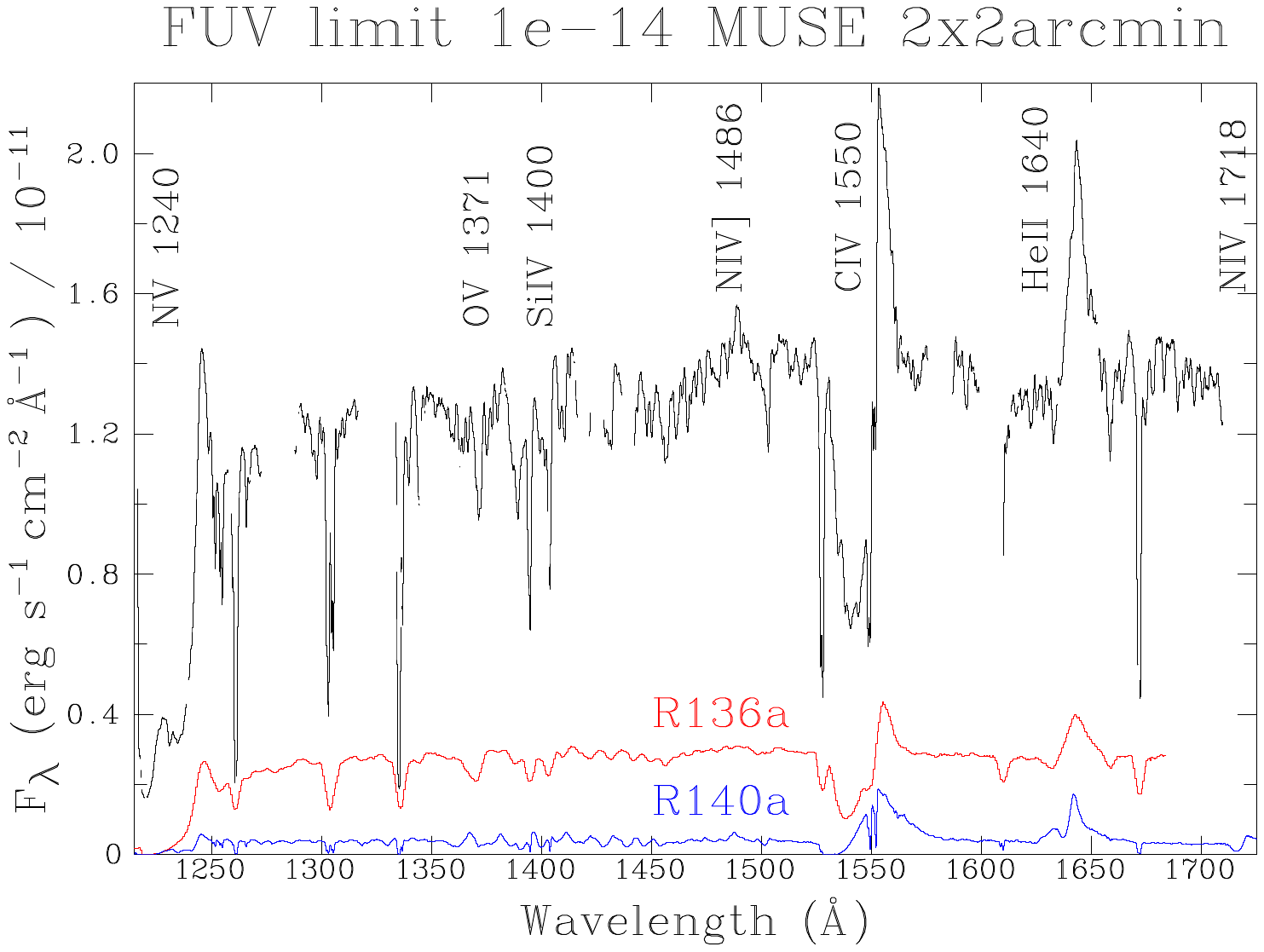}
\caption{Cumulative far-UV spectrum of the central region of NGC~2070 (black) together with the integrated STIS/G140L spectrum of the R136a cluster (red, 3.6$\times$3.6 arcsec$^2$) from \citet{2016MNRAS.458..624C} plus STIS/E140M spectrum of R140a (blue), the UV brightest source in the MUSE field, which is host to classical WN+WC stars. Gaps in spectra arise from incomplete spectral coverage of templates (e.g. COS G130M+G160M).}
\label{fig:far-UV-cluster}
\end{figure}


\begin{table}
\centering
\caption{Primary contributors to the integrated He\,{\sc ii} $\lambda$1640 emission line ($F_{1640}$ units of 10$^{-12}$ erg\,s$^{-1}$\,cm$^{-2}$) in the central 30$\times$30 pc$^{2}$ region of NGC~2070,
including the cumulative line flux from the central cluster R136a which is dominated by WN5h stars R136a1, a2 and a3 \citep{2016MNRAS.458..624C}. Details of templates are provided in Table~\ref{tab:templates2}.}
\begin{tabular}{l @{\hspace{1mm}} l @{\hspace{2mm}} r @{\hspace{2mm}} l @{\hspace{2mm}} r}
\hline
Source &    Sp Type & $F_{1640}$ & Obs & VMS \\
\hline
R136a  & 3$\times$WN5h+O4\,If/WN8+..  & 13.\phantom{0} & STIS/G140L & 8$\checkmark$ \\
R140a & WC4+WN6+..  & 12.\phantom{0}   & STIS/E140M  & $\cdots$ \\ 
R140b & WN5(h)+O           &   9.\phantom{0}   & WN6 template & $\cdots$ \\
R134   & WN6(h)                & 6.5    & WN6 template  & $\cdots$ \\
Mk~53  & WN8(h)                & 2.2    & WN7 template  & $\cdots$ \\
Mk~49 & WN6(h)                &  2.1     & WN6 template & $\cdots$   \\
Mk~33Sb & WC5             & 2.1     & WC4 template   & $\cdots$ \\
R136c   & WN5h+            & 1.8  & WN5h template & $\checkmark$ \\
Mk~34  & WN5h+WN5h    & 1.6  & WN5h template & 2$\checkmark$\\
Mk~39   & O2.5\,If/WN6+ & 1.1   & COS/G130M+G160M & $\checkmark$ \\
Mk~42     & O2\,If*             & 0.8   & STIS/E140M  & $\checkmark$ \\
Mk~35      & O2\,If/WN5    & 0.8   & O2If/WN5 template  & $\checkmark$ \\
Mk~37a    & O3.5\,If/WN7  & 0.5  & O3.5\,If/WN7 template & $\checkmark$ \\
\hline
Total      &           &  57.\phantom{0}    &  $\cdots$  & $\cdots$   \\
\hline
\end{tabular}\par
\label{tab:1640}
\end{table}

Fig.~\ref{fig:far-UV} presents the cumulative far-UV spectrum of the MUSE field, including contributions from OB-type (pink), classical WR (blue) stars plus main sequence WNh and Of/WN stars (red). We have split relative contributions from O and WNh stars to the core of R136a by adjusting their combined HST STIS/G140L spectra from \citep{2016MNRAS.458..624C} to the flux of the HST/GHRS spectrum. He\,{\sc ii} $\lambda$1640 is unusually strong ($W_{\lambda}$ = 4.7$\pm$0.2\AA) and broad (FWHM = 9.3$\pm$0.3\AA). Figs~\ref{fig:far-UV-O}--\ref{fig:far-UV-B} in Appendix~A illustrate the contribution of empirical datasets and templates to the far-UV spectra of individual spectral types.  All subtypes contribute to the strong C\,{\sc iv} $\lambda$1550 P Cygni profile, although it is apparent that He\,{\sc ii} $\lambda$1640 emission is dominated by classical WR stars (56\%) plus WNh and Of/WN stars (31\%). All subtypes contribute to N\,{\sc v} $\lambda$1240, aside from B supergiants, although B supergiants and classical WR stars are the primary contributors to P Cygni Si\,{\sc iv} $\lambda$1400. 

In Table~\ref{tab:fuv-list} we have flagged $\sim$18 Very Massive Stars (VMS) with initial masses in excess of $\sim 100 M_{\odot}$ on the basis of spectroscopic results \citep{2014A&A...570A..38B, 2019MNRAS.484.2692T, 2022A&A...663A..36B}. The majority of these have WN5h, Of/WN or O2--4\,If spectral types, exceptions include R136a7 (O3\,III(f*)), R136a4 (O3\,V) and VFTS 506 (ON2\,V). In spite of their scarcity, 18\% of the NGC~2070 far UV flux arises from VMS. 

\begin{figure}
\includegraphics[angle=-90,width=\columnwidth,bb=13 46 553 757]{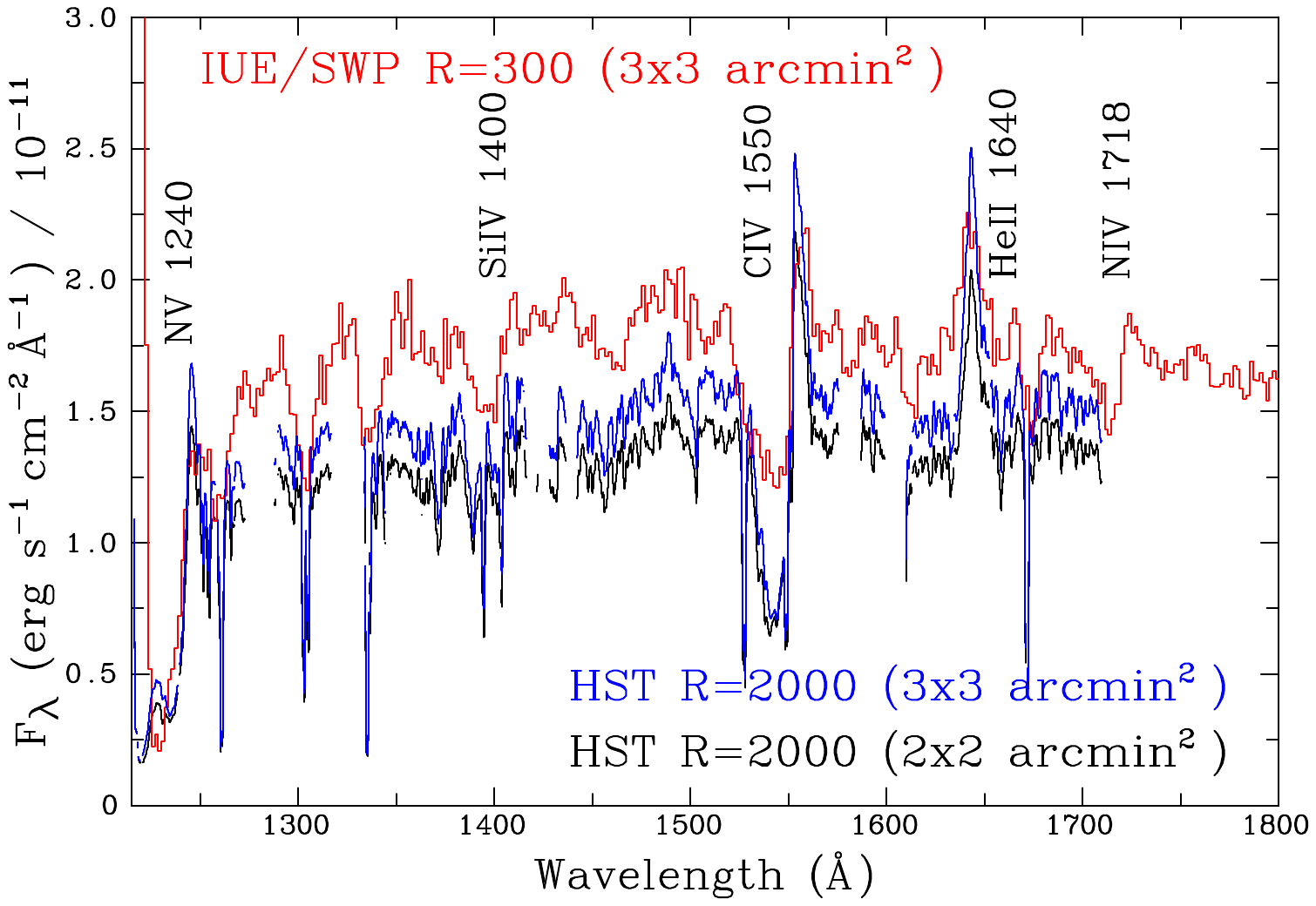}
\caption{Integrated {\it IUE}/SWP spectra of the central $3'\times3'$ (red) region of NGC~2070 from \citet{1995ApJ...444..647V} together with our cumulative far-UV spectrum for the central $2'\times2'$ (black)
and $3'\times3'$ (blue) regions. 
N\,{\sc iv} $\lambda$1718 is prominent in the large aperture {\it IUE} dataset despite its low ($R\sim300$) spectral resolution. {\it IUE} observations suggest 15\% higher far-UV flux levels owing to a combination of flux calibration differences or unresolved stars (primarily B-type) omitted from our study.}
\label{vacca}
\end{figure}

Fig.~\ref{fig:far-UV-cluster} compares the cumulative far-UV spectrum to the integrated STIS/G140L spectrum of R136a from \citet{2016MNRAS.458..624C}, contributing 22\% of the far-UV continuum flux, plus the STIS/E140M spectrum of R140a (VFTS 507), the brightest individual source in the MUSE field (3\% of far-UV continuum flux), host to classical WN+WC stars. Neither of the WR stars within R140a are known to be binaries \citep{2001MNRAS.324...18B, 2019A&A...627A.151S}, although the complete stellar content of R140a remains uncertain. R136a possesses an extremely strong He\,{\sc ii} $\lambda$1640 emission with respect to typical young star clusters, but the richness of the surrounding massive star population in NGC~2070 is such that it only contributes a quarter of the integrated emission line flux of He\,{\sc ii} $\lambda$1640.  Table~\ref{tab:1640} lists the primary contributors to $F_{1640}$, the majority of which originates from classical WR stars, notably R140a.


\citet{1995ApJ...444..647V} have previously scanned NGC~2070 with {\it IUE} with the short wavelength camera (SWP), at low spectral resolution ($R\sim300$) using the large $10''\times20''$ aperture, providing integrated
spectra for the central $20''\times20''$, $1'\times1'$, and $3'\times3'$ regions. Fig.~\ref{vacca} compares the integrated spectrum of the central $3'\times3'$ region (red) with our cumulative far-UV spectrum of the 2$'\times2'$ MUSE region (black). The overall shape of the spectra are similar, but the continuum flux of our far-UV spectrum is only $\sim$80\% of the large {\it IUE} aperture. N\,{\sc iv} $\lambda$1718 is prominent in the large aperture {\it IUE} dataset despite its low spectral resolution.

Table~\ref{tab:iue-list} lists additional stars with known spectral types brighter than $1.0 \times 10^{-14}$ erg\,s$^{-1}$\,cm$^{-2}$\AA$^{-1}$ within the large $3'\times3'$ {\it IUE} aperture. These increase the far-UV continuum flux by $1.7 \times 10^{-12}$ erg\,s$^{-1}$\,cm$^{-2}$\AA$^{-1}$, the majority of which is supplied by R135, R139 and R145. Owing to their relative isolation (Fig.~\ref{30dor-F658N-F336W}), far-UV flux levels for these three bright stars are drawn from low-resolution {\it IUE}/SWP large aperture spectroscopy \citep{1984ApJ...279..578F}. 

The blue spectrum in Fig.~\ref{vacca} additionally incorporates all stars listed in Table~\ref{tab:iue-list}, and reveals a difference of $\sim$15\% in global far-UV flux levels between {\it IUE} and our approach using {\it HST} observations. The difference likely arises from the combination of absolute flux calibration and the omission of a diffuse far-UV background from unresolved stars in our study -- recall 20\% of the integrated far-UV continuum of the rich R136 cluster was from an intra-cluster background \citep{2016MNRAS.458..624C}. Collectively, B-type main sequence stars -- lacking strong UV wind features -- will dominate this population.

\section{Analysis of integrated visual spectroscopy}\label{muse}

\subsection{Nebular properties}

As shown in Fig.~\ref{muse-sp}, the integrated MUSE spectrum of NGC~2070 is dominated by nebular emission lines, plus broad, weak Wolf-Rayet features. Here we undertake an analysis of the nebular spectrum with a focus on the inferred metallicity and age/mass of the ionizing stellar population using commonly used spectral synthesis tools Starburst99 \citep[version 7.0.1,][]{1999ApJS..123....3L, 2014ApJS..212...14L}, Charlot \& Bruzual \citep[CB19,[]{2003MNRAS.344.1000B, 2019MNRAS.490..978P} and 
BPASS \citep[v.2.2.1,][]{2017PASA...34...58E, 2018MNRAS.479...75S}.  A detailed study of the integrated 30 Doradus nebula has been undertaken by  \citet{2011ApJ...738...34P} while \citet{2003ApJ...584..735P} utilise VLT/UVES spectroscopy for a detailed chemical analysis.

\begin{table}
\centering
\caption{Fluxes ($F$), intensities ($I$) and luminosities ($L$) of nebular emission lines and Wolf-Rayet (WR) bumps in the integrated MUSE spectrum of the 2$\times$2 arcmin central region of NGC~2070. Case B recombination theory is adopted together with a standard LMC extinction law \citep{1983MNRAS.203..301H} and an adopted LMC distance of 50 kpc.} 
\begin{tabular}{l @{\hspace{-1mm}} r @{\hspace{1mm}} r @{\hspace{1mm}} r @{\hspace{2mm}} r}
\hline
Line &  \multicolumn{1}{c}{$F$} & \multicolumn{1}{c}{$I$} & \multicolumn{1}{c}{$L$} & Notes \\
         &   $10^{-11}$ erg\,s$^{-1}$\,cm$^{-2}$      & $10^{-10}$ erg\,s$^{-1}$\,cm$^{-2}$                    & $10^{38}$ erg\,s$^{-1}$                      & \\
\hline
N\,{\sc iii} 4634-41  & $0.71\pm0.05$        &  $0.42\pm0.04$   & $0.13\pm0.01$            & WR\\ 
{}[Fe\,{\sc iii}{}] 4658 & $0.31\pm0.02$        &  $0.12\pm0.01$ & $0.04\phantom{\pm0.01}$            & \\
He\,{\sc ii} 4686 & $2.78\pm0.06$        &   $1.18\pm0.03$    & $0.35\pm0.01$            & WR\\ 
{}[Ar\,{\sc iv}{}] 4711  & $0.32\pm0.02$        &  $0.14\pm0.01$    & $0.04\phantom{\pm0.01}$            & \\
{}[Ar\,{\sc iv}{}] 4740 & $0.10\pm0.01$        &   $0.04\pm0.01$   &  $0.01\phantom{\pm0.01}$                    &          \\
H$\beta$           &  $40.0\phantom{0}\pm0.1\phantom{0}$ & $15.9\phantom{0}\pm0.1\phantom{0}$ & $4.8\phantom{0}\pm0.1\phantom{0}$ \\
{}[O\,{\sc iii}{}] 4959 &  $69.7\phantom{0}\pm0.2\phantom{0}$ &  $26.8\phantom{0}\pm0.1\phantom{0}$ & $8.0\phantom{0}\pm0.1\phantom{0}$ \\
{}[Fe\,{\sc iii}{}] 4986 & $0.15\pm0.01$          & $0.06\pm0.01$ & $0.02\phantom{\pm0.01}$ & \\
{}[O\,{\sc iii}{}] 5007 & $213\phantom{.00}\pm1\phantom{.00}$ & $80.7\phantom{0}\pm0.4\phantom{0}$ & $24.1\phantom{0}\pm0.1\phantom{0}$ \\
He\,{\sc ii} 5412 & $0.07\pm0.01$ &   $0.02\phantom{\pm0.01}$  &     $0.01\phantom{\pm0.01}$           & WR \\ 
{}[Cl\,{\sc iii}{}] 5518 & $0.22\pm0.01$ &    $0.07\pm0.01$                   &     $0.02\phantom{\pm0.01}$ \\ 
{}[Cl\,{\sc iii}{}] 5538 & $0.17\pm0.01$ &     $0.06\pm0.01$                  &     $0.02\phantom{\pm0.01}$ \\ 
{}[N\,{\sc ii}{}] 5755 & $0.06\pm0.01$ &       $0.02\phantom{\pm0.01}$  &    $0.01\phantom{\pm0.01}$   \\ 
C\,{\sc iv} 5801--12 & $1.53\pm0.08$ & $0.47\pm0.02$ & $0.14\pm0.01$ & WR \\ 
He\,{\sc i} 5876 & $6.21\pm0.02$ & $1.87\pm0.01$ & $0.56\pm0.01$ & \\
{}[O\,{\sc i}{}] 6300  & $0.46\pm0.02$ & $0.13\pm0.01$ & $0.04\phantom{\pm0.01}$ \\
{}[S\,{\sc iii}{}] 6312  & $0.89\pm0.03$ & $0.24\pm0.01$ & $0.07\pm0.01$ \\
{}[N\,{\sc ii}{}] 6548 & $1.49\pm0.07$ & $0.39\pm0.02$ & $0.12\pm0.01$ \\
H$\alpha$        & $176\phantom{.00}\pm1\phantom{.00}$ & $45.7\phantom{0}\pm0.3\phantom{0}$ & $13.7\phantom{0}\pm0.1\phantom{0}$ \\
{}[N\,{\sc ii}{}] 6584 & $4.48\pm0.20$ & $1.16\pm0.05$ & $0.35\pm0.01$ \\
He\,{\sc i} 6678 & $2.10\pm0.01$ & $0.53\pm0.01$ & $0.16\pm0.01$ & \\
{}[S\,{\sc ii}{}] 6717 & $3.74\pm0.01$ & $0.94\pm0.01$  & $0.28\pm0.01$ \\
{}[S\,{\sc ii}{}] 6731 & $3.12\pm0.01$ & $0.78\pm0.01$  & $0.23\pm0.01$ \\ 
He\,{\sc i} 7065 & $1.88\pm0.01$ &$0.44\pm0.01$ & $0.13\pm0.01$ \\
{}[Ar\,{\sc iii}{}] 7135 & $7.33\pm0.01$ & $1.71\pm0.01$ & $0.51\pm0.01$ \\
{}[S\,{\sc iii}{}] 9069               & $21.3\phantom{0}\pm0.2\phantom{0}$ & $3.82\pm0.04$ & $1.14\pm0.01$ \\
\hline
\end{tabular}\par
\label{tab:muse-lines}
\end{table}





Table~\ref{tab:muse-lines} presents measured MUSE
line fluxes, intensities and luminosities (assuming 50 kpc), the former obtained using the {\tt elf} suite within the Starlink package {\sc dipso} \citep{dipso}.
Although 30 Doradus is known to have a non-standard dust extinction \citep{2014A&A...564A..63M, 2014MNRAS.445...93D, 2023A&A...673A.132B}, we shall adopt a standard LMC extinction law with $R_{\rm V}$ = 3.1 \citep{1983MNRAS.203..301H} for the nebular analysis. We obtain $c(H\beta)$=0.54
from $I$(H$\alpha$)/$I$(H$\beta$) = 2.86 for Case B recombination theory for a standard $N_{e} = 10^{2}$ cm$^{-3}$ and $T_{e} = 10^{4}$ K -- these results are consistent with averages of resolved maps from  \citet{2018A&A...614A.147C}. 

$L$(H$\alpha$) = 1.37$\times 10^{39}$ erg\,s$^{-1}$ equates to an ionizing output of $10^{51}$ ph\,s$^{-1}$, corresponding to a star formation rate of 0.005 $M_{\odot}$\,yr$^{-1}$ using standard conversions from \citet{1998ARA&A..36..189K}. Our MUSE dataset of the central region of NGC~2070 (30$\times$30 pc$^{2}$) lies within  the extended  30 Doradus region. \citet{2010ApJS..191..160P} have investigated a much larger 140$\times$80 pc$^{2}$ region, and highlighted that 50\% of the integrated H$\alpha$ emission originates from relatively low surface brightness regions.

We used the {\tt temden} routine in {\sc iraf} to determine the nebular density
and temperature. We obtain $N_{e} = 240 \pm 10$ cm$^{-3}$  from the standard [S\,{\sc ii}] 6717/6731 diagnostic, and $N_{e} = 310^{+220}_{-250}$  cm$^{-3}$  from the weak [Cl\,{\sc iii}] 5518/5538 diagnostic. Comparable results are obtained from the [Fe\,{\sc iii}] 4658/4986 ratio \citep{2001PNAS...98.9476K}. The usual [O\,{\sc iii}] 4363/5007 temperature diagnostic is not available for MUSE observations at $z$ = 0 , so we use [N\,{\sc ii}] 5755/6584 to obtain $T_{e} = 10,400^{+900}_{-700}$  K and [S\,{\sc iii}] 6312/9069 to obtain $T_{e} = 8300^{+70}_{-50}$  K. For reference, \citet{2003ApJ...584..735P} obtained $N_{e} = 415\pm35~{\rm cm}^{-3}$ ([S\,{\sc ii}]) and $T_{e} = 10\,800\pm300$ K ([N\,{\sc ii}]). 

We are unable to  determine the oxygen abundance from the use of auroral $T_{e}$ diagnostics ($\lambda$4363 [O\,{\sc iii}] is not available to MUSE at zero redshift). Instead, we rely on strong line calibrations \citep{2019A&ARv..27....3M}, which are widely used for extragalactic H\,{\sc ii} regions for which direct temperature determinations are not feasible. The calibration of the N2 index (--1.59) by \citet{2013A&A...559A.114M} implies $12 + \log(O/H) = 8.01 \pm 0.32$, versus $7.90\pm0.16$ following \citet{2017MNRAS.465.1384C}. The 
O3N2 index (+2.3) lies beyond the calibrated range of \citet{2013A&A...559A.114M}, with $12 + \log(O/H) = 8.10 \pm 0.21$ obtained from  \citet{2017MNRAS.465.1384C}, values typical of the present-day Small Magellanic Cloud (SMC) rather than the LMC \citep{1990ApJS...74...93R, 1978MNRAS.184..569P}. Use of the integrated 30 Dor nebular fluxes from \citet{2010ApJS..191..160P} provides N2 and O3N2 indices of --1.42 and 2.09, indicating  $12 + \log(O/H) = 8.10 \pm 0.16$ or $8.19 \pm 0.21$, respectively, following the \citet{2017MNRAS.465.1384C} calibrations. 

Contemporary baseline LMC abundances \citep[][Table~1]{2023A&A...675A.154V} including $12 + \log(O/H) = 8.36$  correspond to a mass fraction of Z=0.008, although the directly determined gas phase oxygen abundance within 30 Doradus ranges from $12 + \log(O/H) $ = 8.26 \citep{2002A&A...391.1081V}  to  8.5 \citep{2003ApJ...584..735P}.  It is clear that standard strong line calibrations significantly underestimate the true oxygen content of NGC~2070, since these are sensitive to  ionization parameter as well as abundance \citep{2002ApJS..142...35K}. Consequently caution should be  used when inferring metallicities of low-redshift star forming galaxies
from diagnostics within the MUSE spectral range \citep[see][]{2023arXiv231103514E}. By way of a test, spectral synthesis calculations are made for both Z=0.008 (well matched to NGC~2070) and Z=0.002 (significantly lower than NGC~2070, albeit inferred from strong line methods).


\subsection{Population synthesis}\label{PopSyn}

In order to estimate the age and mass of the starburst region consistent with MUSE observations, we have employed several widely used population synthesis packages. Age estimates from H$\alpha$ are relatively metallicity insensitive, but WR bumps and UV diagnostics are strongly metallicity-dependent. Starburst99 \citep[v7.0.1,][]{1999ApJS..123....3L, 2014ApJS..212...14L}  involves a $1\times 10^{5} M_{\odot}$ burst of star formation, an  initial mass function (IMF) from \citet{2008ASPC..390....3K} with an upper mass limit of 120 $M_{\odot}$. Modern non-rotating and rotating Geneva evolutionary models at Solar \citep[Z=0.014][]{2012A&A...537A.146E} and SMC \citep[Z=0.002][]{2013A&A...558A.103G} metallicities are available\footnote{Z=0.008 models appropriate for the LMC have only recently become available \citep{2021A&A...652A.137E}}, with the full range of metallicities for historical models available with either standard \citep{1992A&AS...96..269S} or enhanced mass-loss \citep{1994A&AS..103...97M}. 

Starburst99 model atmospheres are obtained from either WM-Basic \citep[O,][]{2001A&A...375..161P}  or PoWR \citep[WR,][]{2002A&A...387..244G}, with ionizing fluxes from
\citet{2002MNRAS.337.1309S} and \citet{2010ApJS..189..309L} and empirical UV spectral templates drawn from either the Milky Way or Magellanic Clouds \citep{1993ApJ...418..749R}.

Observed  rotational velocities  \citep[e.g.][]{2017A&A...600A..81R, 2017A&A...601A..79S} are intermediate between non-rotating and rapidly rotating models (40\% critical), and evidence points to a higher mass limit than 120 $M_{\odot}$  \citep{2010MNRAS.408..731C, 2022A&A...663A..36B}.

We also make use of the latest CB19 grid of  Charlot \& Bruzual models \citep{2003MNRAS.344.1000B, 2019MNRAS.490..978P} which employ non-rotating PARSEC stellar evolutionary tracks from \citet{2015MNRAS.452.1068C} calculated for a range of metallicities, synthetic hot luminous star templates from WM-Basic, TLUSTY (OB), PoWR (WR) \citep{2022ApJS..262...36S} with integrated populations drawn from either a Salpeter, Kroupa or Chabrier IMF for stars up to $M_{\rm up}$ = 100, 300 or 600 $M_{\odot}$ \citep{2022ApJS..262...36S}.

Crucially, binary evolution is neglected in both Starburst99 and Charlot \& Bruzual models, despite observational evidence indicating a high fraction of close binaries amongst massive stars at the LMC metallicity \citep{2013A&A...550A.107S}. Therefore we also consider BPASS \citep[v2.2.1,][]{2017PASA...34...58E, 2018MNRAS.479...75S} models incorporating binary evolution, a Kroupa IMF and an upper mass limit of $M_{\rm up}$ = 100 or 300 $M_{\odot}$. Stellar atmosphere models  incorporated into BPASS that are relevant for young populations are WM-Basic (O), PoWR (WR) and ATLAS (B).

\begin{figure}
\includegraphics[width=\columnwidth,bb=46 23 541 782]{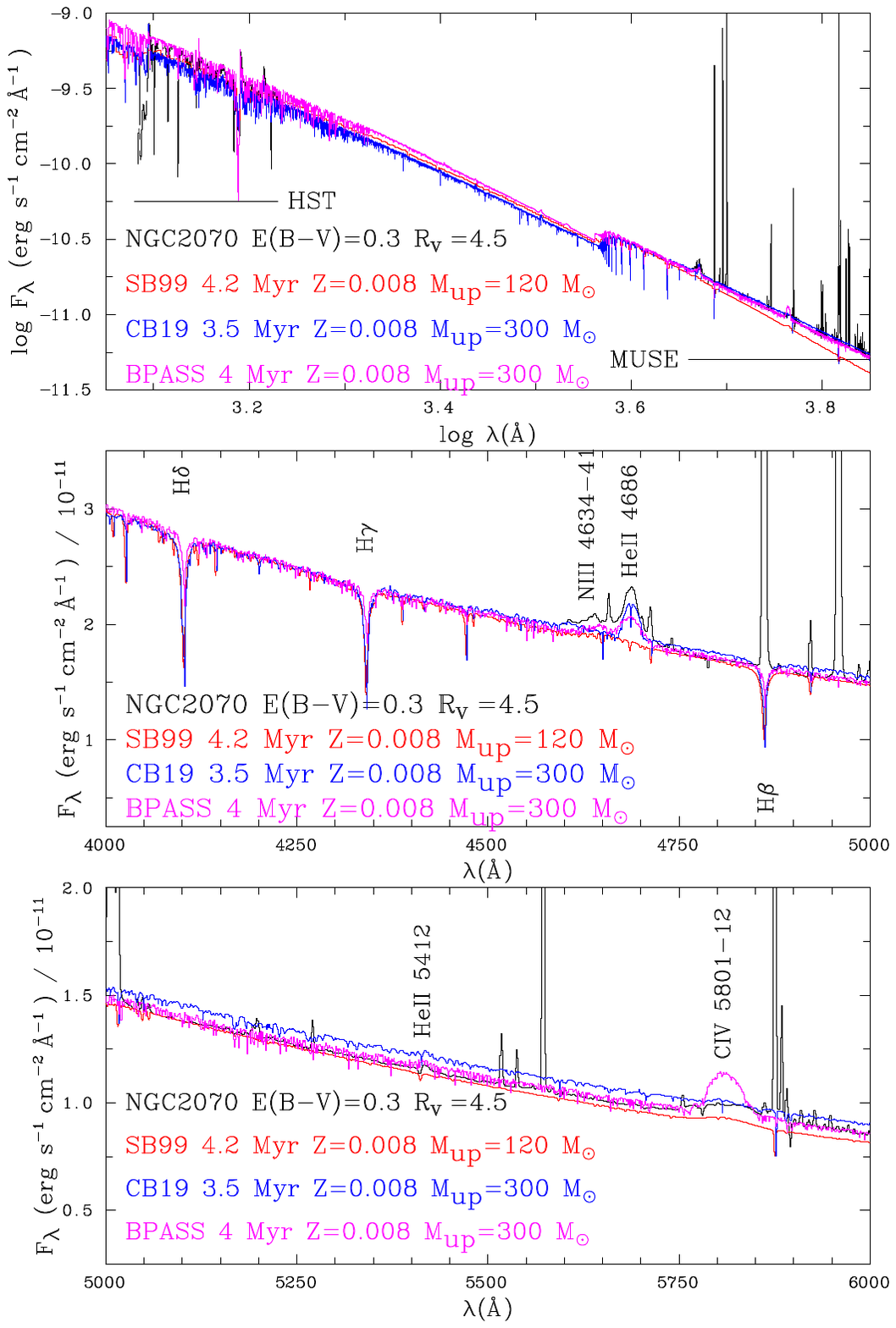}
\caption{Top panel: Comparison between dereddened ($E_{\rm B-V}$ = 0.3, $R_{\rm V}$ = 4.5) integrated UV and MUSE spectrum of the central region of NGC~2070 (black) and the predicted 4.2~Myr Starburst99 spectra (red) based on Z=0.008 metallicity evolutionary models extending to $M_{\rm up}$ = 120 $M_{\odot}$ \citep{1994A&AS..103...97M}, plus predicted 3.5~Myr CB19 spectra (blue) based on Z=0.008 metallicity evolutionary predictions \citep{2015MNRAS.452.1068C} extending to $M_{\rm up}$ = 300 $M_{\odot}$ and predicted 4~Myr BPASS spectra (pink) also based on Z=0.008 predictions \citep{2017PASA...34...58E} for single stars extending to $M_{\rm up}$ = 300 $M_{\odot}$. Central panel:  Zoom including blue WR bump (He\,{\sc ii} $\lambda$4686 and N\,{\sc iii} $\lambda\lambda$4634-41. A weak bump is predicted in the Starburst99 model, with improved agreement obtained from CB19 and BPASS models (similar WR emission is predicted for $M_{\rm up}$ = 100 $M_{\odot}$). Bottom panel: Zoom including yellow WR bump
(C\,{\sc iv} $\lambda\lambda$5801-12). The 4.2~Myr Starburst99 model is a good match to the yellow bump, whereas this feature is too weak in CB19 and too strong in BPASS. Weak stellar
He\,{\sc ii} $\lambda$5412 is also observed in the MUSE dataset. Emission lines not explicitly labelled are nebular (H$\beta$, [O\,{\sc iii}] $\lambda$4959, 5007, He\,{\sc i} $\lambda$5876).}
\label{NGC2070-sb99-hires}
\end{figure}

The H$\alpha$ equivalent width is a sensitive indicator of age for a young star burst. We measure $W_{\lambda}$ (H$\alpha) = 692\pm24$~\AA. We initially focus on Z=0.002 predictions from Starburst99 on the basis of
the strong line calibrations, so adopt contemporary non-rotating models from \citet{2013A&A...558A.103G}.  We obtain an age of 4.2 Myr, such that the H$\alpha$ luminosity (Table~\ref{tab:muse-lines}) corresponds to a stellar mass of $5.5\times 10^{4} M_{\odot}$. Broadly similar results are obtained for historical Z=0.008 models \citep{1994A&AS..103...97M} aside from a higher
mass of $8.5\times 10^{4} M_{\odot}$. Comparable results are obtained from the H$\beta$ equivalent width, $W_{\lambda}$ (H$\beta) = 97\pm1$\AA. Revised ages accounting for underlying absorption in H$\alpha-\beta$ are barely affected owing to the exceptionally strong Balmer emission. 

In reality the stellar population within NGC~2070 is not coeval - see \citep[][their fig.~4]{2018A&A...618A..73S} and \citep[][their fig.~10]{2021A&A...648A..65C}.
In particular the ages of OB stars within 1.2 arcmin of R136 (broadly comparable to MUSE pointing) span 1--7 Myr from comparison with evolutionary predictions for LMC metallicity OB stars \citep{2011A&A...530A.115B}, excluding the R136 star cluster whose age is $\sim$1.5 Myr \citep{2016MNRAS.458..624C, 2022A&A...663A..36B}. Nevertheless, the median age of massive stars in NGC~2070 \citep[3.6 Myr,][]{2018A&A...618A..73S} is in good agreement with the age inferred from H$\alpha$ using Z=0.002 or Z=0.008 metallicity models.

\subsection{WR properties}

The spectral resolution of MUSE prevents the detection of photospheric absorption lines, but the high S/N does reveal broad WR bumps (Fig.~\ref{muse-sp}). The blue bump is dominated by broad He\,{\sc ii} $\lambda$4686 (FWHM $\sim$ 25\AA, $W_{\lambda} \sim 6$\AA), with no nebular component detected. Broad N\,{\sc iii} $\lambda\lambda$4634-41 is also observed, suggesting a dominant late WN population, while the yellow WR bump is dominated by C\,{\sc iv} $\lambda\lambda$5801-12 (FWHM$\sim$80\AA, $W_{\lambda} \sim 4$\AA) from WC stars. The high S/N of our MUSE integrated dataset unusually also permits broad He\,{\sc ii} 5411 emission to be detected. Optical emission line fluxes from individual stars are provided in Table 2 of \citet{2018A&A...614A.147C}. 

Both classical WR stars and main sequence very massive stars (WN5h and Of/WN) 
are major contributors to He\,{\sc ii} $\lambda$4686, with R136a contributing $\sim$20\% of the total,  whereas the blend of N\,{\sc iii} $\lambda\lambda$4634--41 and C\,{\sc iii} $\lambda\lambda$4647--51 and C\,{\sc iv} $\lambda$5801--20 are dominated by classical WR stars (primarily R140a). The presence of WR features in Starburst99, CB19 and BPASS synthetic spectroscopy provide independent age indicators \citep{1998ApJ...497..618S}. We require a non-standard extinction law in order to reconcile the UV and optical spectrophotometry of NGC~2070 with predictions. We adopt an LMC law with $R_{\rm V}$ = 4.5 \citep{2014A&A...564A..63M, 2014MNRAS.445...93D, 2023A&A...673A.132B}, and obtain $E_{\rm B-V}$ = 0.3 mag ($A_{\rm V}$ = 1.35 mag) for spectral energy distribution comparisons in the top panel
of Fig.~\ref{NGC2070-sb99-hires}. Masses inferred from Starburst99, CB19 and BPASS stellar continua are $7.7\times 10^{4} M_{\odot}$, $6.5\times 10^{4} M_{\odot}$ and $11 
\times 10^{4} M_{\odot}$, respectively, in reasonable agreement with $L$(H$\alpha$) derived masses. The nebular continuum is not included in Fig.~\ref{NGC2070-sb99-hires} since it has a negligible contribution to the
far-UV and Paschen continua.

Z=0.002 metallicity models predict extremely weak WR emission for all population synthesis models. Consequently we adopt more realistic Z=0.008 models for age determinations based on WR stars, although
different models have different time resolutions (e.g. 1 Myr intervals for BPASS). The peak strength of optical WR bumps occurs at 4.2~Myr, 3~Myr and 4~Myr for Starburst99, CP19 and BPASS models. Starburst99 
models at solar metallicity \citep[Z=0.014,][]{2012A&A...537A.146E}. allow comparisons between peak WR ages associated with non-rotating (4.5$\pm$0.5 Myr) and rotating (5$\pm$2 Myr) models.

We compare the Starburst99 synthetic spectrum associated with the 4.2~Myr age corresponding to the maximum WR population from the \citet{1994A&AS..103...97M} evolutionary models in the central and bottom panels of Fig.~\ref{NGC2070-sb99-hires}, which coincide with the H$\alpha$-inferred age. Both WR bumps are predicted, albeit significantly weaker than observed for the He\,{\sc ii} $\lambda$4686 + N\,{\sc iii} $\lambda\lambda$4634-41 bump. It is clear that either the number of WR stars predicted from single star models at low metallicity or their line luminosities, or both, are underestimated (stronger
emission is predicted for solar metallicity models). The synthetic spectrum also highlights strong photospheric absorption lines associated with Balmer lines.

In contrast, the 3.5~Myr CB19 synthetic spectrum (blue) based on PARSEC evolutionary models at Z=0.008 with $M_{\rm up} = 300 M_{\odot}$ \citep{2015MNRAS.452.1068C} provides a stronger He\,{\sc ii} $\lambda$4686 emission in the central panel of Fig.~\ref{NGC2070-sb99-hires}, albeit too weak and with negligible N\,{\sc iii} $\lambda\lambda$4634-41 and C\,{\sc iv} $\lambda\lambda$5801-12 emission in lower panel. This age is close to the median 3.6 Myr age of massive stars in NGC~2070 according to \citet{2018A&A...618A..73S}. Comparable WR emission is predicted at this age for CB19 models with $M_{\rm up} = 100 M_{\odot}$, since very massive stars have somewhat shorter lifetimes. 

For the BPASS single star models with  $M_{\rm up} = 300 M_{\odot}$,  WR emission is most prominent at 4 Myr, with comparable C\,{\sc iv} $\lambda\lambda$5801-12 emission predicted (too strong), and a higher He\,{\sc ii} $\lambda$4686 equivalent width, albeit comparable to predictions from CB19 models (central and bottom panels of Fig.~\ref{NGC2070-sb99-hires}). The H$\alpha$ luminosity at this age for the  $M_{\rm up} = 300 M_{\odot}$ model corresponds to a stellar mass of $\sim6 \times 10^{4} M_{\odot}$.


Alternatively, it is possible to empirically estimate WR populations if line luminosity calibrations are available \citep{1998ApJ...497..618S}. 
Assuming the blue bump is dominated by WN stars, one obtains a population of $\sim$20 WN6--8 stars (or 12 WN5--7h stars) based on the latest WR line luminosities at the LMC metallicity from \citet{2023MNRAS.521..585C}. The  yellow bump is dominated by WC stars, from which a population of 4 WC4--5 stars is obtained. In reality, WC stars also contribute 15\% to the blue bump via C\,{\sc iii} $\lambda$4650 and He\,{\sc ii} $\lambda$4686. Consequently the inferred number of WN stars is reduced to 17 WN6--8 stars (or 10 WN5--7h stars). These calibrations are within a factor of two of the known population (11 WN stars, 2 WC4 including individual components of multiple WR systems R140a and Mk~34).

\section{Analysis of integrated far-UV spectroscopy}\label{pop-syn}

In common with previous studies of young extragalactic stellar populations in the ultraviolet, we again utilise predictions of the emergent far-UV spectrum from the population synthesis tool Starburst99 \citep[version 7.0.1,][]{1999ApJS..123....3L, 2014ApJS..212...14L}, Charlot \& Bruzual \citep[CB19,][]{2003MNRAS.344.1000B, 2019MNRAS.490..978P} and BPASS \citep[v2.2.1,][]{2017PASA...34...58E, 2018MNRAS.479...75S} once our empirical dataset has been dereddened according to an LMC extinction law with $R_{\rm V}$ = 4.5 and $E_{\rm B-V}$ = 0.3 mag, as for the optical comparison. The $\lambda$1500 luminosity of the NGC~2070 empirical dataset corresponds to $\log L_{\rm FUV}  = 38.04$ erg\,s$^{-1}$. 

Strong interstellar Ly$\alpha$ absorption impacts the observed N\,{\sc v} $\lambda\lambda$1238-42 profile \citep[e.g.][their fig.~3]{2023MNRAS.523.3949W}. We have estimated $\log N$(H\,{\sc i})/cm$^{2}$ = 21.75$\pm$0.05 from Ly$\alpha$ fits to COS/G130M and STIS/E140M observations of individual stars within the MUSE field-of-view (Table~\ref{calib}). However, application to theoretical predictions leads to a suppressed continuum extending to $\sim$1300\AA, contrary to observations. This is because LMC templates, which dominate the cumulative far-UV spectrum, span a broad range of $N$(H\,{\sc i}) column densities \citep{2012ApJ...745..173W}. Consequently no corrections to Ly$\alpha$ have been applied, of relevance to detailed comparisons between observed and predicted N\,{\sc v} $\lambda\lambda$1238-42 profiles.

Once again, although the optical strong line calibrations favour an unphysical SMC-like metallicity, we discuss both Z=0.002 and Z=0.008 metallicity predictions in the far-UV in order to test predictions. 
We limit our discussion to single bursts, although in reality NGC~2070 and other far-UV observations involve a mixed population. Dual or multiple age populations are occasionally implemented for spectral fitting \citep[e.g.][]{2019ApJ...882..182C, 2022AJ....164..208S}, but NGC~2070 comprises a star cluster embedded within an extended (spatially and temporally) star forming region.

\begin{figure}
\includegraphics[angle=-90,width=0.9\columnwidth,bb=55 67 553 757]{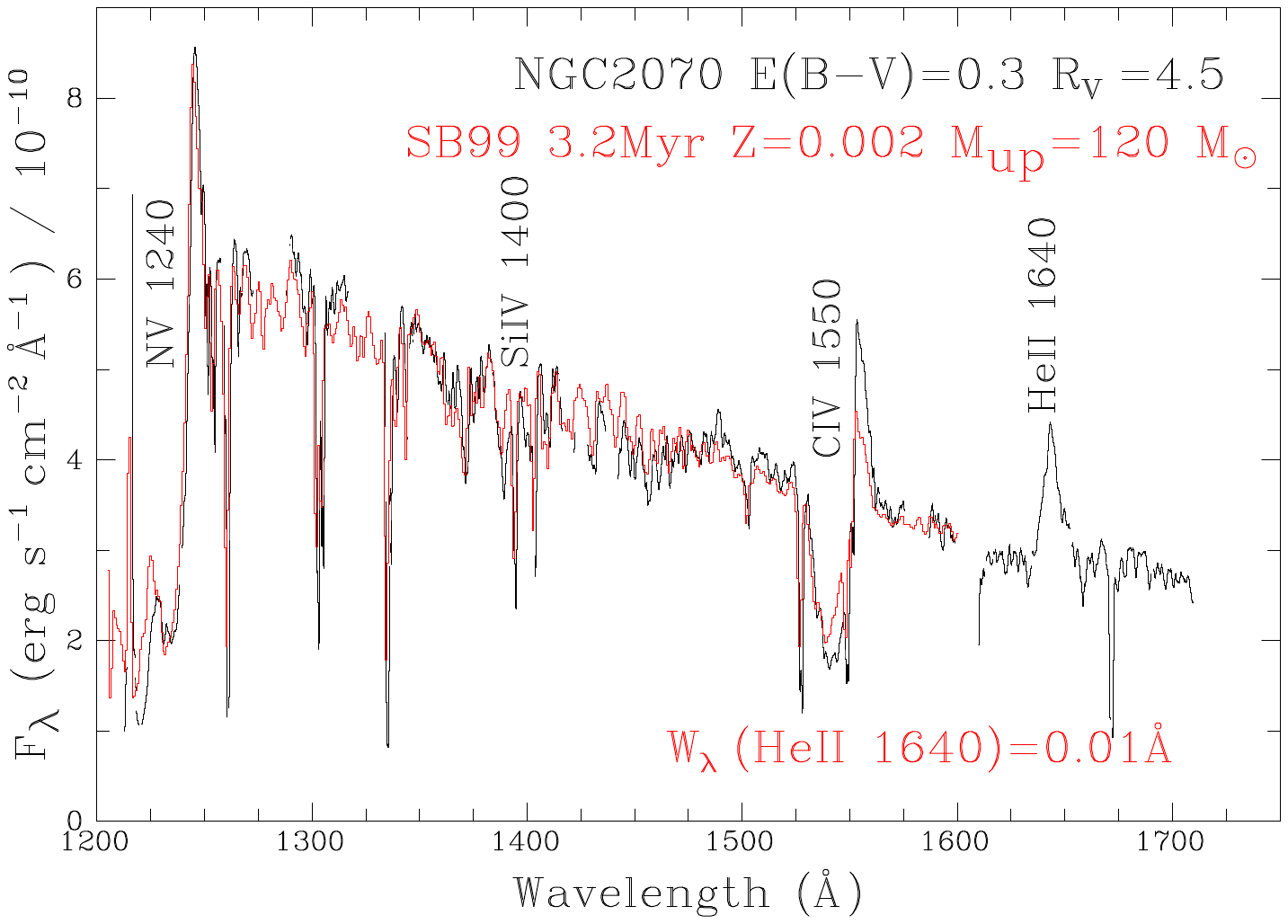}
\caption{Comparison between dereddened ($E_{\rm B-V}$ = 0.3, $R_{\rm V}$ = 4.5) cumulative far-UV spectrum of the central region of NGC~2070 (black) and the predicted Starburst99 spectrum based on Z=0.002 evolutionary models \citep{2013A&A...558A.103G} with LMC/SMC UV templates at 3.2 Myr (red). Far-UV luminosities for this age correspond to stellar masses of $5 \times 10^{4} M_{\odot}$. The cutoff at $\lambda$1600 is due to the use of empirical templates (predicted He\,{\sc ii} $\lambda$1640 emission equivalent width is indicated)}
\label{NGC2070-sb99-z002}
\end{figure}

\begin{figure}
\includegraphics[angle=-90,width=0.9\columnwidth,bb=55 67 553 757]{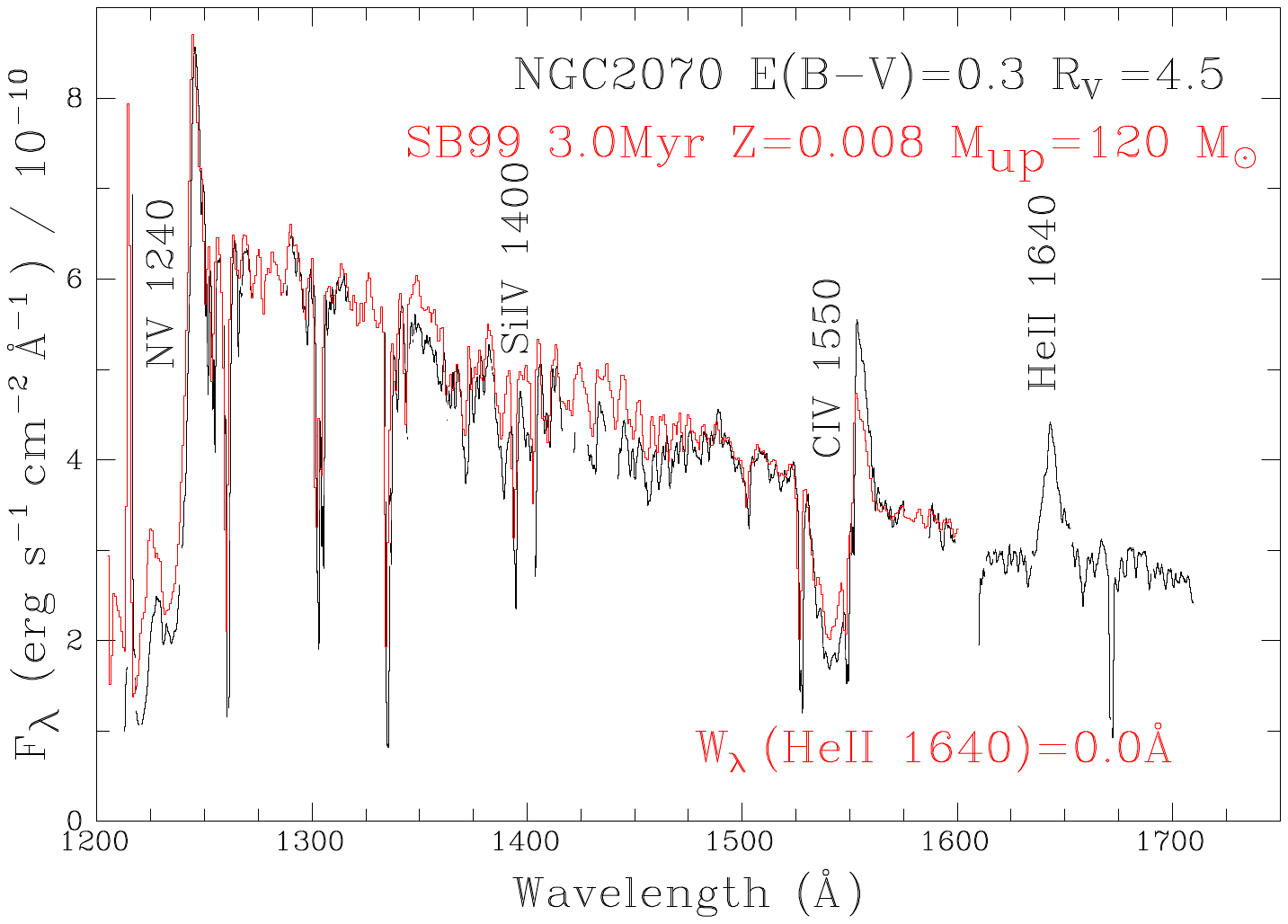}
\includegraphics[angle=-90,width=0.9\columnwidth,bb=55 67 553 757]{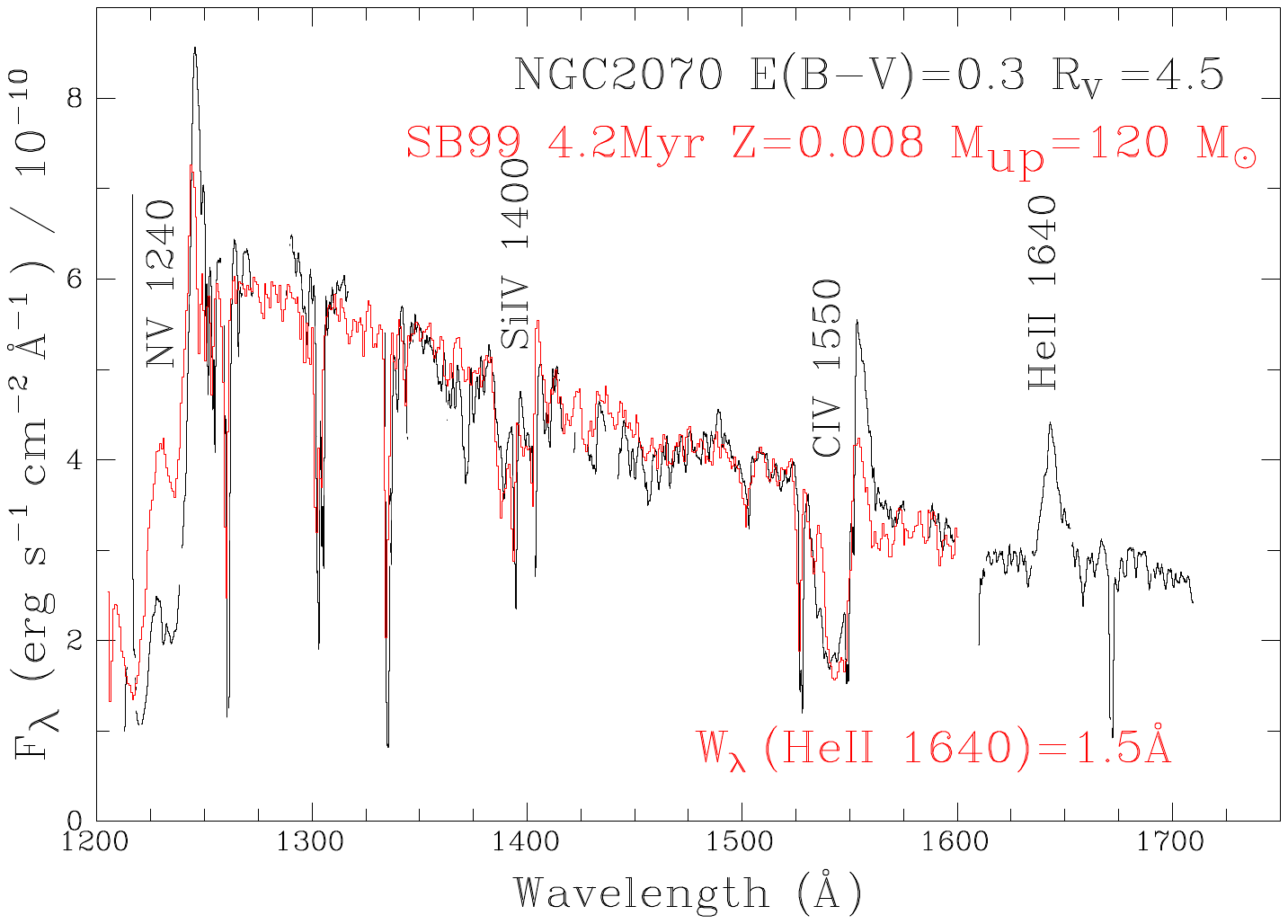}
\caption{Comparison between dereddened ($E_{\rm B-V}$ = 0.3, $R_{\rm V}$ = 4.5) cumulative far-UV spectrum of the central region of NGC~2070 (black) and the predicted Starburst99 spectrum (red) based on Z=0.008 metallicity evolutionary models \citep{1994A&AS..103...97M} at 3.0 Myr (upper panel) and 4.2 Myr (lower panel) with LMC/SMC UV templates. Far-UV luminosities for this range of ages correspond to stellar masses of $5.5-8.5 \times 10^{4} M_{\odot}$. The cutoff at $\lambda$1600 is due to the use of empirical templates (predicted He\,{\sc ii} $\lambda$1640 emission equivalent width is indicated).}
\label{NGC2070-sb99}
\end{figure}

\subsection{Starburst99}

 The primary age indicators in the far-UV are N\,{\sc v} $\lambda\lambda$1238--42, Si\,{\sc iv} $\lambda\lambda$1393--1402 and C\,{\sc iv} $\lambda\lambda$1548--51 for metal-poor populations with Starburst99 since empirical LMC/SMC stellar libraries cutoff at $\lambda$1600, preventing spectral comparisons for He\,{\sc ii} $\lambda$1640, although predicted equivalent widths of several WR lines are available, including He\,{\sc ii} $\lambda$1640. The maximum He\,{\sc ii} $\lambda$1640 emission occurs at an age of 3.2 Myr for the Z=0.002 \citep{2013A&A...558A.103G} Starburst99 model. Fig.~\ref{NGC2070-sb99-z002} provides a comparison between this model and dereddened NGC~2070 spectrum, revealing a good match to N\,{\sc v} $\lambda\lambda$1238--42, a weak P Cygni C\,{\sc iv} $\lambda\lambda$1548--51 profile, and a somewhat too weak Si\,{\sc iv} $\lambda\lambda$1393--1402 wind signature. He\,{\sc ii} $\lambda$1640 represents a major discrepancy, since the predicted emission of $W_{\lambda}$ = 0.01\AA, is negligible with respect to the observed  strength ($W_{\lambda}$ = 4.7$\pm$0.2\AA).
 
In view of the overall poor match to observations at Z=0.002, in Fig.~\ref{NGC2070-sb99} we compare the dereddened
NGC~2070 spectrum to predictions for non-rotating Z=0.008 metallicity models \citep{1994A&AS..103...97M} plus LMC/SMC empirical templates at 3~Myr (start of WR emission) and 4.2~Myr (peak WR emission). At 3 Myr
results are broadly similar to the Z=0.002 case, whereas at 4.2 Myr (H$\alpha$-inferred age from Section~\ref{PopSyn}) the P Cygni Si\,{\sc iv} $\lambda\lambda$1393--1402 is well reproduced owing to OB supergiants being present in sizeable numbers, but the strength of both  the N\,{\sc v} $\lambda\lambda$1238--42 and C\,{\sc iv} $\lambda\lambda$1548--51 P Cygni profiles are underpredicted. The He\,{\sc ii} $\lambda$1640 emission strength remains underestimated, albeit with an improved predicted equivalent width of 1.5\AA.

\begin{figure}
\includegraphics[angle=-90,width=0.9\columnwidth,bb=55 67 553 757]{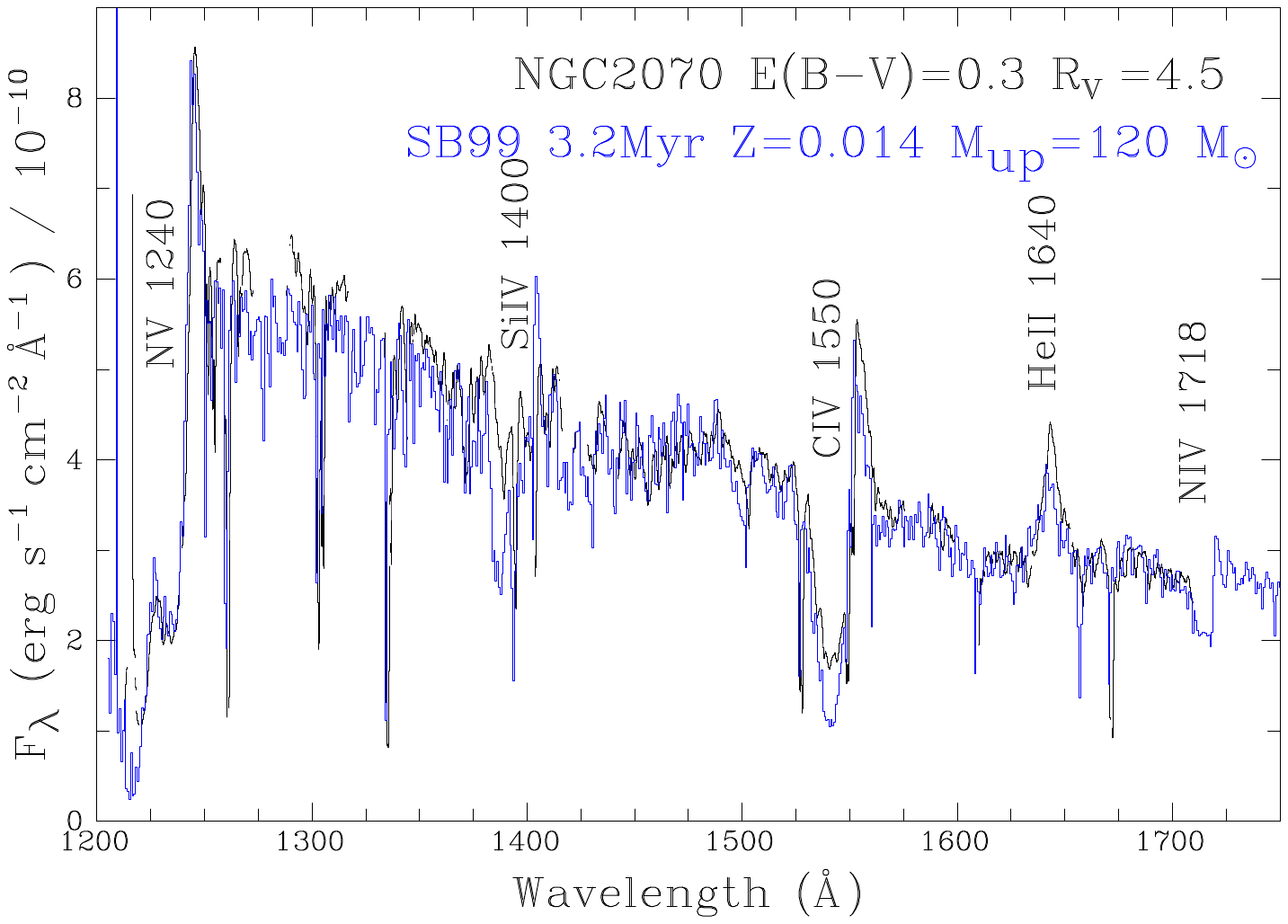}
\caption{Comparison between dereddened ($E_{\rm B-V}$ = 0.3, $R_{\rm V}$ = 4.5) cumulative far-UV spectrum of the central region of NGC~2070 (black) and the predicted Starburst99 spectrum based on contemporary, non-rotating Z=0.014  evolutionary models \citep{2012A&A...537A.146E} plus empirical Milky Way UV templates at 3.2 Myr (blue). The Far-UV luminosity of this model corresponds to a stellar masses of $6 \times 10^{4} M_{\odot}$.}
\label{NGC2070-sb99-solar}
\end{figure}

\begin{figure}
\includegraphics[angle=-90,width=0.9\columnwidth,bb=55 67 553 757]{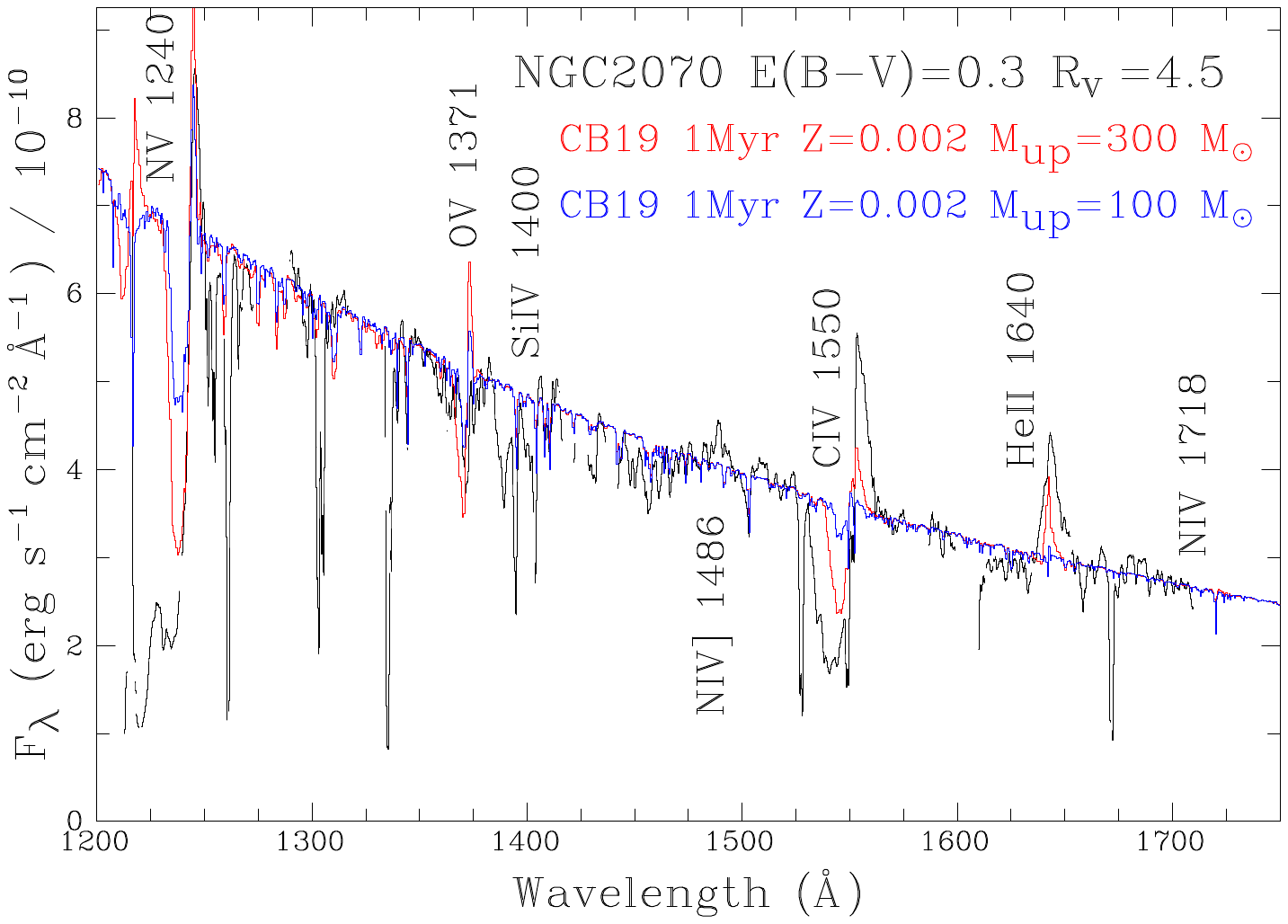}
\includegraphics[angle=-90,width=0.9\columnwidth,bb=55 67 553 757]{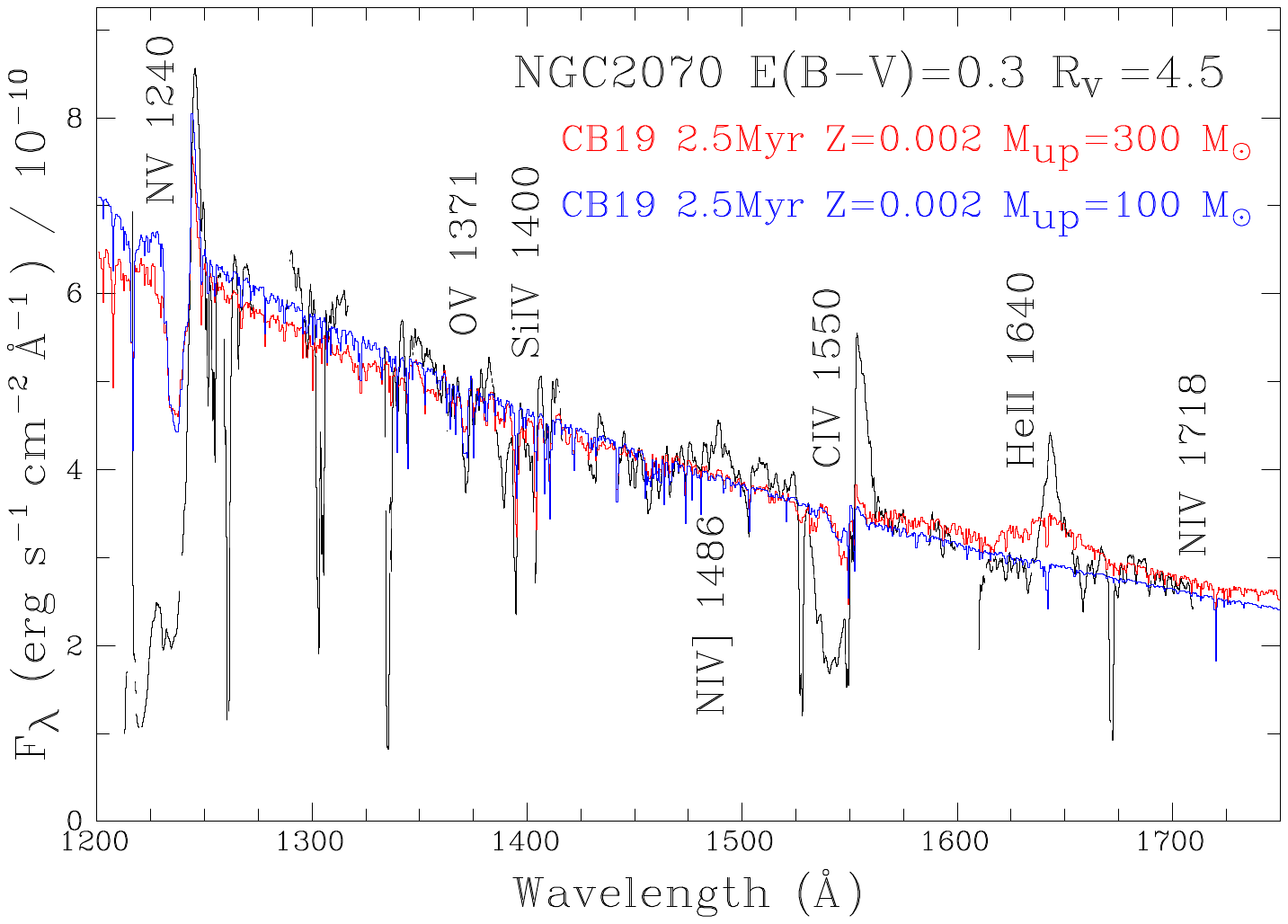}
\caption{Comparison between dereddened ($E_{\rm B-V}$ = 0.3, $R_{\rm V}$ = 4.5) cumulative far-UV spectrum of the central region of NGC~2070 (black) and the predicted CB19 spectrum based on Z=0.002 evolutionary models \citep{2015MNRAS.452.1068C} with $M_{\rm up} = 300 M_{\odot}$ (red) or 100 $M_{\odot}$ (blue) at 1.0 Myr (top panel) and 2.5 Myr (lower panel). Far-UV luminosities for this range of ages correspond to stellar masses of $5-8 \times 10^{4} M_{\odot}$ for the $M_{\rm up} = 300 M_{\odot}$ models.}
\label{NGC2070-cb19-z002}
\end{figure}

Overall, metal poor Starburst99 models fare rather poorly in quantitatively reproducing the empirical dataset. The origin of this disagreement has multiple causes:
\begin{enumerate}
\item The central region of NGC~2070 is not a coeval burst, since it hosts a young, potentially coeval, star cluster R136 contributing 1/3 of the far-UV continuum (Fig.~\ref{fig:far-UV-cluster}) with the remainder spanning ages of 1--7 Myr \citep{2018A&A...618A..73S}. Improved agreement could be obtained by adopting dual age populations representing the cluster and extended star-forming region.
\item NGC~2070 is host to a number of VMS, within R136 and beyond, which possess very strong winds owing to their proximity to the Eddington limit \citep{2020MNRAS.493.3938B, 2022A&A...663A..36B}. These stars produce strong
He\,{\sc ii} $\lambda$1640 emission whilst on the main-sequence \citep{2016MNRAS.458..624C}. In addition we have shown that VMS contribute 18\% of the far-UV continuum of NGC~2070, such that models
with upper mass cutoffs at $\sim 100 M_{\odot}$ will underestimate the far-UV continuum. In order to reproduce He\,{\sc ii} $\lambda$1640 in very young populations one needs to extend the IMF in population synthesis calculations to higher masses, and incorporate revised mass-loss prescriptions \citep{2023MNRAS.523.3949W}. 
\item The empirical UV template spectra at low metallicity in Starburst99 are a mix of SMC and LMC stars, since these were incorporated at a time prior to the availability of large numbers of high quality far-UV templates \citep{2020RNAAS...4..205R}. OB stars in the SMC possess significantly weaker winds than LMC counterparts \citep{1998MNRAS.300..828P, 2007A&A...473..603M}, so the use of SMC templates will yield weaker P Cygni profiles than observations at LMC composition. Alternatively theoretical spectra could lead to improved agreement \citep[e.g.][]{2010ApJS..189..309L}, especially at low metallicities where empirical templates are scarce or unavailable.
\item It is well known that evolutionary models for single stars struggle to reproduce observed WR populations \citep{2019A&A...625A..57H}. In reality, binary channels (including mergers) will increase the production of WR and lower mass He stars, which will enhance the strength of He\,{\sc ii} $\lambda$1640 from all WR subtypes, as well as other far-UV P Cygni profiles (N\,{\sc v} $\lambda\lambda$1238--42 from WN stars, C\,{\sc iv} $\lambda\lambda$1548--51 from WC stars).
\end{enumerate}

Previous analyses of extragalactic star clusters  \citep[e.g.][]{2016ApJ...823...38S, 2022AJ....164..208S} or entire galaxies \citep[e.g.][]{2014ApJ...795..109J, 2019ApJ...882..182C} are often more successful at reproducing far-UV observations than we have achieved for NGC~2070, albeit not universally so \citep{2006MNRAS.370..799S, 2014ApJ...781..122W, 2018ApJ...865...55L}. 
In part, this is achieved since the metallicity is varied, with high metallicities preferred in cases of strong stellar wind signatures, irrespective of nebular results for the regions in question. 

To illustrate the role played by metallicity, we compare the dereddened UV spectrum of NGC~2070 with  prediction of a 3.2 Myr model at Z=0.014  \citep{2012A&A...537A.146E} plus {\it Milky Way} UV templates in Fig.~\ref{NGC2070-sb99-solar}. In contrast to Fig.\ref{NGC2070-sb99}, N\,{\sc v} $\lambda\lambda$1238--42, C\,{\sc iv} $\lambda\lambda$1548--51 and He\,{\sc ii} $\lambda$1640 are now reasonably well predicted, with Si\,{\sc iv} $\lambda\lambda$1393--1402 a little too strong and the far-UV iron forest also over predicted. N\,{\sc iv} $\lambda$1718 is also prominent (recall Fig.~\ref{vacca}), in contrast to metal poor predictions. Nevertheless, the overall match is satisfactory for an age close to the median of OB stars in NGC~2070 \citep{2018A&A...618A..73S}, albeit reliant on unphysical evolutionary models and templates.


\begin{figure}
\includegraphics[angle=-90,width=0.9\columnwidth,bb=55 67 553 757]{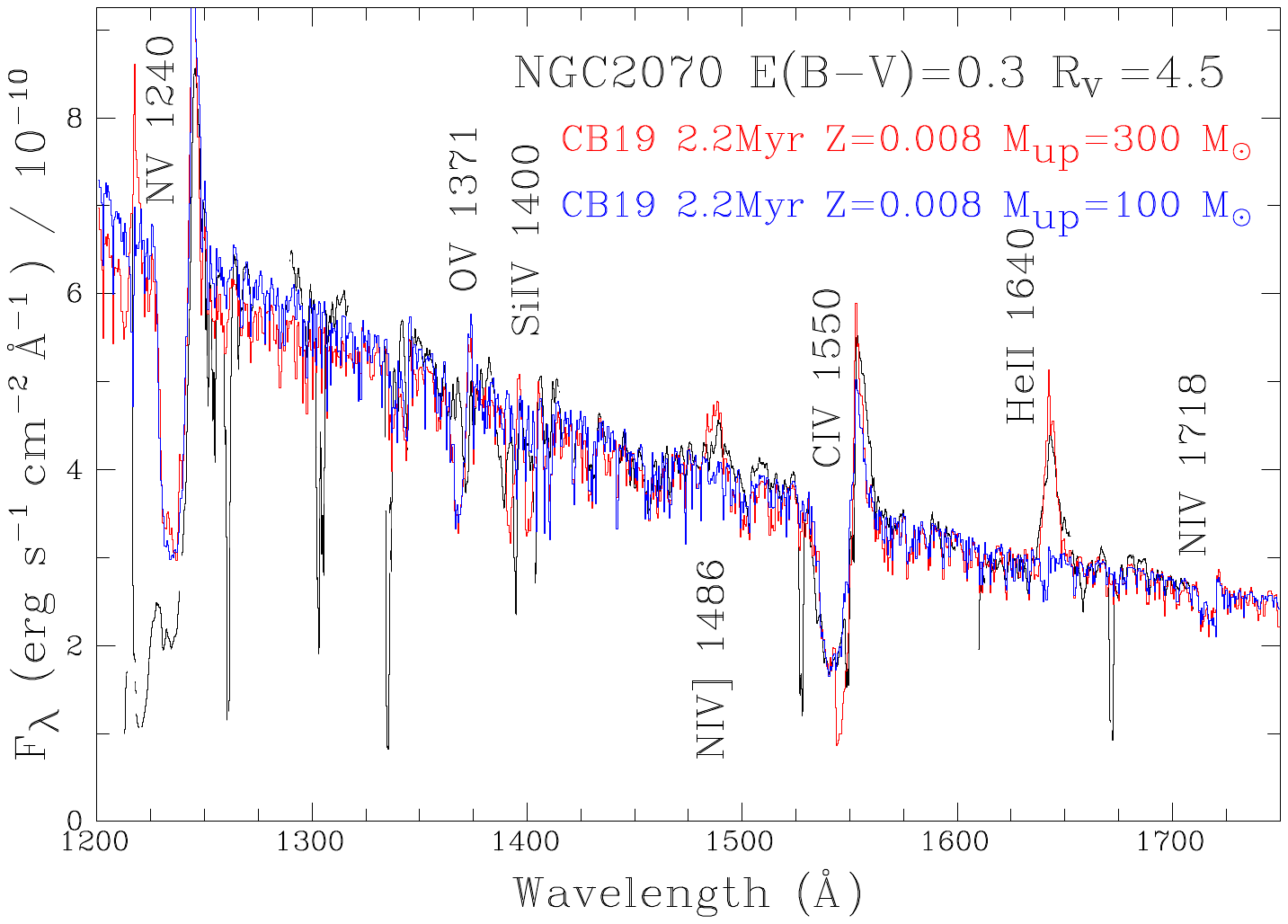}
\includegraphics[angle=-90,width=0.9\columnwidth,bb=55 67 553 757]{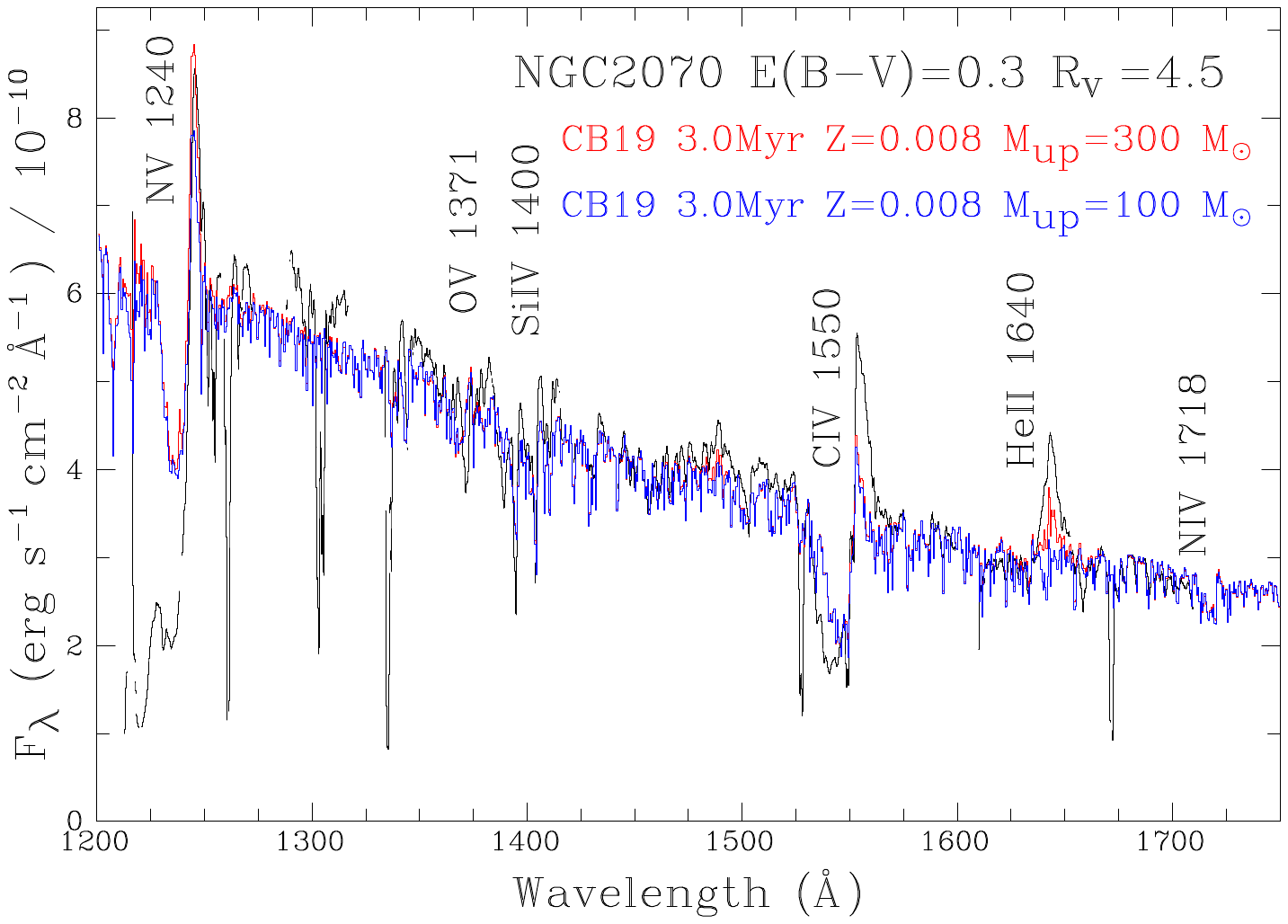}
\includegraphics[angle=-90,width=0.9\columnwidth,bb=55 67 553 757]{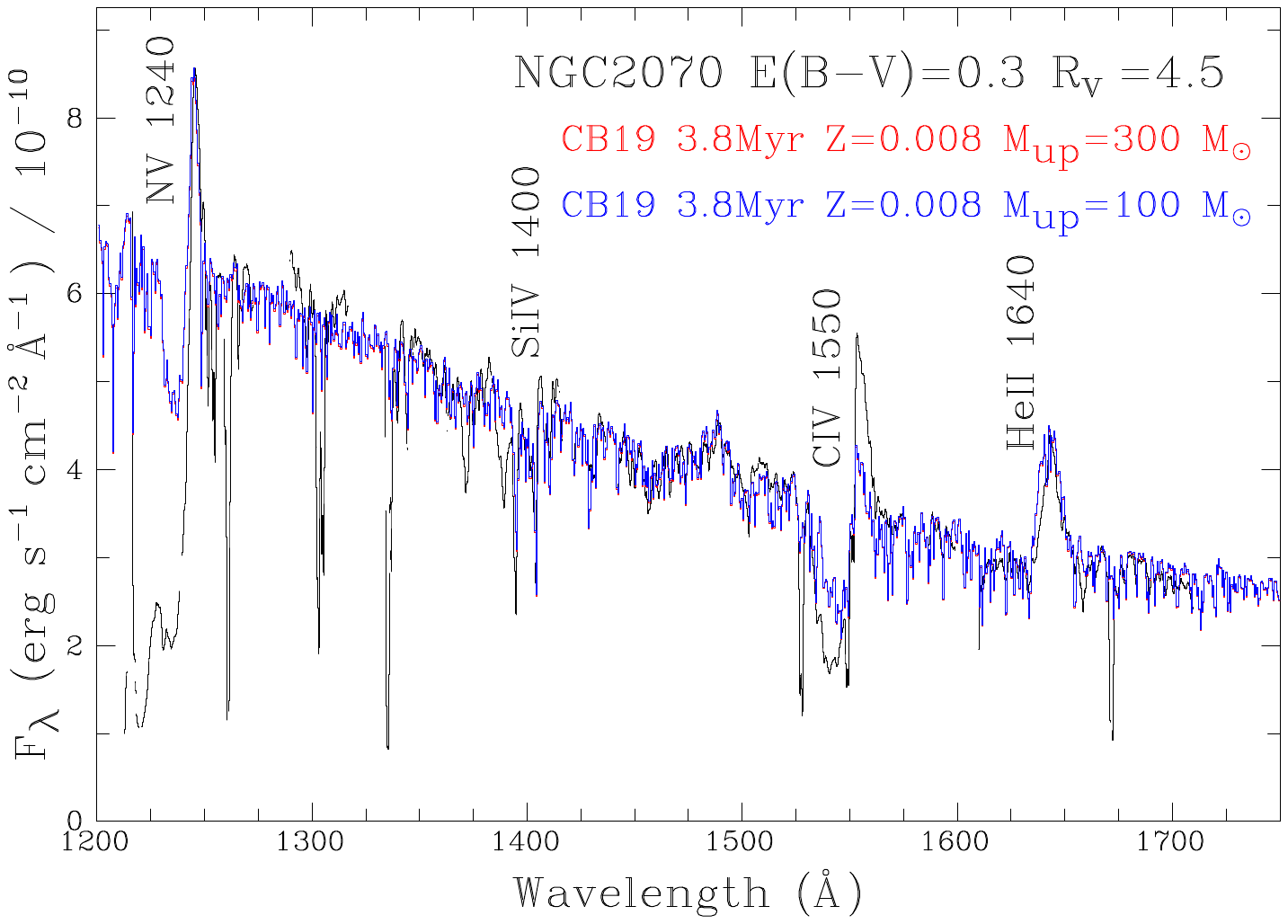}
\caption{Comparison between dereddened ($E_{\rm B-V}$ = 0.3, $R_{\rm V}$ = 4.5) cumulative far-UV spectrum of the central region of NGC~2070 (black) and the predicted CB19 spectrum based on Z=0.008 evolutionary models \citep{2015MNRAS.452.1068C} with $M_{\rm up} = 300 M_{\odot}$ (red) or 100 $M_{\odot}$ (blue) at 2.2 Myr (top panel), 3.0 Myr (central panel) and 3.8 Myr (lower panel). Far-UV luminosities for this range of ages correspond to stellar masses of $4-9 \times 10^{4} M_{\odot}$ for the $M_{\rm up} = 300 M_{\odot}$ models.}
\label{NGC2070-cb19}
\end{figure}

\begin{figure}
\includegraphics[angle=-90,width=0.9\columnwidth,bb=55 67 553 757]{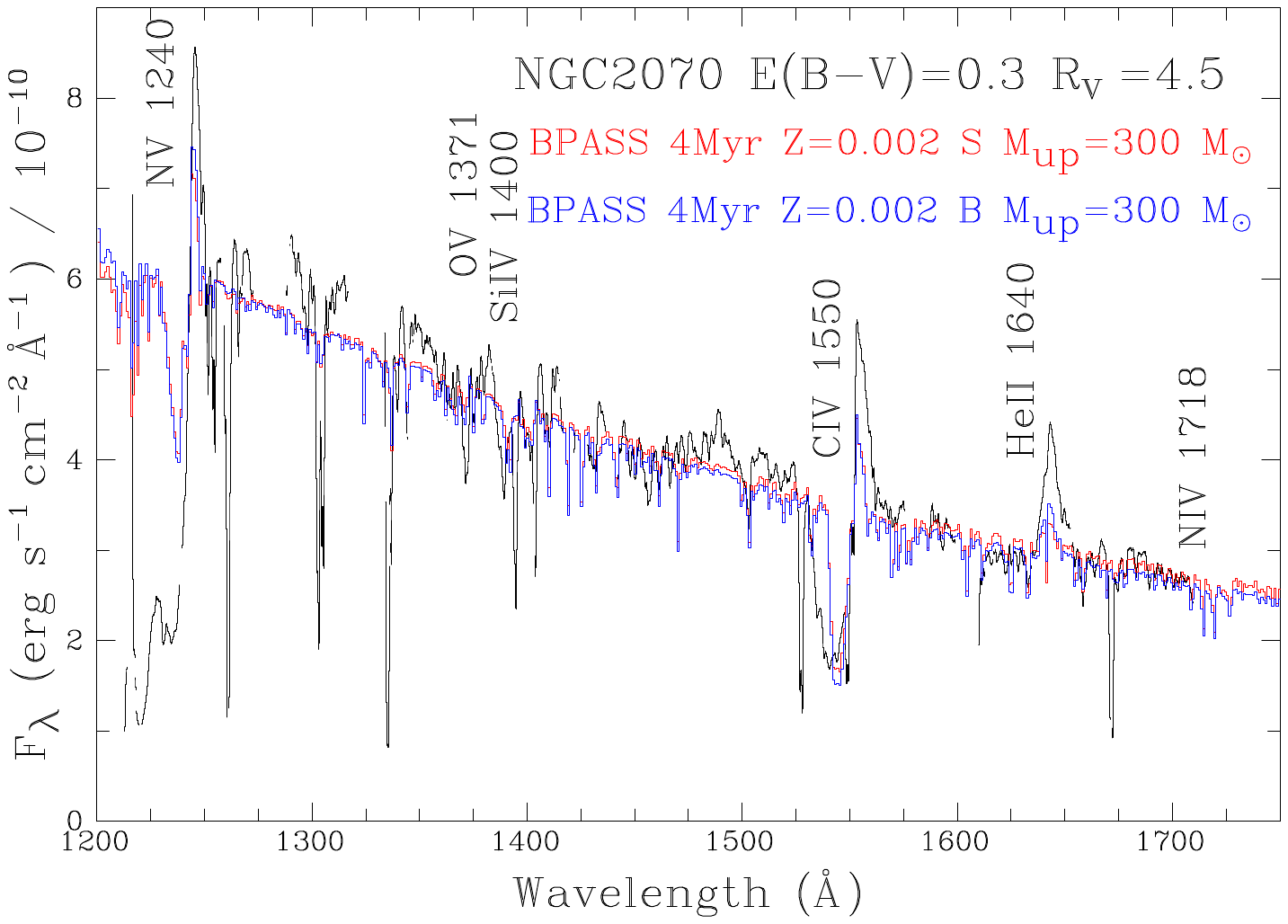}
\caption{Comparison between dereddened ($E_{\rm B-V}$ = 0.3, $R_{\rm V}$ = 4.5) cumulative far-UV spectrum of the central region of NGC~2070 (black) and the predicted BPASS spectrum based on Z=0.002 metallicity evolutionary models \citep[v.2.2.1][]{2017PASA...34...58E} at 4~Myr for single stars (red) and binaries (blue). Far-UV luminosities correspond to stellar masses of $9 \times 10^{4} M_{\odot}$ for the binary models.}
\label{NGC2070-bpass-z002}
\end{figure}

\subsection{Charlot \& Bruzual}

The extension of Charlot \& Bruzual 2019 models to IMFs with high upper mass limits at a wide range of metallicities permits some deficiencies of current Starburst99 models to be addressed. As demonstrated by \citet{2023MNRAS.523.3949W}, the difference in predicted He\,{\sc ii} $\lambda$1640 emission at ages of 1--3 Myr between $M_{\rm up}$ = 100 and 300 $M_{\odot}$ at LMC metallicity (Z=0.008) is striking (their fig 4). This arises from the inclusion of mass-loss prescriptions for very massive stars \citep{2011A&A...531A.132V}. We compare far-UV CB19 (Z=0.002, Kroupa IMF) predictions for  $M_{\rm up}$ = 100 and 300 $M_{\odot}$ at 1.0 Myr and 2.5 Myr with NGC~2070 observations in Fig.~\ref{NGC2070-cb19-z002}. These ages correspond to the maximum He\,{\sc ii} $\lambda$1640 emission due to very massive stars (1~Myr) and classical WR stars (2.5 Myr) for $M_{\rm up}$ = 300 $M_{\odot}$. At 1 Myr the high upper mass limit prediction provides a closer agreement with observations, albeit with O\,{\sc v} $\lambda$1371 too strong, He\,{\sc ii} $\lambda$1640 too weak and C\,{\sc iv} $\lambda\lambda$1548--51 extremely weak. At 2.5 Myr, all wind features are far too weak, with the exception of He\,{\sc ii} $\lambda$1640 for the $M_{\rm up}$ = 300 $M_{\odot}$ case which has a satisfactory emission equivalent width, albeit far broader than observed.

Consequently we also consider Z=0.008 predictions from CB19 with  $M_{\rm up}$ = 100 and 300 $M_{\odot}$. Fig.~\ref{NGC2070-cb19} compares predictions at 2.2, 3.0 and 3.8 Myr with observations, which span the range of ages at which prominent He\,{\sc ii} $\lambda$1640 emission is predicted. At all ages the $M_{\rm up}$ = 300 $M_{\odot}$ model provides a better match to observations. \citet[][their fig.~4]{2023MNRAS.523.3949W} highlight the sharp peak in He\,{\sc ii} $\lambda$1640 at 2.2~Myr once very massive stars enter the late WN phase \citep[see also][]{2023arXiv231003413S}. Consequently the predicted He\,{\sc ii} $\lambda$1640 is in good agreement with observations at this precise age, as is C\,{\sc iv} $\lambda$1550, albeit with O\,{\sc v} $\lambda$1371 over predicted, and Si\,{\sc iv} $\lambda\lambda$1393--1402 under predicted. N\,{\sc iv}] $\lambda$1486 is observed in NGC~2070, though is overpredicted in models at this age that extend to $M_{\rm up}$ =  300 $M_{\odot}$. 

Turning to later ages, O\,{\sc v} $\lambda$1371 has faded in strength after 3 Myr, with Si\,{\sc iv} $\lambda\lambda$1393--1402 now well reproduced, with He\,{\sc ii} $\lambda$1640 emission a little too weak, with C\,{\sc iv} $\lambda\lambda$1548--51 also too weak, alongside Si\,{\sc iv} $\lambda$1640. C\,{\sc iv} $\lambda\lambda$1548--51 continues to weaken at later ages (3.8 Myr) with He\,{\sc ii} $\lambda$1640 from classical WR stars now well reproduced and Si\,{\sc iv} $\lambda\lambda$1393--1402 still matched. Overall, use of CB19 models at Z=0.008  favour an age of 2.2~Myr for NGC~2070 plus a high upper mass limit, in spite of the poor match to O\,{\sc v} $\lambda$1371 and very weak N\,{\sc iv} $\lambda$1718 (in contrast to {\it IUE} observations in Fig.~\ref{vacca}).

\begin{figure}
\includegraphics[angle=-90,width=0.9\columnwidth,bb=55 67 553 757]{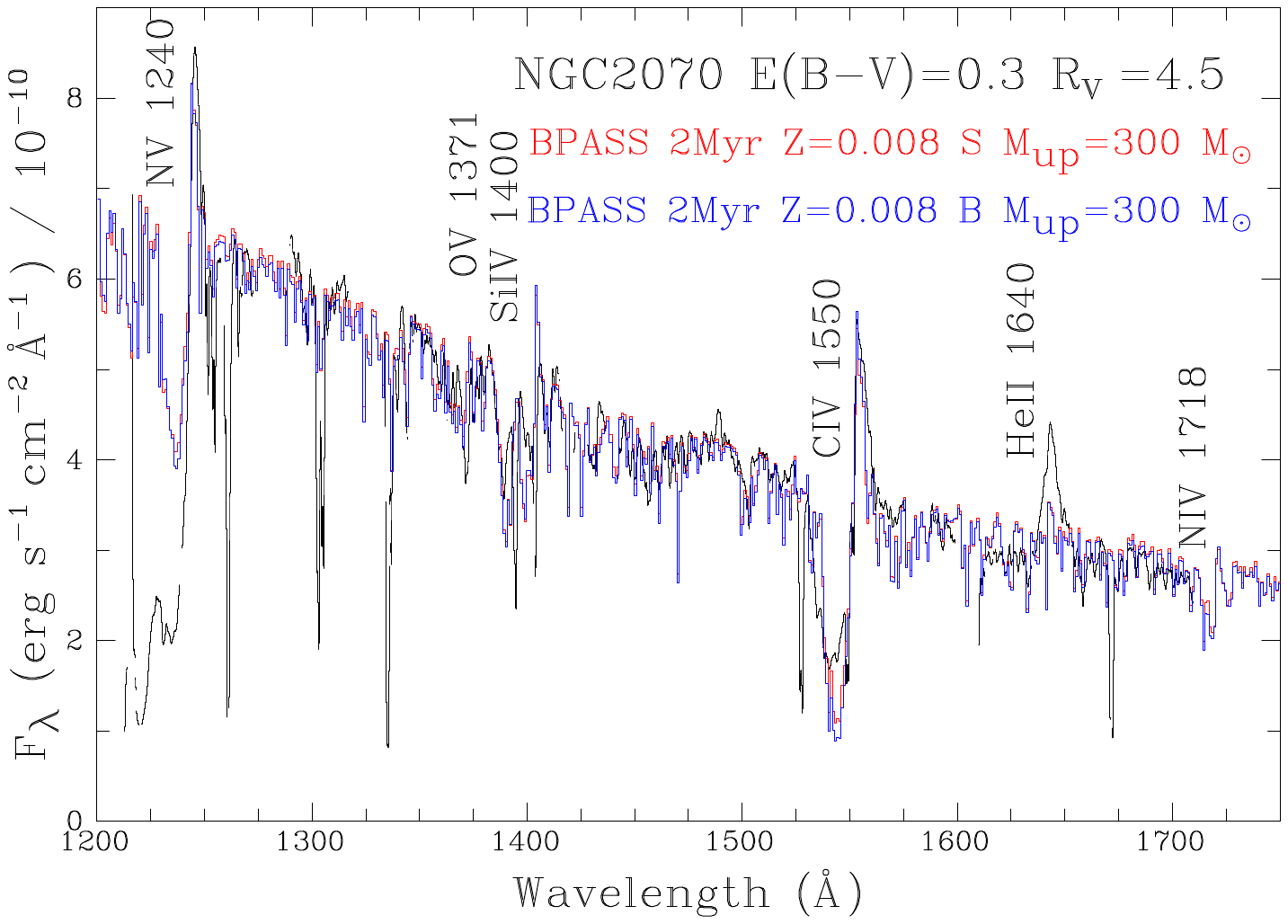}
\includegraphics[angle=-90,width=0.9\columnwidth,bb=55 67 553 757]{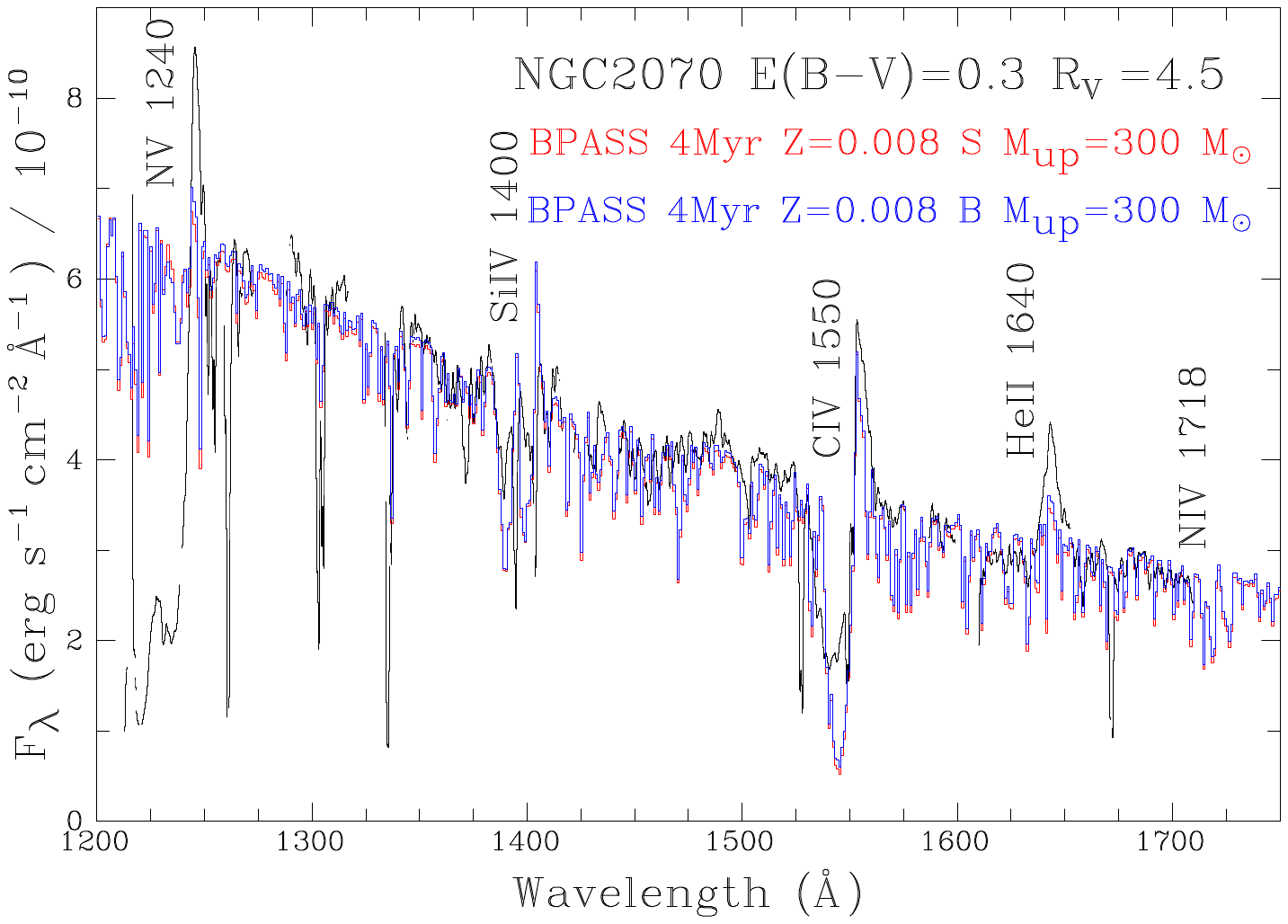}
\caption{Comparison between dereddened ($E_{\rm B-V}$ = 0.3, $R_{\rm V}$ = 4.5) cumulative far-UV spectrum of the central region of NGC~2070 (black) and the predicted BPASS spectrum based on Z=0.008 metallicity evolutionary models \citep[v.2.2.1][]{2017PASA...34...58E} at 2~Myr (upper panel) and 4~Myr (lower panel) for single stars (red) and binaries (blue). Far-UV luminosities for this range of ages correspond to stellar masses of $5-9 \times 10^{4} M_{\odot}$ for the binary models.}
\label{NGC2070-bpass}
\end{figure}

\subsection{BPASS}

Finally, we consider BPASS \citep[v.2.2.1][]{2017PASA...34...58E, 2018MNRAS.479...75S} models for single or binary populations at Z=0.002 and Z=0.008, with a Kroupa IMF and $M_{\rm up}$ = 300 $M_{\odot}$. 
The maximum He\,{\sc ii} $\lambda$1640 emission at Z=0.002 is predicted at an age of 4 Myr (single or binary case), as shown in Fig.~\ref{NGC2070-bpass-z002}. At this metallicity all wind lines are predicted to be too weak, albeit with He\,{\sc ii} $\lambda$1640 only a factor of two weaker than observed, since several H-rich WN stars are predicted for cluster mass of $9 \times 10^{4} M_{\odot}$ which reproduces the far-UV luminosity of NGC~2070.  At Z=0.008, predictions are improved, as shown in Fig.~\ref{NGC2070-bpass} for ages of 2~Myr (upper panel) and 4~Myr (lower panel). For the 2 Myr case C\,{\sc iv} $\lambda\lambda$1548--51, N\,{\sc v} $\lambda\lambda$1238--42 and Si\,{\sc iv} $\lambda\lambda$1393--1402 are reasonably well reproduced, although He\,{\sc ii} $\lambda$1640 emission is only weakly present despite the inclusion of very massive stars,
and N\,{\sc iv} $\lambda$1718 is very weak (in contrast with {\it IUE} observations in Fig.~\ref{vacca}). 

Binary models predict $\sim$6 very massive stars at 2 Myr for a stellar mass of $5\times 10^{4} M_{\odot}$ that reproduces the far-UV luminosity of NGC~2070, categorized as a mix of O, Of and H-rich WN-types.
In contrast with CB19 models, BPASS mass-loss prescriptions fail to account for the proximity of very massive stars to the Eddington limit \citep{2020MNRAS.493.3938B, 2023A&A...673A.132B}. Predictions from single models are very similar to those from binary stars at such ages. At 4 Myr, preferred from predicted optical WR bumps, N\,{\sc v} $\lambda\lambda$1238--42 is now too weak, Si\,{\sc iv} $\lambda\lambda$1393--1402 is too strong, with He\,{\sc ii} $\lambda$1640 again too weak despite arising from classical WR stars. A stellar mass of 9$\times 10^{4} M_{\odot}$ is required for the binary models to reproduce the far-UV luminosity of NGC~2070 at 4~Myr, with $\sim$10 classical WR stars predicted (mix of WN and WC). Overall the 2 Myr BPASS models (single or binary) provide the closest match to UV observations, aside from the weakness of He\,{\sc ii} $\lambda$1640.



\section{Discussion and conclusions}\label{discussion}

We present the integrated VLT/MUSE spectrum of the central 2$\times$2 arcmin$^{2}$ (30$\times$30 pc$^{2}$) region of NGC~2070, the dominant giant H\,{\sc ii} region of the Tarantula (30 Doradus) region in the LMC, and construct an empirical far-UV spectrum of this region by combining observations of individual stars with templates from the ULLYSES survey \citep{2020RNAAS...4..205R}. This region is unique in the sense that we are able to compare results from an individual treatment of stars with an integrated approach, plus both classical Wolf-Rayet stars and VMS have been identified within this region. A summary of UV and optical results is presented in Table~\ref{tab:summary}.

\citet{2023A&A...678A.159M} consider UV and optical spectroscopic indicators of WR and very massive stars in nearby star-forming regions, favouring optical diagnostics to identify the latter. Neither the 
far-UV (HST) nor optical (MUSE) spectroscopy of NGC~2070 permit unique diagnostics of VMS since both populations have similar He\,{\sc ii} $\lambda$1640 morphologies (Fig.~\ref{fig:far-UV}) and 
metal lines in the vicinity of He\,{\sc ii} $\lambda$4686 are dominated by classical WR stars (Fig.~\ref{muse-sp}). Consequently, despite NGC~2070 having the richest VMS population in the Local Group, their presence
is masked by its mixed age population, which includes classical WR stars. Nevertheless, unambigious diagnostics of very young populations (required for VMS) exist in the optical 
\citep{2023A&A...678A.159M} and far-UV \citep{2016MNRAS.458..624C}, the latter involving the presence of P Cygni O\,{\sc v} $\lambda$1371 and He\,{\sc ii} $\lambda$1640 emission with Si\,{\sc iv} $\lambda\lambda$1393-1402 absent.

\begin{figure}
\includegraphics[width=0.8\columnwidth,bb=11 34 566 564]{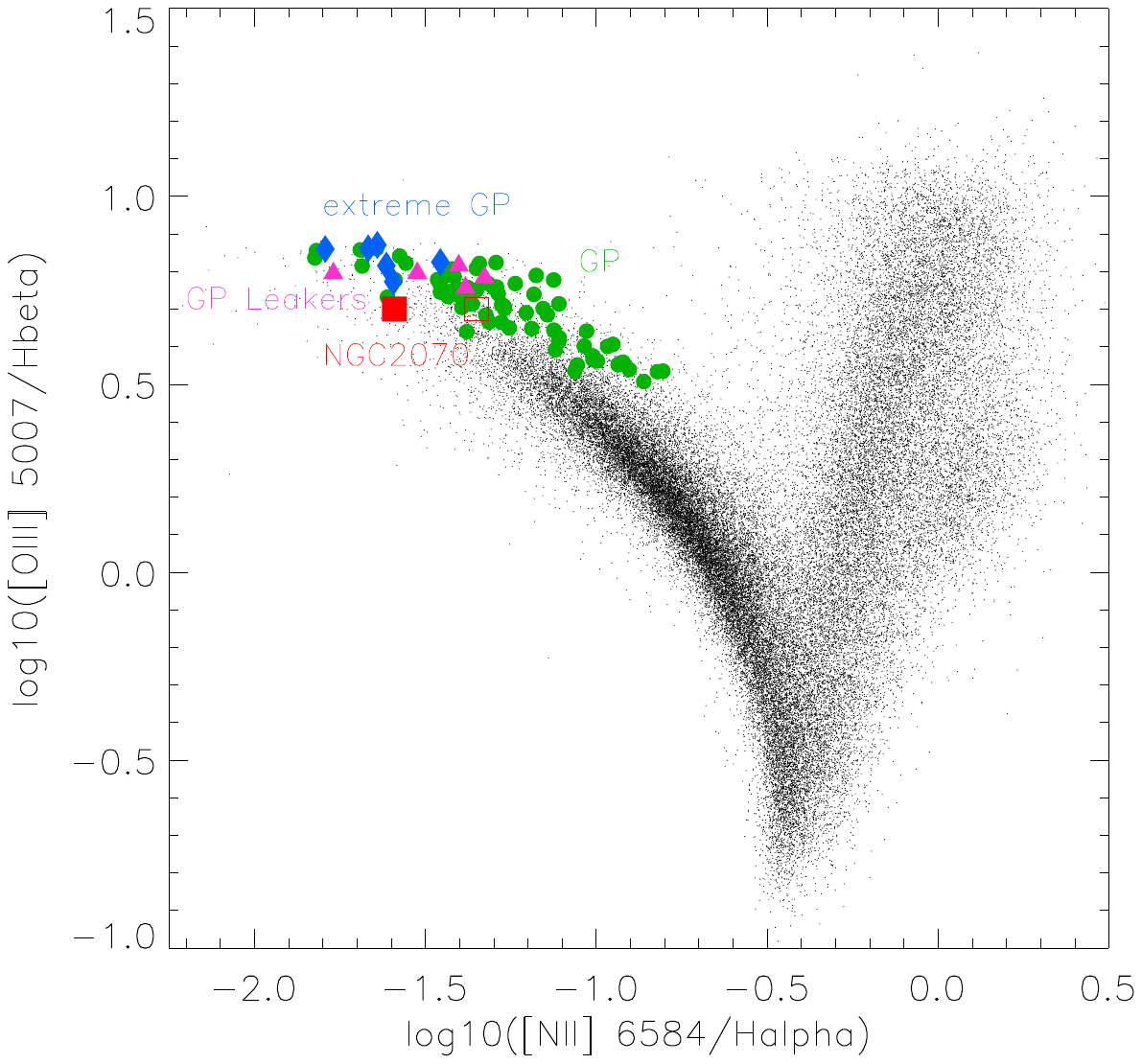}
\caption{BPT diagram \citep{1981PASP...93....5B} illustrating the similarity in integrated strengths between 
NGC~2070/Tarantula (filled/open red square), Green Pea (green circles), 
extreme Green Pea (blue diamonds), and Lyman-continuum emitting Green Pea (pink triangles) galaxies, together with
plus SDSS star-forming galaxies \citep[black dots,][]{2009ApJS..182..543A}, adapted from \citet[][their fig.2]{2017ApJ...845..165M} and \citet{2017Msngr.170...40C}.}
\label{BPT}
\end{figure}

\begin{figure}
\includegraphics[width=0.8\columnwidth,bb=28 33 561 563]{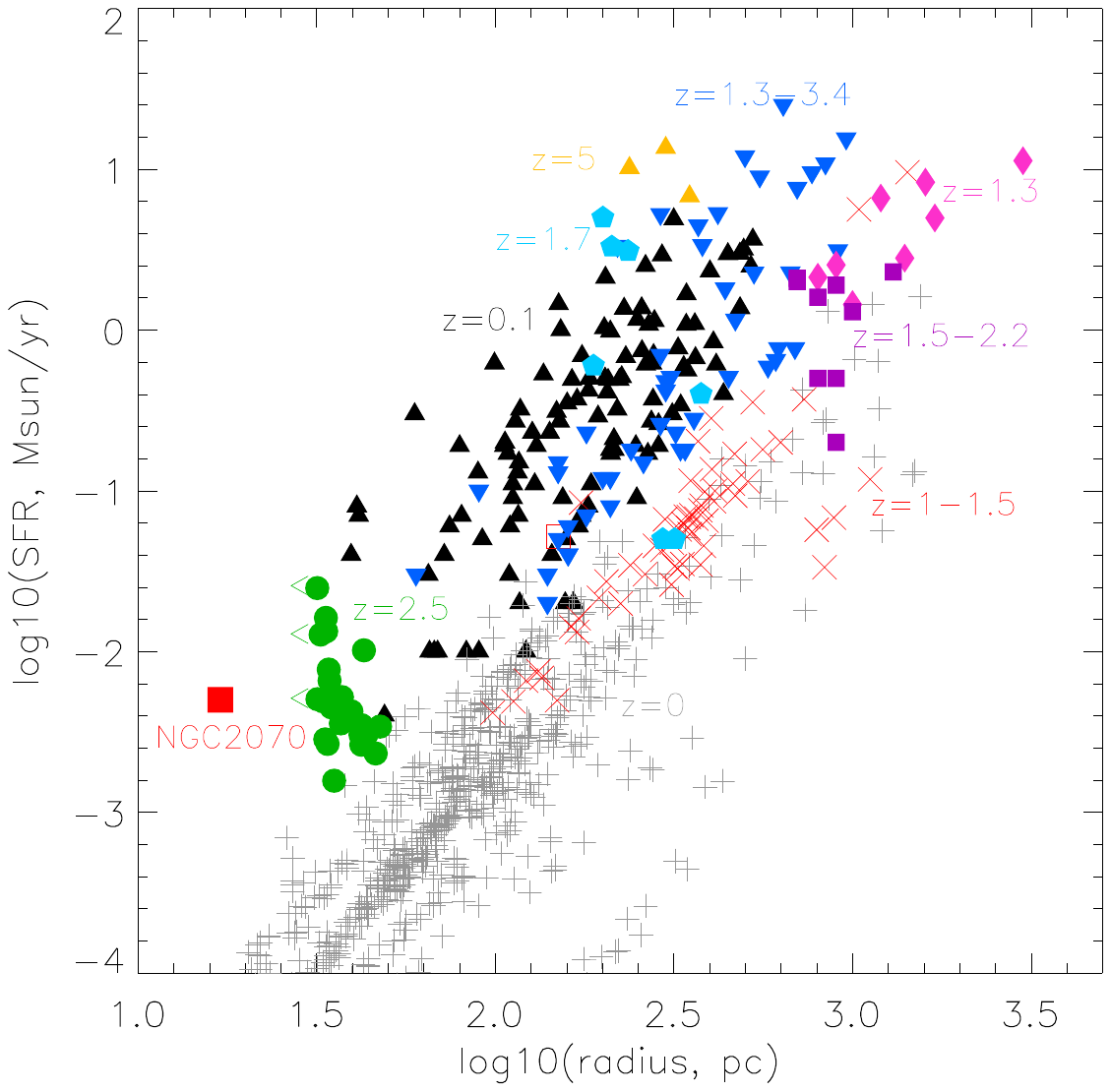}
\caption{Comparison between the integrated star-formation rate of NGC~2070/Tarantula (filled/open red square) and star-forming knots
from  local star-forming galaxies \citep[$z$=0, grey][]{2003PASP..115..928K} to those at a range of redshifts ($z$=0.1 to 5), adapted from \citet[][their fig.2]{2017ApJ...843L..21J}
 and \citet{2017Msngr.170...40C}.}
\label{SFR}
\end{figure}

 If we were reliant solely on integrated visual spectroscopy of NGC~2070 we
would substantially underestimate its metallicity from strong line diagnostics \citep{2013A&A...559A.114M, 2017MNRAS.465.1384C} but obtain an age of the stellar population ($\sim$4 Myr from Starburst99 and BPASS, 3.5 Myr from CB19) in close agreement with the median age  of massive stars within the region. The Starburst99 inferred mass is broadly consistent with that of the star cluster R136 \citep{1995ApJ...448..179H, 2009ApJ...707.1347A} which likely dominates the stellar mass of the region. The strength of WR bumps from historical LMC metallicity models \citep{1994A&AS..103...97M} implemented in Starburst99 are underestimated, although results from contemporary LMC models \citep{2015MNRAS.452.1068C} plus a high $M_{\rm up}$ = 300 $M_{\odot}$ in CB19 and BPASS models are improved for He\,{\sc ii} $\lambda$4686 and C\,{\sc iv} $\lambda\lambda$5801-12. The number of WR stars estimated from line luminosities of blue and yellow WR bumps is in good agreement with latest empirical calibrations for LMC metallicity \citep{2023MNRAS.521..585C}.

Starburst99 predictions from Z=0.008 metallicity evolutionary models \citep{1994A&AS..103...97M} plus Magellanic Cloud empirical UV templates are incapable of reproducing the primary wind features in the integrated far-UV spectrum of NGC~2070 at a single age. For example, the maximum predicted emission $W_{\lambda}$(He\,{\sc ii} $\lambda$1640) = 1.5\AA, a factor of three lower than observed (4.7$\pm$0.2\AA). Starburst99 models using contemporary solar metallicity models \citep{2012A&A...537A.146E} and Milky Way templates at 3.2 Myr provide a better match despite an unphysical metallicity. Various reasons for these deficiencies are set out in Section~\ref{pop-syn} including the mixed age of massive stars contributing to the integrated light (only 1/5 arises from the R136 cluster), plus the lack of high quality empirical templates \citep[e.g. ULLYSES,][]{2020RNAAS...4..205R},  neglect of very massive stars and the lack of binary evolutionary models. 

CB19 models accounting for Z=0.008 evolutionary models \citep{2015MNRAS.452.1068C} and synthetic spectra are more successful at reproducing He\,{\sc ii} $\lambda$1640 emission at 2.2 Myr since they extend to higher  initial masses $M_{\rm up}$ = 300 $M_{\odot}$) and fold in contemporary mass-loss prescriptions for very massive stars, although no single age provides a satisfactory solution for all UV diagnostics.  BPASS \citep{2017PASA...34...58E} models incorporate binary evolution, but do not currently incorporate enhanced mass-loss for very massive stars resulting from their proximity to the Eddington limit. Consequently BPASS models are unable to reproduce strong He\,{\sc ii} $\lambda$1640 in metal poor models even with $M_{\rm up}$ = 300 $M_{\odot}$). 

Fixing Z=0.008 and combining UV and optical spectroscopic diagnostics, Starburst99 models at $\sim$4.2 Myr are capable of reproducing some optical (H$\alpha$, C\,{\sc iv} $\lambda\lambda$5801-12) and UV (Si\,{\sc iv} $\lambda\lambda$1393-1402) features of NGC~2070, although He\,{\sc ii} $\lambda$1640, $\lambda$4686 and C\,{\sc iv} $\lambda\lambda$1548--51 are too weak. CB19 models at 3.8~Myr provide a better
match to He\,{\sc ii} $\lambda$1640 and $\lambda$4686, although C\,{\sc iv} $\lambda\lambda$1548--51 and $\lambda\lambda$5801-12 are too weak, while BPASS models at 4~Myr underpredict He\,{\sc ii} $\lambda$1640, $\lambda$4686, overpredicts C\,{\sc iv} $\lambda\lambda$58012-12, with Si\,{\sc iv} $\lambda\lambda$1393-1402 and C\,{\sc iv} $\lambda\lambda$1548-51 broadly satisfactory.
In view of our results, caution should be given to specific ages and metallicities of young, unresolved stellar populations  \citep[see][]{2022A&A...659A.163M, 2023MNRAS.523.3949W}. Should observed regions include multiple star clusters or clusters embedded within extended star-forming regions, it would be more realistic to incorporate either a dual or multiple populations \citep[e.g.][]{2019ApJ...882..182C, 2022AJ....164..208S}, although the extended population of NGC~2070 is far from coeval \citep{2018A&A...618A..73S}.

\begin{table}
\centering
\caption{Summary of results for individual versus integrated treatment of NGC~2070 in ultraviolet and visual, with integrated results obtained from Starburst99 (SB99) using Z=0.008 evolutionary models \citep{1994A&AS..103...97M} and $M_{\rm up}$ = 120 $M_{\odot}$, Charlot \& Bruzual (CB19) using Z=0.008 evolutionary models \citep{2015MNRAS.452.1068C} and $M_{\rm up}$ = 100  or 300 $M_{\odot}$
or BPASS Z=0.008 single and binary models \citep[v.2.2.1][]{2017PASA...34...58E, 2018MNRAS.479...75S}. LMC WR template luminosities are from \citet{2023MNRAS.521..585C} while N2 and O3N2 strong line calibrations are from  \citet{2017MNRAS.465.1384C}.}
\begin{tabular}{l @{\hspace{0mm}} c @{\hspace{2mm}} l @{\hspace{3mm}}  c @{\hspace{4mm}} l}
\hline
Property & Individual & Ref & Integrated & This study  \\
\hline
                &   \multicolumn{4}{c}{Visual (MUSE)} \\
Age  (Myr)          & 1--7, 1--2.5$\dag$ & a,b & 4.2, 3.0, 4 & SB99, CB19, BPASS \\
Mass ($10^{5} M_{\odot}$) & 0.5--1.0 & c,d & 0.85, 0.6, 1.1 & SB99, CB19, BPASS \\
$M_{\rm up} (M_{\odot}$) & $\sim$270 & b & 100--300 & CB19 \\
N(WN)               &  11   & e    &  10--17 & WR templates \\
N(WC4)                & 2    &e    & 4          & WR templates\\
$\log$(O/H)+12 & 8.26--8.5 & f,g & 7.9--8.1 & N2, O3N2 diagnostics \\
                &   \multicolumn{4}{c}{Far-UV (HST)} \\
Age  (Myr)          & 1.5$\dag$ & h & 4.2, 2.2, 2 & SB99, CB19, BPASS \\
Mass ($10^{5} M_{\odot}$) & $\cdots$ & $\cdots$ & 0.85, 0.4, 0.5 & SB99, CB19, BPASS \\
$M_{\rm up} (M_{\odot}$) & $\sim$250 & h & 300 & CB19 \\
$Z$ & $\cdots$  & $\cdots$ & 0.014 & SB99$\ddag$ \\
\hline
\end{tabular}\par
\label{tab:summary}
a: \citet{2018A&A...618A..73S}; b: \citet{2022A&A...663A..36B};  c: \citet{1995ApJ...448..179H}; d: \citet{2009ApJ...707.1347A}; e: \citet{2013A&A...558A.134D}; f: \citet{2002A&A...391.1081V}; g: \citet{2003ApJ...584..735P}; h: \citet{2016MNRAS.458..624C} \\
 $\ddag$ Z=0.014  models \citep{2012A&A...537A.146E} and Milky Way templates. $\dag$ R136 cluster
\end{table}

Although NGC~2070 is located in the LMC, our nearest extragalactic star-forming galaxy, if one compares its nebular properties in the BPT diagram \citep{1981PASP...93....5B}, it is located close to extreme Green Pea galaxies \citep{2009MNRAS.399.1191C} as shown in Fig.~\ref{BPT}. Green Pea galaxies have received considerable interest since a subset have been established as Lyman continuum leakers \citep{2016MNRAS.461.3683I}. H$\alpha$ observations of NGC~2070 (Fig.~\ref{30dor-F658N-F336W}) suggests a significant fraction of ionizing photons escape the region. By way of example, BPASS 
models at 4 Myr favour a stellar mass of $8.5-11 \times 10^{4} M_{\odot}$ from FUV--optical continua, but only $6 \times 10^{4} M_{\odot}$ from its H$\alpha$ luminosity, favouring a sizeable escape fraction of
ionizing photons \citep[see also][]{2013A&A...558A.134D}. 

The relatively high star-formation rate of NGC~2070 for its size is also more comparable to the properties of star-forming knots at high redshift with respect to local star-forming galaxies \citep{2003PASP..115..928K}, as shown in Fig.~\ref{SFR}. Specifically, NGC~2070 properties are close to clumps in the lensed galaxy SDSS J1110+6459 at $z\sim$2.5 \citep{2017ApJ...843L..21J}. Star-forming knots in the lensed Sunburst arc have a similar size to NGC~2070 but are significantly more intensive \citep{2017A&A...608L...4R, 2019Sci...366..738R, 2022A&A...659A...2V}. 

As we have shown, the ULLYSES survey \citep{2020RNAAS...4..205R} provides high quality far-UV empirical templates for OB stars at half-solar metallicity, which serve as alternatives to widely used theoretical
models  \citep{2010ApJS..189..309L}  in population synthesis tools. High quality optical VLT/Shooter spectroscopy of ULLYSES targets has also been acquired \citep[XshootU,][]{2023A&A...675A.154V}, which complement the present MUSE observations since templates achieve higher spectral resolution, extend to the violet and complement existing optical spectroscopic libraries of massive stars \citep[e.g.][]{2022A&A...661A..50V}. 

 Indeed, ULLYSES/XshootU spectroscopy have been secured for a large number of OB stars in the SMC.  Ideally we would wish to extend this study to the 1/5 solar metallicity of the SMC. The richest young star formation region in the SMC is NGC~346 \citep{1989AJ.....98.1305M}. STIS/G140L spectroscopy has recently been obtained for the O stars within the central region of this giant H\,{\sc ii} region \citep{2022A&A...666A.189R}, whose central 1$\times$1 arcmin$^{2}$ (20$\times$20 pc$^{2}$)
has been observed with VLT/MUSE. Unfortunately, NGC~346 lacks a massive young star cluster at its heart such that any analysis would need to reflect stochasticity in both its star formation history and IMF \citep[e.g.][]{2014MNRAS.444.3275D, 2015MNRAS.452.1447K, 2022MNRAS.509..522O}. 

Nevertheless, there are other large star-forming complexes in the Magellanic Clouds \citep{2006A&A...456..623E}, for which quantitative UV-optical analysis could be undertaken based on ULLYSES templates plus integral field observations from the upcoming SDSS-V Local Volume Mapper \citep[LVM,][]{2017arXiv171103234K}.
 
\section*{Acknowledgements}
This study has been made possible courtesy of: (i) Science Verification MUSE observations of NGC~2070 (proposal led by Jorge Melnick and Chris Evans); (ii) the Director's Discretionary ULLYSES survey, which was implemented by a Space Telescope Science Institute (STScI) team led by Julia Roman-Duval, having been recommended by the Hubble UV Legacy Science Definition Working Group chaired by Sally Oey, convened in 2018 by the then STScI Director Ken Sembach. Thanks to Linda Smith for providing useful comments on a draft version of the manuscript, and the referee for feedback which helped to improve the clarity of the paper. 

Based on observations made with ESO telescopes at the Paranal observatory under programme ID 60.A-9351(A) and observations obtained with the NASA/ESA Hubble Space Telescope, retrieved from the Mikulski Archive for Space Telescopes (MAST) at the STScI. STScI is operated by the Association of Universities for Research in Astronomy, Inc. under NASA contract NAS 5-26555. 
PAC is supported by the Science and Technology Facilities Council research grant ST/V000853/1 (PI. V. Dhillon). NC gratefully acknowledges funding from the Deutsche Forschungsgemeinschaft (DFG) CA 255/1-1. 

This research has made use of the SIMBAD database, operated at CDS, Strasbourg, France. Starlink software is currently supported by the East Asian Observatory. {\sc iraf} was distributed by the National Optical Astronomy Observatory, which was managed by the Association of Universities for Research in Astronomy (AURA) under a cooperative agreement with the National Science Foundation. For the purpose of open access, the author has applied a Creative Commons Attribution (CC BY) license to any Author Accepted Manuscript version arising.

\section*{Data Availability} 

The integrated MUSE and far-UV spectroscopy of NGC~2070 are available in ascii format (wavelength in \AA, flux in erg\,s$^{-1}$\,cm$^{-2}$\,\AA$^{-1}$) at 10.5281/zenodo.10204404.


\bibliographystyle{mnras}
\bibliography{muse} 





\clearpage

\appendix

\section{Template UV spectra}

\begin{figure}
\includegraphics[angle=-90,width=0.9\columnwidth,bb=13 46 553 757]{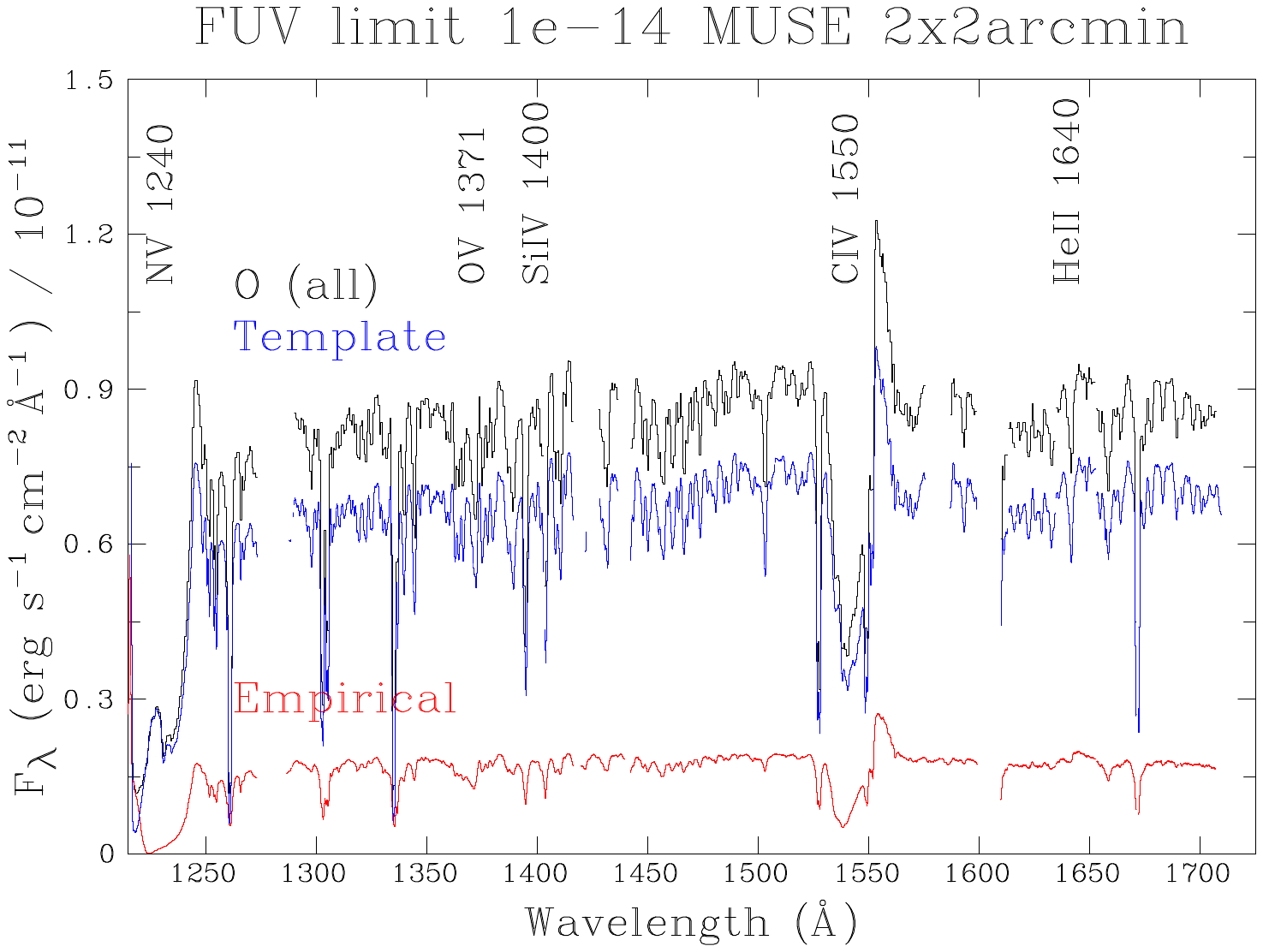}
\caption{Cumulative O-type far-UV spectrum of the central region of NGC~2070 (black) based on empirical data of 34 individual stars (red) and 159 LMC template stars (blue). Gaps in spectra arise from incomplete spectral coverage (e.g. COS G130M+G160M).}
\label{fig:far-UV-O}
\end{figure}

\begin{figure}
\includegraphics[angle=-90,width=0.9\columnwidth,bb=13 46 553 757]{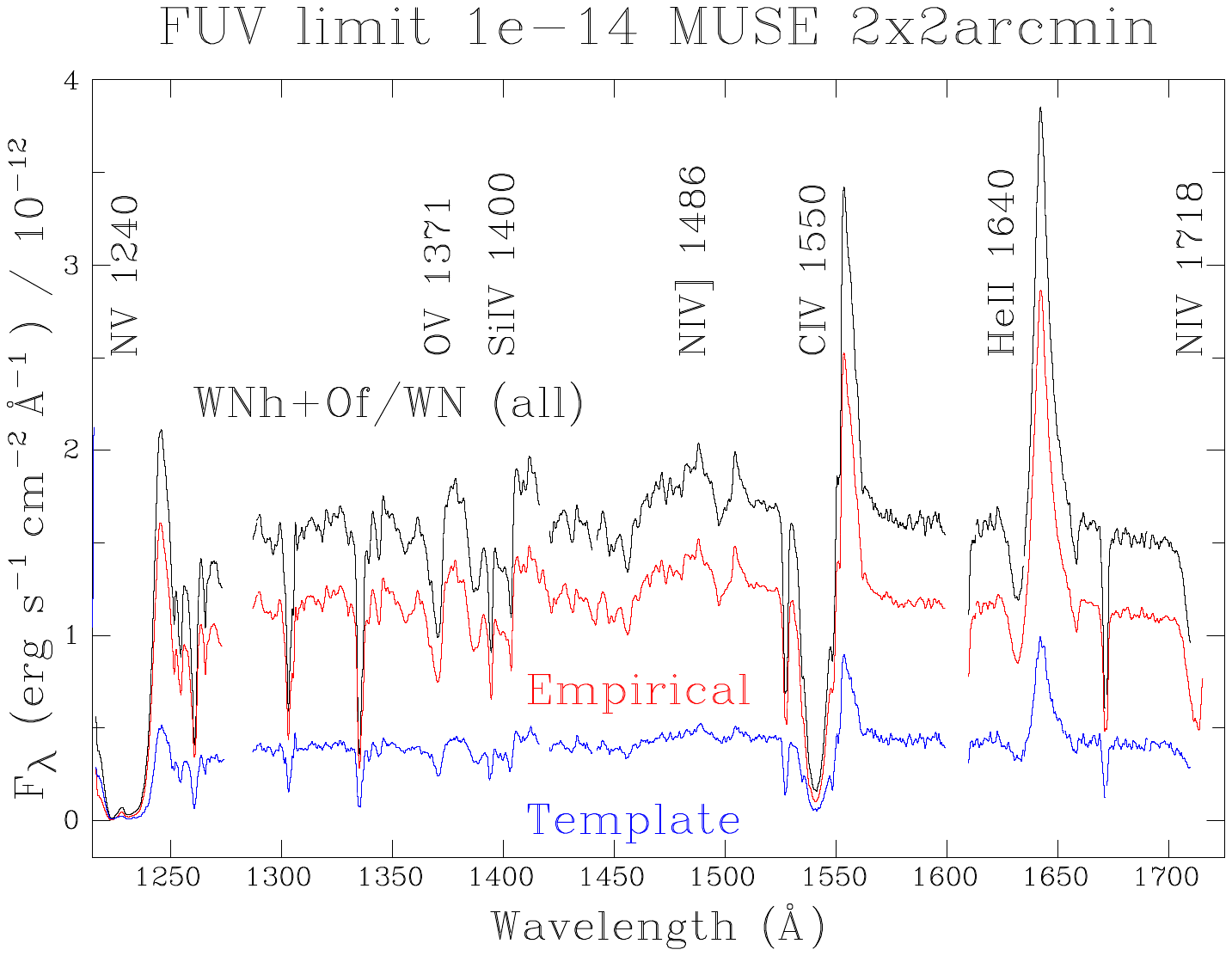}
\caption{Cumulative WNh+Of/WN far-UV spectrum of the central region of NGC~2070 (black) based on empirical data of 7 individual stars (red) and 5 LMC template stars (blue). Gaps in spectra arise from incomplete spectral coverage (e.g. COS G130M+G160M).}
\label{fig:far-UV-OfWN}
\end{figure}

\begin{figure}
\includegraphics[angle=-90,width=0.9\columnwidth,bb=13 46 553 757]{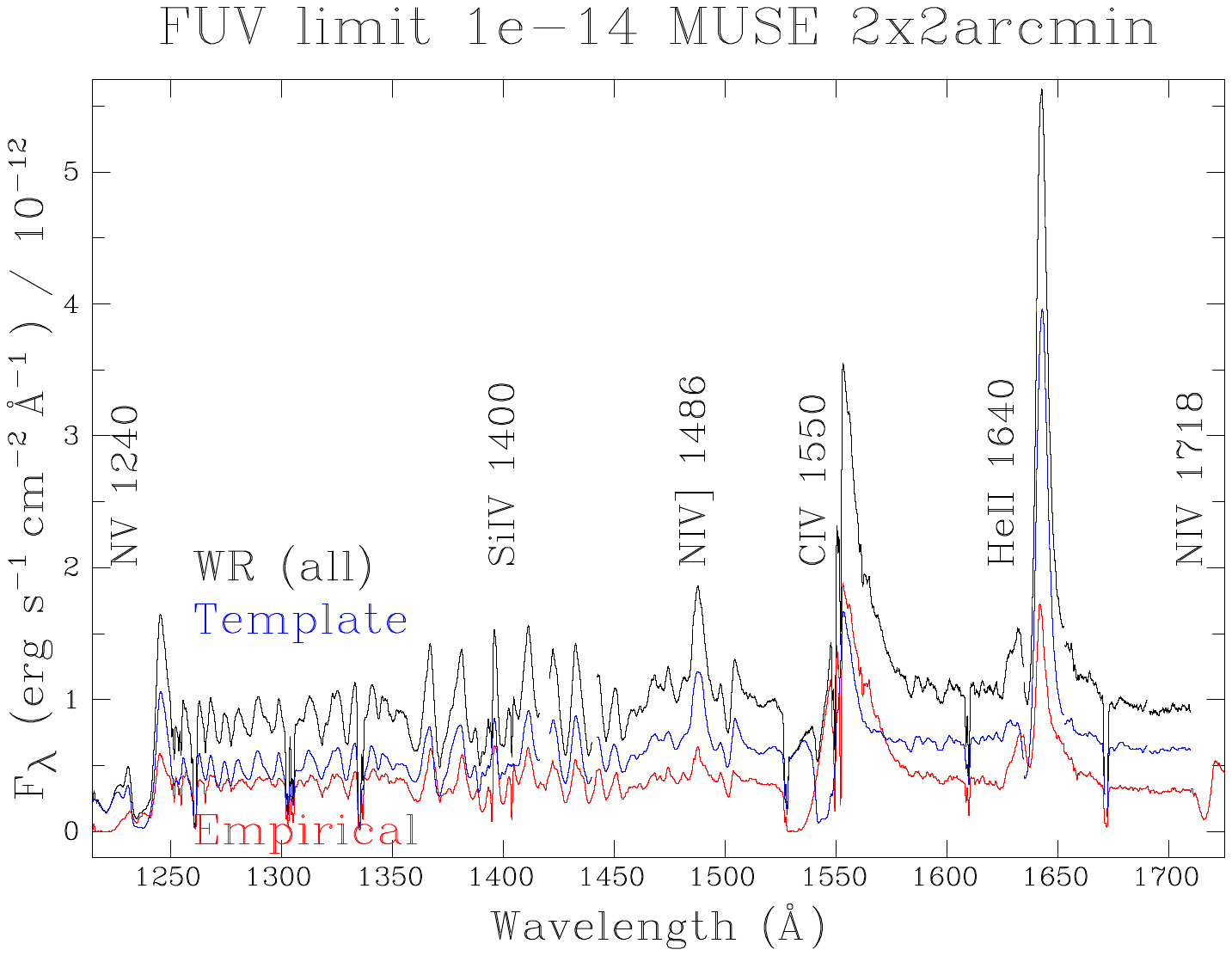}
\caption{Cumulative WR far-UV spectrum of the central region of NGC~2070 (black) based on empirical data of R140a1 (red, WN+WC) and 5 LMC template stars (blue, 4 WN, 1 WC).}
\label{fig:far-UV-WR}
\end{figure}

\begin{figure}
\includegraphics[angle=-90,width=0.9\columnwidth,bb=13 46 553 757]{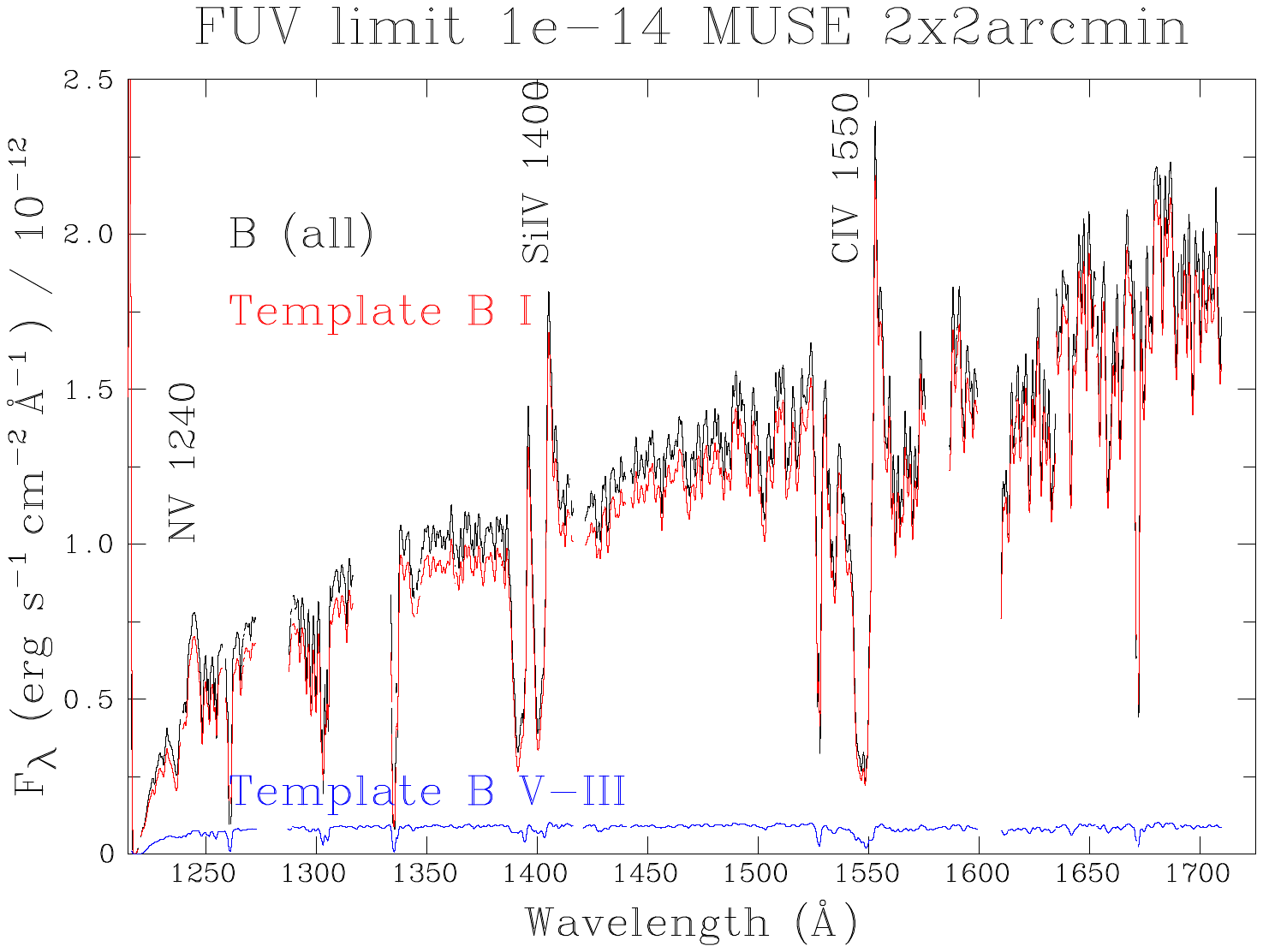}
\caption{Cumulative B-type far-UV spectrum of the central region of NGC~2070 (black) based on LMC templates for 10 supergiants (red) and 6 non-supergiants (blue). Gaps in spectra arise from incomplete spectral coverage (e.g. COS G130M+G160M)}
\label{fig:far-UV-B}
\end{figure}

The NGC~2070 far-ultraviolet spectrum has been constructed from a combination of empirical spectra (of individual stars) and use of template OB stars  (Table~\ref{tab:templates1}) and Of/WN and WR stars (Table~\ref{tab:templates2} for the remainder, primarily drawn from Data Release 6 (DR6) of the ULLYSES survey \citep{2020RNAAS...4..205R}. Figures~\ref{fig:far-UV-O}--\ref{fig:far-UV-B} present cumulative far-UV spectra for O, Of/WN, WR and B stars, respectively, including breakdowns between empirical datasets and templates (separated into supergiants and non-supergiants for B stars).

\begin{table*}
\centering
\caption{LMC template UV spectra of normal OB stars from HST ULLYSES survey \citep{2020RNAAS...4..205R}. COS gratings G130M+G160M are abbreviated as G1\#0M.}
\begin{tabular}{
r @{\hspace{4mm}} 
l @{\hspace{2mm}} l @{\hspace{4mm}} 
l @{\hspace{2mm}} l @{\hspace{4mm}} 
l @{\hspace{2mm}} l }
\hline
                  & \multicolumn{2}{c}{--- Dwarf ---} & \multicolumn{2}{c}{--- Giant ---} &  \multicolumn{2}{c}{--- Supergiant ---} \\                  
Template  & Star (Sp Type)  & Instrument/Grating & Star (Sp Type) & Instrument/Grating & Star (Sp Type) & Instrument/Grating \\
\hline
O2           & BI~237  (O2\,V)      & COS G1\#0M & VFTS 72 (O2\,V--III) & COS G1\#0M & Mk~42 (O2\,If*) & STIS/E140M \\
O2.5        & VFTS 169 (O2.5\,V) & COS G1\#0M & N11 ELS 26 (O2.5\,III)        & COS G1\#0M & $\cdots$ & $\cdots$ \\
O3           & N11 ELS 60 (O3\,V)          & COS G1\#0M & VFTS 267 (O3\,III-I)           & COS G1\#0M & VFTS 180 (O3\,If*) & STIS E140M \\
O3.5        & VFTS 404 (O3.5: V:)              & COS G1\#0M & $\cdots$                & $\cdots$                      & $\cdots$       & $\cdots$ \\
O4           & W61 28-5 (O4\,V)               & COS G1\#0M & Sk --67$^{\circ}$ 69 (O4\,III) & STIS E140M        & Sk --67$^{\circ}$ 166 (O4\,I) & STIS E140M \\
O4.5        & Sk --70$^{\circ}$ 60 (O4--5\,V) & STIS E140M              & $\cdots$              & $\cdots$                       & $\cdots$         & $\cdots$ \\
O5           & PGMW~3120 (O5.5\,V)        & STIS E140M     & N11 ELS 38 (O5\,III)     & COS G1\#0M & [ST92] 4-18 (O5\,If) & COS G1\#0M \\
O6           &  PGMW~3070 (O6\,V)         & STIS E140M       & Sk --71$^{\circ}$ 19 (O6\,III) & COS G1\#0M & Sk --67$^{\circ}$ 111 (O6\,Iafpv) & STIS E140M \\
O6.5        & $\cdots$                    & $\cdots$                      & Sk --71$^{\circ}$ 50 (O6.5\,III) & STIS E140M          & $\cdots$    & $\cdots$ \\
O7            & Sk --67$^{\circ}$ 118 (O7\,V) & STIS E140M      & BI~272 (O7\,II) & STIS E140M &  Sk --69$^{\circ}$ 83 (O7.5\,Iaf) & STIS E140M \\
O8            & BI~184 (O8\,V)                  & COS G1\#0M &  Sk --67$^{\circ}$ 101 (O8\,II) & STIS E140M & PGMW 1363 (O8.5\,Iaf) & STIS E140M \\
O9             & VFTS 66 (O9\,V)     & COS G1\#0M& Sk --71$^{\circ}$ 8 (O9\,II) & STIS E140M & Sk --67$^{\circ}$ 107 (O9\,Ib) & STIS E140M \\
O9.5          & $\cdots$                  & $\cdots$                      & Sk --66$^{\circ}$ 17 (OC9.5\,II) & COS G1\#0M & Sk --67$^{\circ}$ 5 (O9.7\,Ib) & STIS E140M \\
B0             & HV~5622 (B0\,V)      & COS G1\#0M& Sk --70$^{\circ}$ 79 (B0\,III) & STIS E140M           & Sk --68$^{\circ}$ 52 (B0\,Ia) & STIS E140M \\
B0.5          & Sk --67$^{\circ}$ 216 (B0.5\,V) & STIS E140M & $\cdots$       & $\cdots$                          & Sk --68$^{\circ}$ 155 (B0.5\,I) & COS G1\#0M \\
B0.7          & $\cdots$                  & $\cdots$                      & $\cdots$                & $\cdots$                     & Sk --68$^{\circ}$ 140 (B0.7\,Ib-Iab) & COS G1\#0M \\
B1             & Sk --65$^{\circ}$ 2 (B1\,V) & STIS E140M & Sk --71$^{\circ}$ 35 (B1\,II) & COS G1\#0M & Sk --66$^{\circ}$ 35 (BC1\,Ia) & COS G1\#0M \\
B1.5         & $\cdots$                   & $\cdots$                      & $\cdots$                 & $\cdots$                      & Sk --67$^{\circ}$ 14 (B1.5\,Ia) & STIS E140M \\
B2             & $\cdots$                   & $\cdots$                      & $\cdots$                 & $\cdots$                      & Sk --68$^{\circ}$ 26 (BC2\,Ia) & COS G1\#0M \\
B9             & $\cdots$                   & $\cdots$                      & $\cdots$                 & $\cdots$                      & Sk --67$^{\circ}$ 207 (B9\,Ia) & COS G1\#0M \\
\hline
\end{tabular}\par
\label{tab:templates1} 
\end{table*}

\begin{table}
\centering
\caption{LMC template UV spectra of normal Of/WN and WR stars from HST ULLYSES survey \citep{2020RNAAS...4..205R} except where noted}
\begin{tabular}{ r @{\hspace{4mm}} l @{\hspace{2mm}} l @{\hspace{4mm}} }
\hline
Template  & Star (Sp Type)  & Instrument/Grating  \\
\hline
O2\,If/WN5         & Sk -67$^{\circ}$ 22  (O2\,If/WN5) &  STIS/E140M \\
O2.5\,If/WN6         & Mk~39 (O2.5\,If/WN6) & COS G130M+G160M  \\
O3.5\,If/WN7         & Mk 51 (O3.5\,If/WN7) & COS G140L$^{c}$ \\
O4\,If/WN8           & R136b (O4\,If/WN8)           & GHRS G140L$^{a}$ \\
WN5h                    & R136a3 (WN5h)                & GHRS G140L$^{b}$ \\
WN6                      & Sk -71$^{\circ}$ 21 (WN6h) &  STIS E140M \\ 
WN7--8                 & VFTS~108 (WN7h) & COS G140L$^{c}$ \\
WC4                      & Sk -69$^{\circ}$ 191 (WC4)        & STIS E140M \\
\hline
\end{tabular}\par
\label{tab:templates2} 
a: \citet{1998ApJ...509..879D}, b: \citet{1997ApJ...477..792D}; c: GO 15629 (Mahy) \\
\end{table}

\clearpage

\section{Census of far-UV brightest sources of NGC~2070} 

Table~\ref{tab:fuv-list} lists stars of known spectral type within the MUSE footprint, sorted by far-UV flux, based on a calibration of F275W or F336W photometry drawn from HTTP \citep{2013AJ....146...53S, 2016ApJS..222...11S}. HSH95-17 is included since it lies within R136 GHRS/G140L 2$\times$2 arcsec$^{2}$ region and is considered to be an early O star \citep{2022ApJ...935..162K}, although a few other sources exceeding $F_{1500} \geq 10^{-14}$ erg\,s$^{-1}$\,cm$^{-2}$\AA$^{-1}$ have been excluded since their spectral types are unknown. These include HSH95-76 (P 870, $m_{\rm F275W}$ = 13.57 mag, \citet{2016ApJS..222...11S}), SMB~136 ($m_{\rm 275W}$ = 13.88 mag, \citet{2016ApJS..222...11S}), HSH95-87 ($m_{\rm F336W}$ = 14.27 mag, \citet{2016ApJS..222...11S}), HSH95 120 ($m_{\rm F275W}$= 14.55 mag, \citet{2016ApJS..222...11S}), HSH95-139 ($m_{\rm F336W}$ = 14.80 mag, \citet{1995ApJ...448..179H}), HSH95-129 ($m_{\rm F336W}$ = 14.97 mag, \citet{1995ApJ...448..179H}).

Table~\ref{tab:iue-list} lists stars of known spectral type beyond the MUSE footprint but within the $3\times3$ arcmin$^{2}$ region sampled with {\it IUE}/SWP \citep{1995ApJ...444..647V}, sorted by far-UV flux, based on a calibration of F275W or F336W photometry drawn from HTTP \citep{2013AJ....146...53S, 2016ApJS..222...11S}.  Bright sources lacking spectral
types include SMB-183 ($m_{\rm 275W}$ = 14.11 mag, \citet{2016ApJS..222...11S}), SMB-196 ($m_{\rm 275W}$ = 14.27 mag, \citet{2016ApJS..222...11S}) and SMB-245 ($m_{\rm 275W}$ = 14.67 mag, \citet{2016ApJS..222...11S}). 

\begin{table*}
\centering
\caption{Stars within the NGC~2070 MUSE field of view, sorted by far-UV flux ($F_{1500}$  units of erg\,s$^{-1}$\,cm$^{-2}$\AA$^{-1}$), either measured from spectroscopy or estimated from photometry, primarily drawn from HTTP \citep{2013AJ....146...53S, 2016ApJS..222...11S}, the latter indicated in parentheses. Catalogues include R \citep{1960MNRAS.121..337F}, Mk \citep{1985A&A...153..235M}, P \citep{1993AJ....106..560P},  HSH \citep{1995ApJ...448..179H}, SMB \citep{1999A&A...341...98S}, VFTS \citep{2011A&A...530A.108E} and CCE \citep{2018A&A...614A.147C}. Spectra used in our integrated far-UV spectrum of NGC~2070 are of individual sources or templates with the exception of GHRS/G140L spectroscopy of the central 2$\times$2 arcsec of R136a (indicated with $\ddag$) while several templates are adjusted to measured STIS/G140L flux levels from \citet[][indicated with $\dag$]{2005ApJ...627..477M}. Further details of far-UV templates are provided in Tables~\ref{tab:templates1}--\ref{tab:templates2}. Sources considered to be very massive stars (VMS, $\geq$100 $M_{\odot}$) from spectroscopy analyses \citep{2014A&A...570A..38B, 2019MNRAS.484.2692T, 2022A&A...663A..36B} are indicated with $\checkmark$. COS gratings G130M+G160M are abbreviated as G1\#0M.}
\begin{tabular}{
r @{\hspace{1mm}} r @{\hspace{1mm}} r @{\hspace{1mm}} r @{\hspace{1mm}} r @{\hspace{1mm}} r @{\hspace{1mm}} r @{\hspace{1mm}} c @{\hspace{1mm}} c @{\hspace{1mm}} c @{\hspace{1mm}} c @{\hspace{1mm}} c @{\hspace{1mm}} c  @{\hspace{1mm}} c @{\hspace{1mm}} c @{\hspace{1mm}} c @{\hspace{1mm}} c @{\hspace{1mm}} c}
\hline
R       & Mk      &   P  & HSH  & SMB  & VFTS& CCE  & SpT          & Ref & HTTP & $m_{\rm 275W}$ & $m_{\rm F336W}$ & $m_{\rm F555W}$ & Ref & $F_{1500}$ & Spectrum & Ref & VMS \\
          &           &       &          &           &           &           &                 &         &           & mag                     &           mag            & mag                      &        & $10^{-13}$ & (Template) &      \\
\hline
140a  &$\cdots$& 877 & $\cdots$ &  6 & 507    & 3191   & WC4+WN6+ &  Do13       & 053841.601-690513.43 & 10.97        & 11.28       & 12.20                  &   S     & 3.72  & STIS/E140M & 1 & $\cdots$ \\
136a1 &$\cdots$& $\cdots$ &         3   & 7 &$\cdots$&$\cdots$& WN5h           &  CD98       &    $\cdots$             & $\cdots$            & 11.20      & 12.28                 &    D    & (3.32)  & GHRS/G140L$\ddag$  &  4 & $\checkmark$ \\
140b   &$\cdots$& 880  &  $\cdots$&   8& 509   & 3174  & WN5(h)+O  & Ev11          & 053841.613-690515.17  & 11.09     & 11.43       & 12.47                  &    S     & (3.04)    & (WN6) & $\cdots$ & $\cdots$ \\
$\cdots$ & 42  & 922   &          2  & 10& $\cdots$& 2102 & O2\,If           &  CW11       & 053842.104-690555.29 & 11.08        & 11.51       & 12.82                 &    S    & 3.01         & STIS/E140M & 3 & $\checkmark$ \\
136a2 &$\cdots$& $\cdots$&   5  &$\cdots$&$\cdots$&$\cdots$& WN5h       & CD98        & $\cdots$                   & $\cdots$         &  11.33       & 12.34                  &    D     & (2.93)  & GHRS/G140L$\ddag$ & 4 & $\checkmark$ \\ [2pt]
 $\cdots$ & 39 & 767  & 7     & 14 & 482 & 2003 & O2.5\,If/WN6+ & Cr22  & 053840.214-690559.86 & 11.38 & 11.66         & 12.95                   &  S      & 2.74          & COS/G1\#0M & 3 & $\checkmark$ \\
142      &$\cdots$ & 987 & 1 & 3 &  533  & 2912  & B1.5\,Ia$^{+}$  &  Ev15 & 053842.738-690542.57 & 10.84      & 11.13         & 11.79                  &   S       & (2.50)        & (B1.5\,I) & $\cdots$ & $\cdots$ \\
134      &$\cdots$ & 786 & 4 & $\cdots$& 1001 & 1978 & WN6(h)       & CS97 & 053840.539-690557.18        & 11.45     & 11.62         & 12.70                   & S        &(2.18)        & (WN6)  & $\cdots$ & $\cdots$ \\
137      &$\cdots$ & 548 & $\cdots$ & 5 & 431  & 2889   & B1.5\,Ia         &  Ev15 & 053836.959-690507.84     & 11.04    & 11.30        & 12.08                    & S        & (2.08)         & (B1.5\,I) & $\cdots$ & $\cdots$ \\
$\cdots$ & 25 & 871 & $\cdots$ & 19 & 506  & 2395  & ON2\,V        & Wa14 & 053841.545-690519.43    & 11.74    & 12.01       & 13.32                     & S         & 2.04       & COS/G1\#0M & 3 & $\checkmark$ \\ [2pt]
141  &$\cdots$ & 1253 & $\cdots$& 9 & 590 & 2190    & B0.7\,Iab      & Ev15  & 053845.579-690547.80      & 11.42   & 11.60        & 12.58                     & S        & (1.92)       & (B0.7\,I) & $\cdots$ & $\cdots$ \\
136a3 & $\cdots$ & $\cdots$ & 6 & $\cdots$ & $\cdots$  & $\cdots$ & WN5h           & CD98  & $\cdots$                   & $\cdots$       & 11.82       & 12.97                      & D        & (1.87)       & GHRS/G140L$\ddag$ & 4 & $\checkmark$ \\
$\cdots$ & 30  & 1018 &        15 & 24 & 542  & 2999   & O2\,If/WN5  & CW11 & 053843.080-690546.86   & 11.91       & 12.22  & 13.48                         & S           & 1.85        & COS/G1\#0M & 3 & $\cdots$ \\
$\cdots$ & 12 & 1257 & $\cdots$ & 11 & 591  & 1279    & B0.2\,Ia       & Ev15  & 053845.687-690622.49   & $\cdots$ & 11.68           & 12.56                     & S       & (1.80)       & (B0\,I)        & $\cdots$  &$\cdots$ \\   
$\cdots$ & 35   & 1029 &      12 & 23  & 545  & 1474   & O2\,If/WN5  & CW11 & 053843.202-690614.44   & 11.93     & 12.23    & 13.46                      & S &        (1.40) & (O2\,I/WN5) & $\cdots$ & $\checkmark$ \\ [2pt]
$\cdots$ & 32    & 1130 &     13 & 21 & 1034 & 3043  & O7.5\,II        & WB97 & 053844.192-690547.06    & 11.96     & 12.23     & 13.40                   & S  &   (1.37)      & (O8\,III)        & $\cdots$ & $\cdots $\\
$\cdots$ & 34    & 1134 &      8 & 17  & $\cdots$ & 1766 & WN5h+WN5h &  Te19 & 053844.252-690605.93      & 12.04     & 12.11      & 13.15                 & S    & (1.28)       & (WN5)        & $\cdots$ & $\checkmark$ \\
138   & $\cdots$ &   499 & $\cdots$ &  4 & 424  & $\cdots$ & B9\,I+p      & Ev15  & 053836.132-690558.01       & $\cdots$   & 11.50 & 11.79                    & S     & (1.25)      & (B9\,I)           & $\cdots$ & $\cdots$ \\
$\cdots$ & 37Wa & 917 & 11     & 25 & 1021 &  1349   & O4\,If$^{+}$  &  MH98  & 053842.072-690614.32   & 12.16         & 12.35    & 13.36               & S  & (1.14)         & (O4\,I)          & $\cdots$ &   $\checkmark$ \\ [2pt]
$\cdots$ & 11       & 1500 & $\cdots$&15 & 641 & 762   & B0.5:\,I      & Ev15   & 053849.723-690642.95     & $\cdots$    & 12.19    & 13.18               & S    & (1.13)       & (B0.5\,I)      & $\cdots$ & $\cdots$ \\
$\cdots$ & 35Sa  & 1036 & 23 & 37 & 1028  & 1423 & O4--5\,V       & WB97   & 053843.263-690616.51     & 12.18        & 12.53  & 13.85                  & S    & (1.12)      & (O4.5\,V) & $\cdots$ & $\cdots$ \\
$\cdots$ & $\cdots$ & 1080 & 25 & 39 & 1031 & 2186 & O3--4\,V       & Bo99        & 053843.684-690547.89      & 12.20       & 12.56   & 14.23                & S   & (1.10)        & (O3\,V)    & $\cdots$ & $\cdots$ \\
$\cdots$ & 13       & 1311 & $\cdots$ & 40 & 599 & 1433    & O3\,III          & Wa14      & 053846.177-690617.39       & 12.28      & 12.63   & 13.85                & S    & (1.02)  & (O3\,III)      & $\cdots$ & $\cdots$ \\
$\cdots$ & 26        & 1150 & $\cdots$ & 32 &  562  & 2819    & O4\,III         & WB97       & 053844.406-690536.22       & 12.28      & 12.46    & 13.70                & S & (1.02)     & (O4\,III)      & $\cdots$ & $\cdots$ \\ [2pt]
136a5  & $\cdots$ & $\cdots$ & 20 & $\cdots$ & $\cdots$ & $\cdots$ & O2\,If          & Cr16      &   $\cdots$                             & $\cdots$ & 12.49   & 13.71                & D & (1.01)  & GHRS/G140L$\ddag$ & 4 & $\checkmark$ \\
136c  & $\cdots$ & 998 & 10 & 27 & 1025 & 1737    & WN5h+?     & Cr10     & $\cdots$                               & $\cdots$ & 12.54   & 13.43                & D & (0.97)          & (WN5)      & $\cdots$ & $\checkmark$ \\
$\cdots$ & 47  & 607 & $\cdots$ & 29  & 440 & 2417  & O6--6.5\,III & Wa14   & 053837.729-690521.03       & 12.15  & 12.44 & 13.69                    & S & 0.96            & STIS/E140M    & 2 & $\cdots$ \\
$\cdots$ & 33Na & 1140 & 16  & 33 & $\cdots$ & 1943 & OC2.5\,If+O4\,V &  Br22 & 053844.329-690554.66 & 12.37 & 12.55 & 13.64 & S & (0.94) & (O3\,I) & $\cdots$ & $\cdots$ \\
$\cdots$ & 54      & 488 & $\cdots$ & 16 & 420  & 1689 & B0.5\,Ia                 & Ev15 &  053835.941-690609.23 & 12.20 & 12.26 & 13.10 & S & (0.93) & (B0.5\,I) & $\cdots$ & $\cdots$ \\ [2pt]
$\cdots$ & 27       & 850 & $\cdots$ & 38 & 502  & 2653 & O9.7\,II                  & Wa14 & 053841.271-690532.44 & 12.39 & 12.67 & 13.82 & S & (0.92) & (O9.5\,III) & $\cdots$ & $\cdots$ \\
$\cdots$ & $\cdots$ & 1014 & $\cdots$ & 41 & $\cdots$ & 3062  & O8:                         & Ca21 & 053843.030-690540.44 & 12.43 & 12.57 & 13.60 & S & (0.89) & (O8\,III) & $\cdots$ & $\cdots$ \\
136a7    & $\cdots$& $\cdots$ &  24  & 20 & $\cdots$ & $\cdots$ & O3\,III(f*)              & Be20 & $\cdots$                        & $\cdots$ & 12.64 & 13.97 & S & (0.88) & GHRS/G140L$\ddag$   & 4 & $\checkmark$ \\
$\cdots$ & $\cdots$ & 860 & 28  & 53    & $\cdots$ & 1912       &   O3\,V                  &  MH98 & 053841.490-690556.90 & 12.45 & 12.80 & 14.04 & S & (0.87) & (O3\,V) & $\cdots$ & $\cdots$ \\
$\cdots$ & $\cdots$ & 1231 & $\cdots$ & 31  & 585 & 2193  & O7\,V                     & Wa14 & 053845.279-690546.53 & $\cdots$ & 12.66 & 13.78 & S & (0.87) & (O7\,V) & $\cdots$ & $\cdots$ \\ [2pt]
136b & $\cdots$ & 985   & 9      & 18 & $\cdots$ & 1669        & O4\,If/WN8 & Cr16 &   $\cdots$                           & $\cdots$    & 12.27 & 13.24                  & D     & 0.86       & GHRS/G140L & 5 & $\checkmark$ \\
$\cdots$ & 37a  & 949    &  14 & 28 & 1022 & 1442 & O3.5\,If/WN7        & CW11 & 053842.397-690615.08  & 12.33 & 12.48 & 13.52 & S & 0.82$\dag$ & (O3.5\,If/WN7) & 6 & $\checkmark$ \\
$\cdots$ & 24    & 1260 &$\cdots$ &  47  & $\cdots$ & 2760   & O3\,V                    & WB97 & 053845.687-690539.02  & 12.53 & 12.73 & 13.96 & S & (0.81) & (O3\,V) & $\cdots$ & $\cdots$ \\
140c  & $\cdots$ &  908 & $\cdots$ & 55 &  519 & 3112 & O3--4((f))+OB       & Wa14 & 053841.934-690513.02 & 12.53 & 12.81 & 14.18 & S & (0.81) & (O3.5\,V) & $\cdots$ & $\cdots$ \\
$\cdots$ & $\cdots$ & 863 &  29  & 56  &1014 & 1956: & O3\,V                 & MH98 & 053841.507-690600.92 & 12.55 & 12.90 & 14.18 & S & (0.80) & (O3\,V) & $\cdots$ & $\cdots$ \\
$\cdots$ & 33Sa     & 1120 & 18 & 44   &$\cdots$ & 2177 & O3\,III                 & Ma15 & 053844.123-690556.63 & 12.48 & 12.76 & 13.81 & S & 0.79$\dag$ & (O3\,III) & 6 &$\cdots$ \\ [2pt]
$\cdots$ & 50        & 643 & $\cdots$ & 34 & 450 & 1293 & O9.7\,III:+O7::      & Wa14 & 053838.476-690621.96 & 12.57 & 12.75 & 13.69 & S & (0.78) & (O9.5\,III) & $\cdots$ & $\cdots$ \\
136a4 & $\cdots$ & $\cdots$ &  21 & $\cdots$ & $\cdots$ & $\cdots$ & O3\,V              & Be20 & $\cdots$                              & $\cdots$ & 12.81 & 13.96 & H & (0.75) & GHRS/G140L$\ddag$    &  4 & $\checkmark$ \\
$\cdots$ & $\cdots$ & 923  & 33 & 60  & $\cdots$ & 1793 & O3\,V               & Ma05 & 053842.119-690600.73 & $\cdots$ & 13.09 & 14.33 & S & 0.75$\dag$ & (O3\,V) & 6 & $\cdots$ \\
136a6 \#1 & $\cdots$ & $\cdots$  & 19 & $\cdots$ & $\cdots$ & $\cdots$ & O2\,I(n)f*p+?  & Be20 & $\cdots$                             & $\cdots$ & 12.86 & 13.92 & H & (0.72) & GHRS/G140L$\ddag$  &  4 & $\cdots$ \\
$\cdots$ & 49 & 691 & $\cdots$ & 30 & $\cdots$ & 1261    & WN6(h)           & CS97 & 053839.143-690621.24 &  12.66 & 12.62 & 13.41 & S & (0.71) & (WN6) & $\cdots$ & $\cdots$ \\ [2pt]
136a6 \#2   & $\cdots$ & $\cdots$  & 26 & $\cdots$ & $\cdots$ & $\cdots$ & O2\,I(n)f*p+?  & Be20 & $\cdots$                             & $\cdots$ & 12.89 & 14.19 & H & (0.70) & GHRS/G140L$\ddag$  &  4 & $\cdots$ \\
$\cdots$ & $\cdots$ & 912 & 38 & 70 & 1019 & 1608 & O3\,V+O6\,V & Ma02 & 053842.004-690607.56     & 12.72 & 13.02 & 14.30 & S & (0.68) & (O3\,V) & $\cdots$ & $\cdots$ \\
$\cdots$ & 6  & 1563 & $\cdots$ & 61  & 656 & 1979 & O7.5\,IIIp       & Wa14 & 053851.200-690559.28      & 12.73 & 13.11 & 14.36 & S & (0.68) & (O7\,III) & $\cdots$ & $\cdots$ \\
$\cdots$ & 8  & 1531  & $\cdots$ & 58 & 648 & 2780 & O5.5\,IV            & Wa14 & 053850.400-690538.17      & 12.73 & 12.96 & 14.23 & S & (0.67) & (O5.5\,V) & $\cdots$ & $\cdots$ \\
$\cdots$ & 38 & 930    & $\cdots$ & 45 & 525  & 1184 & B0\,Ia              &  Wa14 & 053842.209-690625.56       & 12.55 & 12.79 & 13.83 & S & (0.67) & (B0\,I)   & $\cdots$ & $\cdots$ \\ [2pt]
136a8    & $\cdots$ & $\cdots$ & 27           & $\cdots$ & $\cdots$ & $\cdots$ &  O2--3\,V            &  Cr16           & $\cdots$                   & $\cdots$ & 12.93 & 14.22 & H & (0.67) & GHRS/G140L$\ddag$  &  4 & $\cdots$ \\
$\cdots$& $\cdots$ & $\cdots$ & 17           & $\cdots$ & $\cdots$ & $\cdots$  & O                       & Ka22 & $\cdots$                    & $\cdots$ & 13.00 & 13.78 & H & (0.63) &  GHRS/G140L$\ddag$ &  4 & $\cdots$ \\
$\cdots$ & $\cdots$ & 1273      & $\cdots$ & 59          & $\cdots$ & 1223        & O7:                  & Ca21            & 053845.842-690620.83 & 12.82 & 13.05 & 14.05 & S & (0.62) & (O7\,III) & $\cdots$ & $\cdots$ \\
$\cdots$ & $\cdots$ & $\cdots$ & 30          & $\cdots$ & $\cdots$ & $\cdots$  & O6.5\,Vz          & Be20          & $\cdots$                    & $\cdots$  & 13.02 & 14.21 & D & (0.62) & GHRS/G140L$\ddag$ &  4 & $\cdots$ \\
$\cdots$ & $\cdots$ & $\cdots$ & 31          & 35           & $\cdots$ & $\cdots$ & O2\,V               & Be20           & 053842.471-690604.53 & $\cdots$ & 12.89 & 14.05 & S & 0.61$\dag$ & (O2\,V) & $\cdots$ & $\cdots$ \\ [2pt]
\hline
\end{tabular}\par
\label{tab:fuv-list}
\end{table*}

\addtocounter{table}{-1}

\begin{table*}
\centering
\caption{(continued)}
\begin{tabular}{
r @{\hspace{1mm}} r @{\hspace{1mm}} r @{\hspace{1mm}} r @{\hspace{1mm}} r @{\hspace{1mm}} r @{\hspace{1mm}} r @{\hspace{1mm}} c @{\hspace{1mm}} c @{\hspace{1mm}} c @{\hspace{1mm}} c @{\hspace{1mm}} c @{\hspace{1mm}} c  @{\hspace{1mm}} c @{\hspace{1mm}} c @{\hspace{1mm}} c @{\hspace{1mm}} c @{\hspace{1mm}} c}
\hline
R       & Mk      &   P  & HSH  & SMB  & VFTS& CCE  & SpT          & Ref & HTTP & $m_{\rm 275W}$ & $m_{\rm F336W}$ & $m_{\rm F555W}$ & Ref & $F_{1500}$ & Spectrum & Ref & VMS \\
          &           &       &          &           &           &           &                 &         &           & mag                     &           mag            & mag                      &        & $10^{-13}$ & (Template) &      \\
\hline
$\cdots$ & $\cdots$ &  713       & $\cdots$ & 72           & $\cdots$ & 2245       & O5\,V               & Bo99            & 053839.478-690510.36 & 12.85 & 13.20 & 14.60 & S & (0.60) & (O4.5\,V) & $\cdots$ & $\cdots$ \\
$\cdots$ & $\cdots$ & $\cdots$ & 36          & $\cdots$ & $\cdots$ & $\cdots$  & O2\,If              & Be20            & $\cdots$                    & $\cdots$ & 13.06 & 14.41 & S & (0.59) & (O2\,I) & $\cdots$ & $\checkmark$ \\
$\cdots$ & $\cdots$ & $\cdots$ & 35          & $\cdots$ & $\cdots$ & $\cdots$  & O3\,V              & Be20            & $\cdots$                    & $\cdots$ & 13.12 & 14.43 & D & 0.55$\dag$ & GHRS/G140L$\ddag$ & 4 & $\cdots$ \\
$\cdots$ & $\cdots$ & 975         & $\cdots$ & 71          & $\cdots$ & 2748      & O6--7\,V         & Bo99         & 053842.616-690536.74 & 12.93 & 13.18 & 14.50 & S & (0.56) & (O6\,V) & $\cdots$ & $\cdots$ \\
$\cdots$ & 15           & 1312       & $\cdots$ & 43         & $\cdots$ & 2165       & O7\,V             & WB97           & 053846.158-690551.37 & $\cdots$ & 13.14 & 14.13 & S & (0.55) & (O7\,V) & $\cdots$ & $\cdots$ \\ [2pt]
$\cdots$ & $\cdots$ & 1113       & 53            & 78        & 1032       & 2977       & O8\,III             & Le21        & 053844.063-690544.82  & 12.97 & 13.32 & 14.63 & S & (0.54) & (O8\,III) &$\cdots$ & $\cdots$ \\
140d      & $\cdots$  & $\cdots$ & $\cdots$ & 81          & 497       & 2231        & O3.5\,Vz+OB & Wa14          & 053841.126-690513.17 & 12.97 & 13.28 & 14.65 & S & (0.54) & (O3.5\,V) & $\cdots$ & $\cdots$ \\
$\cdots$ & $\cdots$  & $\cdots$ & 42          &$\cdots$ & $\cdots$ & $\cdots$  & O3\,V+O3\,V &  Ma02         & 053842.119-690600.73 & 12.98 & 13.34 & 14.71 & S,D & (0.54) & (O3\,V) & $\cdots$& $\cdots$ \\ 
$\cdots$ & $\cdots$  & $\cdots$ & 75          &$\cdots$ & $\cdots$   & $\cdots$ & O6\,V           & Be20            & 053842.178-690601.90 & 13.01 & 13.86 & 15.08 & S,D & (0.52) & (O6\,V) & $\cdots$ & $\cdots$ \\ 
$\cdots$ & $\cdots$  & 288      & $\cdots$   &  84         & 385          & 2451     & O4--5V         & Wa14           & O53832.293-690523.85  & 13.22 & 13.38 & 14.65 & S & 0.51 & COS/G140L & 2 & $\cdots$ \\ [2pt]
$\cdots$ & 33Nb       & 1152    & 32            & 66          & $\cdots$  & 1896      & O6.5\,V        & Be20            & 053844.467-690555.50 & 13.03 & 13.28 & 14.36 & S & (0.51) & (O6\,V) & $\cdots$& $\cdots$ \\
$\cdots$ & $\cdots$ & 1034     & 61            & 85         & 1027        & 2987       & O5\,V             & Le21 &  053843.195-690542.61 & 13.05 & 13.38 & 14.68 & S & (0.50) & (O4.5\,V) & $\cdots$ & $\cdots$ \\  
$\cdots$ & $\cdots$ & $\cdots$ & 39           & $\cdots$ & 1005     & $\cdots$  & O3\,V+O5.5\,V & Ma02        & $\cdots$                       & $\cdots$ & 13.26 & 14.50 & D & (0.50) & (O3\,V) & $\cdots$ & $\cdots$ \\ 
$\cdots$ & 53           & $\cdots$ & $\cdots$ & 51        &   427      & 389           & WN8(h)             & Ev11        & 053836.407-690657.48 & $\cdots$ & 13.27 & 13.77 & S & (0.49)  & (WN7--8) & $\cdots$ & $\cdots$ \\ 
$\cdots$ & 35N         & 1013     & 41           & 76         & 1026     & 1494         & O3\,III(f*)              & MH98         & 053843.075-690611.28  & $\cdots$ & 13.29 & 13.97 & S,D & (0.48) & (O3\,III) & $\cdots$& $\cdots$ \\ [2pt] 
$\cdots$ & $\cdots$ & 1042       & 56          & 87         & 1029        & 2128     & O3.5I+OB      & Wa14       & 053843.343-690547.55 & 13.10 & 13.39 & 14.71 & S,D & (0.48) & (O3\,I) & $\cdots$ & $\cdots$ \\
$\cdots$ & $\cdots$  & $\cdots$ & 40          & $\cdots$ & $\cdots$  & $\cdots$ & O3\,V              & Be20          & $\cdots$                          & $\cdots$ & 13.30 & 14.56 & D &  (0.48) & (O3\,V) & $\cdots$ & $\cdots$ \\ 
$\cdots$ & $\cdots$ & 1195         &  43        & 83        & $\cdots$ & 2112       & O3\,V              & MH98         & 053844.950-690554.11 & 13.12 & 13.37 & 14.57 & S & (0.47) & (O3\,V) & $\cdots$ & $\cdots$ \\
$\cdots$ & $\cdots$ & 1123         & $\cdots$ & 74      &  1033          & 2913     & O7\,III             & Le21        & 053844.172-690542.17 & 13.13 & 13.42 & 14.53 & S & (0.46) & (O7\,III) & $\cdots$& $ \cdots$ \\
$\cdots$ & $\cdots$ & 885          & $\cdots$ & 68       &  512            & 1199   & O2\,V-III         & Wa14        & 053841.734-690625.01 & 13.15 & 13.30 & 14.34 & S & (0.46) & (O2\,V) & $\cdots$ & $\cdots$ \\ [2pt] 
$\cdots$ & 51            & 666        & $\cdots$  & 50        &  457      & 603           & O3.5\,If/WN7 & CW11            & 053838.838-690649.49 & $\cdots$ & 12.98 & 13.81 & S & 0.45 & COS/G140L & 2 & $\checkmark$ \\
$\cdots$ & $\cdots$ & 724            & $\cdots$ & 75      & $\cdots$     & 1274  & O7\,III             & Bo99         & 053839.692-690624.01 & 13.17 & 13.37 & 14.41 & S & (0.45) & (O7\,III) & $\cdots$ & $\cdots$ \\
$\cdots$ & 14S        & 1350          & $\cdots$ & 64     & 608             & 1827  & O4\,III              & Wa14       & 053846.785-690603.10 & 13.01 & 13.19 & 14.29 & S & 0.43 & COS/G140L  & 2 & $\cdots$ \\
$\cdots$ & $\cdots$ &  723           & $\cdots$ & 92     & $\cdots$    & 2570    & O5:                  & Ca21        & 053839.638-690526.37 & 13.21 & 13.53 & 14.77 & S & (0.43) & (O5\,III) & $\cdots$ & $\cdots$ \\
$\cdots$ & 36            & 706           & $\cdots$ & 86    & 468            & 1749    & O2\,V            & Wa14          & 053839.369-690606.49 & 13.21 & 13.39 & 14.58 & S & (0.43) & (O2\,V) & $\cdots$& $\cdots$ \\ [2pt]
$\cdots$ & $\cdots$  & 812           &  51         & 89      & $\cdots$ & $\cdots$  & O3\,V             & MH98          & 053840.895-690555.93 & 13.23 & 13.48 & 14.70 & S & (0.43) & (O3\,V) & $\cdots$ & $\cdots$ \\
$\cdots$ & 27E          & 858          & $\cdots$ & 100    & 503       & $\cdots$ & O9\,III            & Wa14         & 053841.367-690532.44 & 13.23 & 13.57 & 14.87 & S,D & (0.43) & (O9\,III) & $\cdots$ & $\cdots$ \\
$\cdots$ & $\cdots$   & 1267        & $\cdots$ & 90      & $\cdots$ & 2911    & O7:\,V               & Pa93       & 053845.757-690540.82 & 13.24 & 13.52 & 14.72 & S & (0.42) & (O7\,V) & $\cdots$& $\cdots$ \\
$\cdots$ & 28             & 805         & $\cdots$   & 80      & $\cdots$ &  2447    & O5--6\,V           & Bo99        & 053840.798-690525.10 & 13.24 & 13.40 & 14.61 & S & (0.42) & (O5.5\,V) &$\cdots$ & $\cdots$ \\
$\cdots$ & $\cdots$    & 787         & $\cdots$ & 103      & $\cdots$ & 2233   & O9--B0\,V & Bo99             & 053850.435-690534.56  & 13.25 & 13.60 & 14.98 & S & (0.42) & (O9\,V) & $\cdots$ & $\cdots$ \\ [2pt]
$\cdots$ & 4                & 1607       & $\cdots$  & 65      & 664          & 774     & O7\,II         & Wa14         & 053852.724-690643.13  & $\cdots$ & 13.37 & 14.38 & S & 0.41 & COS/G140L & 2 & $\cdots$ \\      
$\cdots$ & $\cdots$  & $\cdots$   & 52           & $\cdots$ & $\cdots$  & $\cdots$ & O3--4\,Vz        &  Be20      & $\cdots$                        & $\cdots$  & 13.46 & 14.72 & D & (0.41) &  GHRS/G140L$\ddag$ & 4 & $\cdots$ \\
$\cdots$ & $\cdots$  & $\cdots$ & 45           &  $\cdots$ & $\cdots$  & $\cdots$ & O4:\,Vz         & Be20         & $\cdots$                       & $\cdots$ & 13.48 & 13.65 & D & (0.40)   &  (O4\,V) &$\cdots$ & $\cdots$ \\
$\cdots$ & $\cdots$  & 900          &  37        & 77         & 1018         & 1459    & O2--4.5         & He12        & 053841.874-690612.52 & 13.29 & 13.42 & 14.49 & D & (0.40) & (O3\,III) & $\cdots$ & $\cdots$ \\
$\cdots$ & $\cdots$  & $\cdots$   & 49        & $\cdots$ & $\cdots$  & $\cdots$   & O3\,V            & Be20       & $\cdots$                       & $\cdots$ & 13.49 & 14.75 & D & (0.40) & (O3\,V) & $\cdots$& $\cdots$ \\ [2pt]
$\cdots$ & $\cdots$  & $\cdots$   & 50            & $\cdots$ & $\cdots$  & $\cdots$   & O3--4\,V       & Be20        & $\cdots$                       & $\cdots$ & 13.51 & 14.65 & D & (0.39) & GHRS/G140L$\ddag$ &  4 & $\cdots$ \\
$\cdots$ & $\cdots$  & $\cdots$   & 55            & $\cdots$ & $\cdots$  & $\cdots$   & O2\,Vz           & Be20         & $\cdots$                      & $\cdots$ & 13.52 & 14.74 & D & (0.39) & (O2\,V) & $\cdots$& $\cdots$ \\
$\cdots$ & 5              & 1552        & $\cdots$  & 54          &  652       & 1405      & B2\,Ip+O9III:      & Wa14          & 053851.043-690620.40 & $\cdots$  & 13.06 & 14.15 & S & (0.39) & (B2\,I) & $\cdots$ & $\cdots$ \\
$\cdots$ & $\cdots$  & $\cdots$   & 46            & $\cdots$ & $\cdots$  & $\cdots$   & O2--3\,III        & Be20            & $\cdots$                       & $\cdots$ & 13.53 & 14.56 & S & (0.39) & (O2.5\,III) & $\cdots$ & $\cdots$ \\
$\cdots$ & $\cdots$ &  506          & $\cdots$   & 82         & $\cdots$ & 1632     & O8:                & Ca21            & 053836.237-690608.42 & 13.35 & 13.52 & 14.55 & S & (0.38) & (O8\,III) & $\cdots$ & $\cdots$ \\ [2pt]
$\cdots$ & 14N        & 1317       & $\cdots$     & 91         & 601    & 1890      & O5--6\,V       & Wa14           & 053846.280-690559.32 & 13.36 & 13.56 & 14.68 & S & (0.38) & (O5.5\,V) &$\cdots$& $\cdots$ \\
$\cdots$ & 7            & 1553        & $\cdots$    & 94         & 651    & 2057      & O7\,V           & Wa14           & 053851.029-690554.70 & 13.36 & 13.59 & 14.74 & S & (0.38) & (O7\,V) &$\cdots$ & $\cdots$ \\
$\cdots$ & $\cdots$  & $\cdots$   & 47            & 111       & $\cdots$ & $\cdots$ & O2\,V          & Be20            & 053842.630-690601.92 & 13.39 & 13.66 & 14.72 & S & (0.37) & (O2\,V) &$\cdots$& $\cdots$ \\
$\cdots$ & $\cdots$  & $\cdots$   & 48            & $\cdots$ & $\cdots$  & $\cdots$   & O2--3\,III    & Be20           & $\cdots$                     & $\cdots$ & 13.60 & 14.75 & D & (0.36) & (O2.5\,III) & $\cdots$& $\cdots$ \\
$\cdots$ & $\cdots$  & $\cdots$   & 58             & $\cdots$ & $\cdots$  & $\cdots$   & O2--3\,V    & Be20           & $\cdots$                     & $\cdots$ & 13.61 & 14.80 & D & (0.36) & GHRS/G140L$\ddag$  &  4 & $\cdots$ \\ [2pt]
$\cdots$ & $\cdots$  & 467           & $\cdots$    & 93       &  416            & 1700        & O8.5\,V       & Ma12         & 053835.570-690606.65  & 13.41 & 13.57 & 14.74 & S & (0.36) & (O8\,V) &$\cdots$ & $\cdots$ \\
$\cdots$ & $\cdots$  & $\cdots$   & $\cdots$      & 106    & 583           & 2107          & O8\,V+O8.5\,V & Wa14 & 053845.211-690548.48 & 13.44 & 13.71 & 14.92 & S & (0.35) & (O8\,V) &$\cdots$ & $\cdots$ \\
$\cdots$ & $\cdots$  & $\cdots$   & 62             & $\cdots$ & $\cdots$  & $\cdots$   & O2--3\,V       & Be20        & $\cdots$                        & $\cdots$ & 13.65 & 14.91 & D & (0.35) &  GHRS/G140L$\ddag$  &  4& $\cdots$ \\
$\cdots$ & 52           & 493         & $\cdots$       & 48    & 423        & 660   & B1\,Ia           & Ev15          & 053836.053-690646.50 & $\cdots$ & 13.19 & 13.64 & S & (0.34) & (B1\,I) &$\cdots$ & $\cdots$ \\
$\cdots$ & $\cdots$  & 1340         & $\cdots$       & 110    & 604        & 2884   & O8.5\,V        & Wa14       & 053846.567-690537.10 & 13.47 & 13.71 & 14.95 & S & (0.34) & (O8\,V) &$\cdots$ & $\cdots$ \\ [2pt]
$\cdots$ & $\cdots$ & 1248   & $\cdots$ & 112 & $\cdots$ & 3034      & O6:    & Ca21         & 053845.494-690543.99 & 13.48      & 13.73 & 15.00 & S & (0.34) & (O6\,III)                & $\cdots$ & $\cdots$ \\
$\cdots$ & $\cdots$ & 1614  & $\cdots$ &  118 & 667        & 1699       & O6\,V    & Wa14       & 053852.832-690612.01 & $\cdots$  & 13.95 & 15.08 & S & 0.33 & COS/G140L              & 2 & $\cdots$ \\
$\cdots$ & $\cdots$  & $\cdots$   &  86  & $\cdots$ & $\cdots$  & $\cdots$   &  O5:\,V   & Be20        & $\cdots$                         & $\cdots$ & 13.72 & 14.73 & D   & (0.33) &  GHRS/G140L$\ddag$ & 4& $\cdots$ \\ 
$\cdots$ & $\cdots$ & 827            &  60  & 95     & 1007     & 1763     & O6.5\,V-III & He12   & 053841.066-690601.89 & $\cdots$ & 13.72 & 14.86 & S,D & (0.32)  & (O6.5\,III)             & $\cdots$    & $\cdots$ \\
$\cdots$ & $\cdots$  & $\cdots$   &  67  & 96      & $\cdots$ & 1857:    & O6:       & Ca21         & 053841.324-690557.59 & 13.52      & 13.86 & 15.09 & S,D & (0.32) & (O6\,III)                & $\cdots$    & $\cdots$ \\ [2pt]
$\cdots$ & $\cdots$ & 661 & $\cdots$ & 108       & 455         & 1572     & O5:V:n   & Wa14      & 053838.759-690613.22 & 13.55      & 13.75 & 15.08 & S   & (0.32) & (O4.5\,V)              & $\cdots$    & $\cdots$ \\ 
$\cdots$ & $\cdots$ & 1281 & $\cdots$ & 114      & $\cdots$  & 2033    & O7:       &  Ca21        & 053845.907-690550.77 & 13.56     & 13.76 & 14.90  & S    & (0.31) & (O7\,III)                & $\cdots$    & $\cdots$ \\
$\cdots$  & 37Wb     & 897           & 44   &  88     & 1017     & 1374 &  O2\,If/WN5 & CW11  & 053841.862-690614.41 & 13.56      & 13.60 & 14.53 & S    & (0.31) & (O2\,If/WN5)         & $\cdots$ & $\checkmark$ \\
$\cdots$ & $\cdots$ & 761            & 63    &  105  & $\cdots$ & 2077  & O3--6\,V      & WB97   & 053840.143-690551.26 & 13.56     & 13.76 & 14.86 & S     & (0.31) & (O4.5\,V)               & $\cdots$ & $\cdots$ \\
$\cdots$ & $\cdot$ &  781             & 72     &  113   & $\cdots$ & 1974 & O6:             & Ca21   & 053840.477-690553.42 & 13.57     & 13.83 & 14.99 & S    & (0.31) & (O6\,III)                  & $\cdots$& $\cdots$ \\ [2pt]
\hline
\end{tabular}\par
\label{tab:fuv-list}
\end{table*}

\addtocounter{table}{-1}

\begin{table*}
\centering
\caption{(continued)}
\begin{tabular}{
r @{\hspace{1mm}} r @{\hspace{1mm}} r @{\hspace{1mm}} r @{\hspace{1mm}} r @{\hspace{1mm}} r @{\hspace{1mm}} r @{\hspace{1mm}} c @{\hspace{1mm}} c @{\hspace{1mm}} c @{\hspace{1mm}} c @{\hspace{1mm}} c @{\hspace{1mm}} c  @{\hspace{1mm}} c @{\hspace{1mm}} c @{\hspace{1mm}} c @{\hspace{1mm}} c @{\hspace{1mm}} c}
\hline
R       & Mk      &   P  & HSH  & SMB  & VFTS& CCE  & SpT          & Ref & HTTP & $m_{\rm 275W}$ & $m_{\rm F336W}$ & $m_{\rm F555W}$ & Ref & $F_{1500}$ & Spectrum & Ref & VMS \\
          &           &       &          &           &           &           &                 &         &           & mag                     &           mag            & mag                      &        & $10^{-13}$ & (Template) &      \\
\hline
$\cdots$ & $\cdots$  & $\cdots$   & 70      & $\cdots$ & $\cdots$  & $\cdots$   & O5\,Vz         & Be20  & $\cdots$                        & $\cdots$ & 13.81 & 14.96 & D    & (0.30) & GHRS/G140L$\ddag$ &  4 & $\cdots$ \\
$\cdots$ & $\cdots$  & $\cdots$   & 74      &  121        & $\cdots$ & $\cdots$    & O6\,V          & MH98  & $\cdots$                         & $\cdots$ & 13.82 & 15.11 & D   & (0.30) & (O6\,V)                 & $\cdots$ & $\cdots$ \\
$\cdots$ & $\cdots$ &  1329  & $\cdots$ & 122      & $\cdots$ & 2718  & O9:              & Ca21 & 053849.107-690547.18  & 13.63      & 13.86 & 15.23 & S  & (0.29) & (O9\,III)                 & $\cdots$ & $\cdots$ \\
$\cdots$ & $\cdots$ & 1023         & 59       &  99     & $\cdots$ & 1703  & O3\,III            & MH98 & 053843.168-690603.65 & 13.63     & 13.73 & 14.76 & S,D & (0.29) & (O3\,III)                 & $\cdots$ & $\cdots$ \\
$\cdots$ & $\cdots$ & 974          & $\cdots$ & 104  & 532        & 963    & O3\,V(n)z+OB & Wa14 & 053842.657-690635.83 & $\cdots$ & 13.85 & 14.81 & S & (0.29) & (O3\,V)                 & $\cdots$ & $\cdots$ \\ [2pt]
$\cdots$ & $\cdots$ & 978           & 54        &  109  & $\cdots$ & 1963   & O4:                & Ca21 & 053842.679-690556.30 & 13.67       & 13.80 & 14.79 & S & (0.28) & (O4\,III)                 & $\cdots$& $\cdots$ \\ 
$\cdots$ & $\cdots$  & $\cdots$   & 71  & $\cdots$ & $\cdots$  & $\cdots$   & O2--3\,V        & Be20 & 053842.365-690604.90 & 13.70      & 13.98 & 15.16 & S,D & (0.28) & (O2.5\,V)               & $\cdots$ & $\cdots$ \\
$\cdots$ & $\cdots$  & $\cdots$   & 57  & $\cdots$ & $\cdots$  & $\cdots$   &O3\,III(f*)              & MH98 & $\cdots$                        & $\cdots$ & 13.90 & 14.80 & D & (0.28) & (O3\,III)                 & $\cdots$ & $\cdots$ \\
$\cdots$ & $\cdots$ & 776       & $\cdots$   & 124       & 484 & 3081    & O6--7\,V          & Wa14 & 053840.354-690543.79 & 13.71 & 13.94 & 15.09 & D & (0.27) & (O6\,V)                & $\cdots$ & $\cdots$ \\   
$\cdots$ & $\cdots$ & 921       & 82           & 134     & 522      & 3030   & O6\,II-Iab+O5.5\,V & Wa14 & 053842.082-690545.47 & 13.72 & 14.01 & 15.26 & S & (0.27) & (O6\,III)                  & $\cdots$& $\cdots$ \\ [2pt]
$\cdots$ & $\cdots$  & $\cdots$   & 77  &  $\cdots$ & $\cdots$  & $\cdots$   &O5.5\,V+O5.5\,V & Ma02  & $\cdots$               & $\cdots$ & 13.94 & 15.21 & D & (0.27) & (O5.5\,V)                 & $\cdots$& $\cdots$ \\ 
$\cdots$ & $\cdots$ & 1196      & $\cdots$ & 130     & $\cdots$ & 2763   & O6:                   & Ca21 & 053844.958-690538.63 & 13.75 & 13.89 & 15.14 & S & (0.26) &  (O6\,III)                  & $\cdots$ & $\cdots$ \\
$\cdots$ & $\cdots$  & $\cdots$   & 80  & $\cdots$ & $\cdots$  & $\cdots$   & O8\,V                & Be20 & $\cdots$                    & $\cdots$ & 13.95 & 15.17 & D & (0.26) & (O8\,V)                    & $\cdots$ & $\cdots$ \\
$\cdots$ & $\cdots$ &  973      & $\cdots$ & 129     & $\cdots$ & 2545      & O8:                    & Ca21 & 053842.607-690522.21 & 13.75 & 14.05 & 15.25 & S   & (0.26) & (O8\,III)                   & $\cdots$ & $\cdots$ \\
$\cdots$ & $\cdots$  & $\cdots$   & 66      & $\cdots$ & $\cdots$  & $\cdots$   &O2\,V-III            & Be20 & $\cdots$                    & $\cdots$ & 13.95 & 15.06 & D & (0.26) & GHRS/G140L$\ddag$ &  4 & $\cdots$ \\ [2pt]
$\cdots$ & 33Sb       & 1111      &    34       & $\cdots$ & $\cdots$ & $\cdots$      & WC5                  & MH98 & 053844.062-690555.64 & 13.76 & 13.82 & 14.49 & S    & (0.26) & (WC4)                   & $\cdots$ & $\cdots$ \\
$\cdots$ & $\cdots$ &  600     & $\cdots$ & 123       & $\cdots$ & 2946       & O3--5\,V             & Bo99 & 053837.658-690542.06 & $\cdots$ & 13.96 & 15.04 & S   & (0.26) & (O4\,V)                    & $\cdots$& $\cdots$ \\
$\cdots$ & 15S         & 1306    & $\cdots$ & 117       & $\cdots$ & 2053      & O8\,III               & WB97 & 053846.113-690554.44 & 13.79 & 13.93 & 14.95 & S & (0.25) & (O8\,III)                    & $\cdots$& $\cdots$ \\
$\cdots$ & $\cdots$  & $\cdots$   & 89     & $\cdots$ & $\cdots$  & $\cdots$   & O4\,V               &  Cr16 & $\cdots$                        & $\cdots$ & 13.99 & 14.76 & D & (0.25) & GHRS/G140L$\ddag$ &  4 & $\cdots$ \\
$\cdots$ & $\cdots$  & 621     & $\cdots$ &  97             & 445    & 2981 & O3--4\,V+O4--7\,V & Wa14 & 053838.026-690543.30 & 13.80 & 13.58 & 14.79 & S & (0.25) & (O3.5\,V)                 & $\cdots$ & $\cdots$ \\ [2pt]
$\cdots$ & $\cdots$ & $\cdots$ & 69         & $\cdots$ & $\cdots$ & $\cdots$ & O4--5\,V         & Be20  & $\cdots$                            & $\cdots$ & 14.02 & 15.05 & D,H & (0.25) & GHRS/G140L$\ddag$ & 4 & $\cdots$ \\
$\cdots$ & $\cdots$ & $\cdots$ & 65         & $\cdots$ & $\cdots$ & $\cdots$ & O4\,V              & Cr16 & $\cdots$                            & $\cdots$ & 14.03 & 15.18 & D & (0.24) & (O4\,V)& $\cdots$ & $\cdots$ \\
$\cdots$ & $\cdots$ & $\cdots$ & 64         & $\cdots$ & $\cdots$ & $\cdots$ & O4--5\,V           & Be20 & 053842.646-690601.05   & 13.83 & 13.95 & 14.53 & S & (0.24) & (O4\,V)&  $\cdots$ & $\cdots$ \\
$\cdots$ & $\cdots$ & $\cdots$ & 78         & $\cdots$ & $\cdots$ & $\cdots$ & O4:\,V              & Be20 & $\cdots$                            & $\cdots$ & 14.04 & 15.26 & D & (0.24) & GHRS/G140L$\ddag$ &  4 & $\cdots$ \\
$\cdots$ & $\cdots$ & 324        & $\cdots$ & 131       & 393          & 2256     & O9.5(n)           & Wa14  & 053832.992-690513.05 & 13.87 & 14.04 & 15.26 & S & (0.24) & (O9.5\,III) & $\cdots$ & $\cdots$ \\ [2pt]
$\cdots$ & $\cdots$ & 977         & $\cdots$ & 139      & $\cdots$ & 2893      & O6:\,V             &  Bo99   & 053842.636-690538.69 & 13.88 & 14.12 & 15.36 & S & (0.23) & (O6\,V)    & $\cdots$ & $\cdots$ \\
$\cdots$ & $\cdots$ & 1222      & $\cdots$ & 116       & $\cdots$ & 3180      & O3--6\,V         & WB97 & 053845.150-690508.34 & 13.88 & 14.08 & 14.99 & S & (0.23) & (O4.5\,V) & $\cdots$ & $\cdots$ \\
$\cdots$ & $\cdots$ & $\cdots$ & 73         & $\cdots$ & $\cdots$ & $\cdots$ & O9.7--B0\,V    & Be20 & $\cdots$                            & $\cdots$ & 14.12 & 15.13 & D & (0.23) & GHRS/G140L$\ddag$ &  4 & $\cdots$ \\
$\cdots$ & $\cdots$ & 1594      & $\cdots$ & 125       & 661         & 2691     & O6.5\,V+O9.7\,V & Wa14 & 053852.049-690533.79 & 13.94 & 14.00 & 15.19 & S & (0.22) & (O6\,V) & $\cdots$ & $\cdots$ \\
$\cdots$ & $\cdots$ & 1295      & $\cdots$ & 137       & 596         & 1325      & O7--8\,V          &  Wa14 & 053846.059-690615.55 & 13.97 & 14.25 & 15.35 & S & (0.21) & (O7\,V) & $\cdots$ & $\cdots$ \\ [2pt]
$\cdots$ & $\cdots$ & 884        & $\cdots$ & 143       & 511         & 1008      & O5\,Vz               & Wa14 & 053841.717-690628.13 & 13.98 & 14.21 & 15.31 & S & (0.21) & (O4.5\,V) & $\cdots$ & $\cdots$ \\
$\cdots$ & $\cdots$ & $\cdots$ & 95         & 155        & $\cdots$ & 1920      & O4:                    & Ca21 & 053840.949-690555.14  & 14.00 & 14.26 & 15.43 & S & (0.21) &  (O4\,IIII) & $\cdots$& $\cdots$ \\
$\cdots$ & $\cdots$ & $\cdots$ & 84          & 145        & $\cdots$ & 1956:     & O5:                    & Ca21 & $\cdots$                          & $\cdots$ & 14.20 & 14.80 & H & (0.21) & (O5\,III) & $\cdots$ & $\cdots$ \\
$\cdots$ & $\cdots$ & $\cdots$ & 92         & $\cdots$ & $\cdots$ & $\cdots$ & O6\,Vz               & Be20 & $\cdots$                            & $\cdots$ & 14.20 & 15.46 & D & (0.21) & GHRS/G140L$\ddag$ &  4 & $\cdots$ \\
$\cdots$ & $\cdots$ & 1026       & 68         & 127        & $\cdots$ & 1787      & O4--5\,V              & Be20 & 053843.180-690601.73 & 14.01 & 14.10 & 14.80 & S & (0.21) & (O4.5\,V) & $\cdots$ & $\cdots$ \\ [2pt]
$\cdots$ & $\cdots$ & 1141        &109 & 169      & 1035      & 2129      & O8.5\,I-II             & He12 & 053844.312-690545.10  & 14.01 & 14.29 & 15.57 & S & (0.21) &  (O8.5\,I) & $\cdots$ & $\cdots$ \\
$\cdots$ & $\cdots$ & 970        & 83         & 142        & 1023      & 1580      & O8\,III-V             & He12 & 053842.618-690610.00  & 14.02 & 14.22 & 15.31 & S & (0.20) & (O8\,III) & $\cdots$ & $\cdots$ \\
$\cdots$ & $\cdots$ & $\cdots$ & 90         & $\cdots$ & $\cdots$ & $\cdots$ & O4:\,V:               & Be20 & $\cdots$                            & $\cdots$ & 14.24 & 15.48 & D & (0.20) & GHRS/G140L$\ddag$ &  4 & $\cdots$ \\
$\cdots$ & $\cdots$ & $\cdots$ & $\cdots$ & 141       & 373        & 2090      & O9.5n                & Wa14 & 053831.224-690553.03   & $\cdots$ & 14.24 & 15.27 & S & (0.20) & (O9.5\,III) & $\cdots$ & $\cdots$ \\
$\cdots$ & $\cdots$ & 1191       & $\cdots$ & 128      & 575         & 2804       & B0.7\,III            & Ev15 & 053844.904-690533.13      & 13.76 & 13.99 & 15.12 & S & (0.20) & (B1\,III) & $\cdots$ & $\cdots$ \\ [2pt]
$\cdots$ & $\cdots$ & $\cdots$ & 88          & 165      & 1009       & 1732       & O6.5\,V--IIII      & He12 &  053841.147-690602.91    & 14.07 & 14.23 & 15.42 & S & (0.20) & (O6.5\,III) & $\cdots$ & $\cdots$ \\
$\cdots$ & $\cdots$ & $\cdots$ & 108         & $\cdots$ & $\cdots$ & $\cdots$ & O7--8\,V         & Cr16 & $\cdots$                            & $\cdots$ & 14.27 & 15.44 & D & (0.20) & (O7\,V) & $\cdots$ & $\cdots$ \\
$\cdots$ & $\cdots$ & 729         & 138        & 144      & $\cdots$ & 1535       & O5:                   & Ca21 & 053839.700-690608.63 & 14.08 & 14.20 & 15.31 & S & (0.19) & (O5\,III) & $\cdots$ & $\cdots$ \\
$\cdots$ & $\cdots$ & $\cdots$  & 96          & 163      & 1008      & 1852        & ON6.5\,II--I      & He12 & 053841.093-690558.40 & 14.09 & 14.32 & $\cdots$ & S & (0.19) & (O6\,I) & $\cdots$ & $\cdots$ \\
$\cdots$ & $\cdots$ & 957          & 99          & 167     & $\cdots$ & 1527        & O8\,V                 & MH98 & 053842.445-690609.22    & 14.11 & 14.32 & 15.41 & S & (0.19) & (O8\,V) & $\cdots$& $\cdots$ \\ [2pt] %
$\cdots$ & $\cdots$ & 1527       & $\cdots$ & 119         & 646         & 1951      & B0.5\,III(n)          & Ev15      & 053850.263-690604.37 & 13.84      & 14.03 & 14.99 & S & (0.19) & (B0\,III) & $\cdots$ & $\cdots$ \\
$\cdots$ & $\cdots$ & $\cdots$ & 93          & $\cdots$ & $\cdots$ & $\cdots$ & O4--5\,V              & Cr16$^{a}$ & $\cdots$                         & $\cdots$ & 14.34 & 15.60 & H & (0.18) & GHRS/G140L$\ddag$ &  4& $\cdots$ \\
$\cdots$ & $\cdots$ & 531        & $\cdots$ & 166        & $\cdots$  & 1941       & O8\,V                  & Bo99      & 053836.728-690556.46 & 14.15      & 14.32 & 15.54 & S & (0.18) & (O8\,V) & $\cdots$ & $\cdots$ \\
$\cdots$ & $\cdots$ & 740        & $\cdots$ & 159        & $\cdots$  & 1537       & O6:                     & Ca21       & 053839.837-690607.99 & 14.15     & 14.29 & 15.41 & S & (0.18) & (O6\,III) & $\cdots$ & $\cdots$ \\
$\cdots$ & $\cdots$ & $\cdots$ & 94          & $\cdots$ & $\cdots$ & $\cdots$ & O4--5\,Vz            & Be20      & $\cdots$                        & $\cdots$ & 14.36 & 15.57 & D & (0.18) & GHRS/G140L$\ddag$ &  4 & $\cdots$ \\ [2pt]
$\cdots$ & $\cdots$ & 1031       & $\cdots$ & 152        & 543        & 2521       & O9\,IV+O9.7:\,V & Wa14       & 053843.190-690527.52 & 14.17    & 14.26 & 15.44 & S & (0.18) & (O9\,V) & $\cdots$ & $\cdots$ \\ 
$\cdots$ & $\cdots$ & 1288       & $\cdots$ & 175        & 597        & 375         & O8--9\,V(n)           & Wa14    & 053846.063-690656.16 & $\cdots$ & 14.38 & 15.62 & S & (0.18) & (O9\,V) & $\cdots$ & $\cdots$ \\
$\cdots$ & $\cdots$ & 1468       & $\cdots$ & 174        & 635        & 1334       & O9.5\,IV                & Wa14    & 053849.039-690619.57 & $\cdots$ & 14.39 & 15.56 & S & (0.18) & (O9\,V) & $\cdots$ & $\cdots$ \\
$\cdots$ & $\cdots$ & $\cdots$ & 101         & 178        & 1020      & 1401       & O3--4                    & He12    & 053842.012-690616.83 & 14.21     & 14.40 & 15.53 & S & (0.17) & (O3.5\,V) & $\cdots$ & $\cdots$ \\
$\cdots$ & $\cdots$ & 1145        & $\cdots$ & 168       & 561        & 2301      & O9:(n)                    & Wa14     & 053844.368-690514.36 & $\cdots$ & 14.41 & 15.46 &S & (0.17) & (O9\,III) & $\cdots$ & $\cdots$ \\ [2pt]
$\cdots$ & $\cdots$ &  901         & $\cdots$ & 138        & 518       & 1068      & O3.5\,III(f*)              & Wa14     & 053841.934-690629.70 & 14.21     & 14.24 & 15.15 & S & (0.17) & (O3\,III) & $\cdots$ & $\cdots$ \\
$\cdots$ & $\cdots$ & 887         & 85          & 154        & 1016      & 1371     & O8\,V                      & Bo99      & 053841.755-690619.06 & 14.22      & 14.33 & 15.31 & S & (0.17) & (O8\,V) & $\cdots$ & $\cdots$ \\
$\cdots$ & $\cdots$ & 1336       & $\cdots$ & 176       & $\cdots$ & 2945     & O8:                        & Ca21     & 053846.481-690542.20   & 14.22      & 14.39 & 15.54 & S & (0.17) & (O8\,III) &$\cdots$ & $\cdots$ \\
$\cdots$ & $\cdots$ & $\cdots$ & 114          & $\cdots$ & $\cdots$ & $\cdots$ & O5--6\,V            & Be20      & $\cdots$                        & $\cdots$ & 14.43 & 15.68 & D & (0.17) & (O5.5\,V) &$\cdots$ & $\cdots$ \\ 
$\cdots$ & $\cdots$ & $\cdots$ & $\cdots$ & 198        & $\cdots$ & 3167     & O8:                       & Ca21     & 053841.818-690507.18   & 14.23      & 14.48 & 15.76 & S & (0.17) & (O8\,III) & $\cdots$ & $\cdots$ \\ [2pt]
\hline
\end{tabular}\par
\label{tab:fuv-list}
\end{table*}

\addtocounter{table}{-1}

\begin{table*}
\centering
\caption{(continued)}
\begin{tabular}{
r @{\hspace{1mm}} r @{\hspace{1mm}} r @{\hspace{1mm}} r @{\hspace{1mm}} r @{\hspace{1mm}} r @{\hspace{1mm}} r @{\hspace{1mm}} c @{\hspace{1mm}} c @{\hspace{1mm}} c @{\hspace{1mm}} c @{\hspace{1mm}} c @{\hspace{1mm}} c  @{\hspace{1mm}} c @{\hspace{1mm}} c @{\hspace{1mm}} c @{\hspace{1mm}} c @{\hspace{1mm}} c}
\hline
R       & Mk      &   P  & HSH  & SMB  & VFTS& CCE  & SpT          & Ref & HTTP & $m_{\rm 275W}$ & $m_{\rm F336W}$ & $m_{\rm F555W}$ & Ref & $F_{1500}$ & Spectrum & Ref & VMS \\
          &           &       &          &           &           &           &                 &         &           & mag                     &           mag            & mag                      &        & $10^{-13}$ & (Template) &      \\
\hline
$\cdots$ & $\cdots$ & 841         & 98          & 157        & $\cdots$ & 2016      & O4--6(n)(f)p         & WB97      & 053841.186-690552.12  & 14.25       & 14.42 & 15.45 & S & (0.17) & (O5\,III) & $\cdots$ & $\cdots$ \\
$\cdots$ & $\cdots$ & $\cdots$ & 100          & $\cdots$ & $\cdots$ & $\cdots$ & B0\,V            & MH98      & 053842.206-690614.82  & 14.27 & 14.52 & 15.64 & S,H & (0.17) & (B0\,V) & $\cdots$ & $\cdots$ \\ %
$\cdots$ & $\cdots$ & 992        & $\cdots$ & 182        & $\cdots$ & 2665      & O7:                       & Ca21     & 053842.838-690530.36  & 14.29      & 14.44 & 15.56 & S & (0.16) & (O7\,III) & $\cdots$ & $\cdots$ \\
$\cdots$ & $\cdots$ & 485        & $\cdots$ & 158        & 419       & 2897        & O9:\,V(n)                & Wa14    & 053835.906-690534.95 & 14.32       & 14.33 & 15.41 & S & (0.16) & (O9\,V) & $\cdots$ & $\cdots$ \\
$\cdots$ & $\cdots$ & $\cdots$ & 113        & 190        & $\cdots$ & 1503        & O9\,V                       & MH98     & 053842.243-690612.23 & 14.32        & 14.53 & 15.62 & S & (0.16) & (O9\,V) & $\cdots$ & $\cdots$ \\ [2pt]
$\cdots$ & $\cdots$ & 796        & $\cdots$ & 185        & $\cdots$ & 2565        & O6:                      & Ca21    & 053840.665-690531.16 & 14.32       & 14.42   & 15.59 & S & (0.16) & (O6\,III) & $\cdots$ & $\cdots$ \\ 
$\cdots$ & $\cdots$ & 670        & $\cdots$  & 172        & 456        & 2385         & Onn(f)                  & Wa14     & 053838.818-690525.56 & 14.34      & 14.42 & 15.45 & S & (0.15) & (O6\,III) & $\cdots$ & $\cdots$ \\
$\cdots$ & $\cdots$ & 994        & $\cdots$   & 146      & $\cdots$ & 2270        & B0:\,V                  & Bo99   & 053842.841-690514.80 & 14.07            & 14.19 & 15.44 & S & (0.15) & (B0\,V) & $\cdots$ & $\cdots$ \\
$\cdots$ & $\cdots$ & 1201      & $\cdots$ & 171       & 579          & 3116        & O9:((n))               & Wa14 & 053844.969-690507.65 & 14.36         & 14.56   & 15.47 & S & (0.15) & (O9\,V) & $\cdots$ & $\cdots$ \\
$\cdots$ & $\cdots$ & $\cdots$ & 141         & $\cdots$ & $\cdots$ & $\cdots$ & O5--6\,V            & Cr16      & $\cdots$                        & $\cdots$ & 14.56 & 15.82 & D & (0.15) & GHRS/G140L$\ddag$ &  4 & $\cdots$ \\ [2pt]
$\cdots$ & $\cdots$ & $\cdots$ & 116          & $\cdots$ & $\cdots$ & $\cdots$ & O7\,V            & Be20      & $\cdots$                        & $\cdots$ & 14.57 & 15.79 & D & (0.15) & (O7\,V) & $\cdots$ & $\cdots$ \\  
$\cdots$ & $\cdots$ & $\cdots$ & 115          & $\cdots$ & $\cdots$ & $\cdots$ & O9+\,V            & Cr16     & $\cdots$                        & $\cdots$ & 14.57 & 15.76 & D & (0.15) & (O9\,V) & $\cdots$ & $\cdots$ \\ 
$\cdots$ & $\cdots$ & 1359      & $\cdots$  & 202         & $\cdots$ & 2782       & O8:                  & Ca21      & 053846.901-690536.82 & 14.37     & 14.59 & 15.78 & S & (0.15) & (O8\,III) & $\cdots$ & $\cdots$ \\
$\cdots$ & $\cdots$ & 1560     & $\cdots$    & 200       & 654          & 3010       & O9\,Vnn           & Wa14   & 053851.159-690541.81  & 14.38 & 14.62    & 15.77 & S & (0.15) & (O9\,V) & $\cdots$ & $\cdots$ \\
$\cdots$ & $\cdots$ & $\cdots$ & $\cdots$ & 179         & 570         & 3134       & O9.5ne+           & Wa14    & 053844.684-690545.19 & 14.39 & 14.42    & $\cdots$ & S & (0.15) & (O9.5\,III) & $\cdots$ & $\cdots$ \\ [2pt]
$\cdots$ & $\cdots$ & 578       & $\cdots$  & 209         & 436         & 2223       & O7--8\,V            & Wa14    & 053837.348-690521.29  & 14.39 & 14.58 & 15.93 & S & (0.15) & (O7\,V) & $\cdots$ & $\cdots$ \\  
$\cdots$ & $\cdots$ & $\cdots$ & 112         & $\cdots$ & $\cdots$ & $\cdots$ & O7--9\,Vz            & Be20      & $\cdots$                        & $\cdots$ & 14.61 & 15.74 & D & (0.14) & GHRS/G140L$\ddag$ &  4 & $\cdots$ \\ 
$\cdots$ & $\cdots$ & $\cdots$ & $\cdots$  & 224        & 660        & 3027       & O9.5\,Vnn          & Wa14       & 053851.812-690546.85 & 14.43    & 14.76 & 16.00 & S & (0.14) & (O9\,V) & $\cdots$ & $\cdots$ \\
$\cdots$ & $\cdots$ & $\cdots$ & 132          & $\cdots$ & $\cdots$ & $\cdots$ & O7:\,V            & Be20      & $\cdots$                        & $\cdots$ & 14.64 & 15.86 & D & (0.14) & (O7\,V) & $\cdots$ & $\cdots$ \\ 
$\cdots$ & $\cdots$ & 348     & $\cdots$ & 227        & 400        & 2607         & O9.7                   & Be20      & 053833..536-690521.73 & 14.45 & 14.79 & 16.05 & S & (0.14) & (O9.5\,III) &$\cdots$ & $\cdots$ \\ [2pt]
$\cdots$ & $\cdots$ & $\cdots$ & $\cdots$ & 212      & $\cdots$ & 2776       & O8:                    & Ca21        & 053843.127-690537.76 & 14.51 & 14.67 & 15.89 & S & (0.13) & (O8\,III) & $\cdots$ & $\cdots$ \\ 
$\cdots$ & $\cdots$ & 490        & $\cdots$  & 140      & 422        & 1373       & O4\,III(f)             & Wa14        & 053835.993-690616.91 & 14.15 & 14.23 & 15.15 & S & 0.13 & COS/G140L & 2 & $\cdots$ \\
$\cdots$ & $\cdots$ & $\cdots$ & 123          & $\cdots$ & $\cdots$ & $\cdots$ & O6\,V            & Be20      & 053842.043-690604.58 & 14.52 & 14.71 & 15.76 & S & (0.13) & (O6\,V) & $\cdots$ & $\cdots$ \\ 
$\cdots$ & $\cdots$ & $\cdots$ & 121          & $\cdots$ & $\cdots$ & $\cdots$ & O9.5\,V            & Be20      & $\cdots$                  & $\cdots$ & 14.73 & 15.85 & D & (0.13) & (O9\,V) & $\cdots$ & $\cdots$ \\ 
$\cdots$ & $\cdots$ & $\cdots$ & 134          & $\cdots$ & $\cdots$ & $\cdots$ & O7\,Vz           & Be20      & 053842.182-690604.95 & 14.52 & 14.74 & 15.44 & S & (0.13) & (O7\,V) & $\cdots$ & $\cdots$ \\ [2pt]
$\cdots$ & $\cdots$ & $\cdots$ & $\cdots$  & 206         & 476        & 2789      & O((n))             & Wa14      & 053839.734-690539.27 & 14.53 & 14.73 & 15.72 & S & (0.13) & (O6\,III) & $\cdots$ & $\cdots$ \\
$\cdots$ & $\cdots$ & $\cdots$ & $\cdots$ & $\cdots$ & 444        & $\cdots$ & O9.7               & Wa14       & 053838.011-690508.53 & 14.55 & 14.86 & 16.12 & S & (0.13) & (B0\,V) & $\cdots$ & $\cdots$ \\
$\cdots$ & $\cdots$ & $\cdots$ & $\cdots$ & 218        & 615         & 1398      & O9.5\,IIInn     & Wa14        & 053847.320-690617.74 & $\cdots$ & 14.78 & 15.64 & S & (0.12) & (O9.5\,III) & $\cdots$ & $\cdots$ \\
$\cdots$ & $\cdots$ & 881         & $\cdots$ & 215        & $\cdots$ & 2389      & O9:              & Ca21         & 053841.643-690523.77 & 14.58 & 14.69 & 15.87 & S & (0.12) & (O9\,III) & $\cdots$ & $\cdots$ \\ 
$\cdots$ & $\cdots$ & $\cdots$ & 118         & $\cdots$ & $\cdots$ & $\cdots$ & O7--8\,V     &Cr16         & $\cdots$                         & $\cdots$ & 14.79 & 15.89 & D & (0.12) & (O8\,V) & $\cdots$ & $\cdots$ \\ [2pt]
$\cdots$ & $\cdots$ & $\cdots$ & $\cdots$ & 211        & $\cdots$ & 1415       & O7:             & Ca21         & 053843.317-690611.65 & 14.61 & 14.80 & 15.92 & S & (0.12) & (O7\,III) & $\cdots$ & $\cdots$ \\ 
$\cdots$ & $\cdots$ & 743         & $\cdots$ & 151       & $\cdots$ & 1523       & O6:             & Ca21          & 053839.846-690609.50 & 14.61 & 14.83 & 15.40 & S & (0.12) & (O6\,III) & $\cdots$ & $\cdots$ \\ 
$\cdots$ & $\cdots$ & 918:        & $\cdots$ & 311:      & $\cdots$ & 1522       & O7:             & Ca21           & 053842.067-690608.95 & $\cdots$ & 14.82 & 15.81 & S & (0.12) & (O7\,III) & $\cdots$ & $\cdots$ \\
$\cdots$ & $\cdots$ & $\cdots$ & $\cdots$ & 194        & 446       & 1339        & Onn           &  Wa14         &  053838.245-690617.36 & 14.62     & 14.63 & 15.46 & S & (0.12)  & (O9.5\,V) & $\cdots$ & $\cdots$ \\
$\cdots$ & $\cdots$ & 915         & 144        & 238        & $\cdots$ & 2130      & O8:             & Ca21         & 053842.061-690552.23    & 14.62     & 14.85 & 15.97 & S & (0.12) & (O8\,III)  & $\cdots$ & $\cdots$ \\ [2pt]
$\cdots$ & $\cdots$ & $\cdots$ & 103        & 189         & 1010      & 1346       & O7\,V--III    & Ma12       & 053841.255-690617.02   & 14.65     & 14.69 & 15.55 & S & (0.11) & (O7\,III) & $\cdots$ & $\cdots$ \\ 
$\cdots$ & $\cdots$ & $\cdots$ & $\cdots$ & 226        & 382       & 2901       & O4--5\,V       & Wa14        & 053832.253-690544.58  & 14.66     & 14.82 & 15.93 & S & (0.11) & (O4.5\,V) & $\cdots$ & $\cdots$ \\
$\cdots$ & $\cdots$ & $\cdots$ & 140        & 187       & 1015       & 1734      & O8:\,V           & Le21       & 053841.637-690603.25   & 14.66     & 14.83 & 15.76  & S & (0.11) & (O8\,V) & $\cdots$ & $\cdots$ \\
$\cdots$ & $\cdots$ & $\cdots$ & $\cdots$ & 208      & 443        & 1557       & O7:\,V           & Wa14       & 053837.972-690615.30   & 14.67     & 14.75 & 15.73 & S & (0.11) & (O7\,V) & $\cdots$ & $\cdots$ \\
$\cdots$ & $\cdots$ & 547        & $\cdots$ & 214       & 432         & 466        & O8--9\,V       & Wa14        & 053837.030-690650.70  & $\cdots$ & 14.88 & 15.73 & S & (0.11) & (O8\,V) & $\cdots$ & $\cdots$ \\ [2pt]
$\cdots$ & $\cdots$ & 1369      & $\cdots$ & 219       & 613         & 2005      & O8.5\,Vz       & Wa14        & 053847.161-690554.46 & 14.68      & 14.80 & 15.80 & S & (0.11) & (O8\,V) & $\cdots$ & $\cdots$ \\ 
$\cdots$ & $\cdots$ & 853        & $\cdots$ & 177       & 1011       & 1447      & O5\,V           & Le21       & 053841.348-690614.04 & 14.70       & 14.75 & 15.39 & S & (0.11) & (O4.5\,V) & $\cdots$ & $\cdots$ \\
$\cdots$ & $\cdots$ & $\cdots$  & 126         & 193       & 1004      & 1725      & O9.5\,V-III     & He12        & 053840.831-690604.60 & 14.70      & 14.82 & 15.59 & S & (0.11) & (O9.5\,III) & $\cdots$ & $\cdots$ \\
$\cdots$ & $\cdots$ & 803         & $\cdots$ & 213         & 491      & 276         & O6\,V            & Wa14       & 053840.846-690657.51 & $\cdots$ & 14.96 & 15.69 & S & (0.10) & (O6\,V) & $\cdots$ & $\cdots$ \\
$\cdots$ & $\cdots$ & $\cdots$ & $\cdots$ & 239         & 649       & 2038       & O9.5\,V        & Wa14       & 053850.610-690554.57 & 14.77       & 15.02 & 16.14 & S & (0.10) & (O9\,V) & $\cdots$ & $\cdots$ \\ [2pt]
$\cdots$ & $\cdots$ & 1224      & $\cdots$ & 231         & $\cdots$ & 1472     & O6:               & Ca21        & 053845.213-690613.62 & 14.78        & 14.90 & 15.99 & S & (0.10) & (O6\,III) & $\cdots$ & $\cdots$ \\ 
$\cdots$ & $\cdots$ & $\cdots$ & $\cdots$ & 277        & $\cdots$ & 2453      & O8:              & Ca21        & 053840.166-690520.63 & 14.78         & 15.04 & 16.34 & S & (0.10) & (O8\,III) & $\cdots$ & $\cdots$ \\
$\cdots$ & $\cdots$ & 1401      & $\cdots$ & 234        & 619        & 689         & O7--8\,V(n)   & Wa14       &  053847.717-690644.94 & $\cdots$    & 14.99 & 16.07 & S & (0.10) & (O8\,V) & $\cdots$& $\cdots$ \\
$\cdots$ & $\cdots$ & $\cdots$ & 102       & 207         & $\cdots$ & $\cdots$ & O2--3\,III       & Cr16        & 053843.165-690600.85 & 14.79        & 14.79 & 15.52 & S & (0.10) & (O2.5\,III) & $\cdots$ & $\cdots$ \\
$\cdots$ & $\cdots$ & 1087      & $\cdots$ & 272        & $\cdots$ & 1275      & O8:                & Ca21       & 053843.749-690621.89 & 14.79         & 15.06 & 16.27 & S & (0.10) & (O8\,III) & $\cdots$& $\cdots$ \\ [2pt]
$\cdots$ & $\cdots$ & 1086      & $\cdots$ & 247        & $\cdots$ & 2454      & O7:                & Ca21       & 053843.709-690521.64 & 14.80          & 14.97 & 16.16 & S & (0.10) & (O7\,III) & $\cdots$ & $\cdots$ \\
$\cdots$ & $\cdots$ & 982        & 160         & 258        & $\cdots$ & 2044      & O4:\,V          & Pa93      & 053842.721-690552.80 & 14.80          & 14.98  & 16.14 & S & (0.10) & (O4\,V) & $\cdots$ & $\cdots$ \\
\hline
\end{tabular}\par
{\bf Sp Types:}
Be20: \citet{2020MNRAS.499.1918B}; 
Be22: \citet{2022MNRAS.510.6133B}; 
Bo99: \citet{1999A&AS..137...21B}; 
Ca21 \citet{2021A&A...648A..65C}; 
Cr10: \citet{2010MNRAS.408..731C};  
Cr16: \citet{2016MNRAS.458..624C}; 
Cr22: \citet{2022MNRAS.515.4130C}; 
CD98: \citet{1998MNRAS.296..622C}; 
CS97: \citet{1997A&A...320..500C}; 
CW11: \citet{2011MNRAS.416.1311C};  
De97: \citet{1997ApJ...477..792D};
Do13:\citet{2013A&A...558A.134D}  
Ev11: \citet{2011A&A...530A.108E}; 
Ev15: \citet{2015A&A...574A..13E}; 
He12: \citet{2012A&A...546A..73H}; 
Ka22 \citet{2022ApJ...935..162K};
Le21 : D.~Lennon (priv comm, 2021); 
Ma02: \citet{2002ApJ...565..982M}; 
Ma05: \citet{2005ApJ...627..477M};
Ma12: \citet{2012ApJ...748...96M};
MH98 : \citet{1998ApJ...493..180M}; 
Pa93: \citet{1993AJ....106..560P}; 
Te19: \citet{2019MNRAS.484.2692T}; 
Wa14: \citet{2014A&A...564A..40W}; 
WB97: \citet{1997ApJS..112..457W} \\
{\bf Spectroscopy:} 1. GO~16272 (Shenar); 2. GO~15629 (Mahy); 3. \citet{2020RNAAS...4..205R}; 4. \citet{1992ESOC...44..347H}; 5.  \citet{1998ApJ...509..879D}; 6. \citet{2005ApJ...627..477M}.
Note: (a) spectral type of HSH 119 (fainter component of blend with $m_{\rm F336W}$ = 15.0 mag) \\
{\bf Photometry:} D: \citet{2011ApJ...739...27D}; H: \citet{1995ApJ...448..179H}; S: \citet{2016ApJS..222...11S} \\
\end{table*}




\begin{table*}
\centering
\caption{Stars within the {\it IUE} 3$\times$3 arcmin$^{2}$  \citet{1995ApJ...444..647V} field of view which are exterior to MUSE, sorted by far-UV flux ($F_{1500}$  units of erg\,s$^{-1}$\,cm$^{-2}$\AA$^{-1}$), measured from {\it IUE}/SWP spectroscopy or estimated from photometry, the latter indicated in parentheses. See Table~\ref{tab:fuv-list} for references in common.}
\begin{tabular}{
r @{\hspace{1mm}} r @{\hspace{1mm}} r @{\hspace{1mm}} r @{\hspace{1mm}} r @{\hspace{1mm}} r @{\hspace{1mm}} r @{\hspace{1mm}} c @{\hspace{1mm}} c @{\hspace{1mm}} c @{\hspace{1mm}} c @{\hspace{1mm}} c @{\hspace{1mm}} c  @{\hspace{1mm}} c @{\hspace{1mm}} c @{\hspace{1mm}} c @{\hspace{1mm}} c @{\hspace{1mm}} c}
\hline
R           & Mk        &   P     & HSH     & SMB       & VFTS & CCE  & SpT                              & Ref & HTTP & $m_{\rm 275W}$ & $m_{\rm F336W}$ & $m_{\rm F555W}$ & Ref & $F_{1500}$ & Spectrum & Ref & VMS \\
              &             &           &             &                &           &           &                                      &         &           & mag                     &           mag            & mag                      &        & $10^{-13}$ & (Template) &      \\
\hline
139        &$\cdots$& 952   & $\cdots$ &  2           & 527    & $\cdots$ & O6.5\,Iafc+O7\,Iaf    &Wa14   & 053842.351-690458.19 & 10.97        & 10.86       & 12.02           &   S     & 6.21  & (O7\,I)  & 1 & $\cdots$ \\
145        &$\cdots$& 1788 & $\cdots$ &$\cdots$ & 695   & $\cdots$ & WN6h+O3/5\,If/WN7 & Sh17  & 053857.072-690605.58  & $\cdots$   & 10.88      & 11.94            &    S    & 4.21  & (WN6) & 1 & $\checkmark$ \\
135        &$\cdots$& 355   &  $\cdots$ &12          & 402   & $\cdots$ & WN5:+WN7              & Sh19    & 053833.615-690450.47  & 11.63     & 11.68       & 12.82             &    S     & 2.34    & (WN6) & 1 & $\cdots$ \\
$\cdots$ & 23        & 1163 &  $\cdots$ & 42         & 566   & $\cdots$& O3\,III(f*)                    & Wa14   & 053844.558-690451.19 & 12.40        & 12.68       & 14.03           &    S    & (0.91)    & (O3\,III) & $\cdots$ & $\cdots$ \\
$\cdots$ &$\cdots$& 1035&  $\cdots$ &$\cdots$&$\cdots$&$\cdots$& O3--6\,V                  & WB97    & 053843.159-690441.82 & 12.93        &  13.35       & 14.75          &    S     & (0.56)  & (O4.5\,V)& $\cdots$ & $\cdots$ \\[2pt]
$\cdots$ & 57        & 541   & $\cdots$ & 79          & 429     & $\cdots$ & O7:\,V+B1:\,V        &  Sh22     & 053836.854-690458.28 & 12.95       & 13.32        & 14.70           &   S      & (0.55) & (O7\,V) & $\cdots$ & $\cdots$ \\
$\cdots$ & $\cdots$ & 809  & $\cdots$ & 133      & $\cdots$ & $\cdots$ & O8--9\,V             & Bo99       & 053840.773-690452.83 & 13.67      & 14.01        & 15.34           &   S      & (0.28) & (O9\,V) & $\cdots$ & $\cdots$ \\
$\cdots$ & $\cdots$ & 1077& $\cdots$ &$\cdots$& 550        & $\cdots$ & O5\,V((f))z          & Wa14     & 053843.599-690442.44 & 13.78       & 14.12         & 15.42         &  S       & (0.26) & (O5\,V) & $\cdots$ & $\cdots$ \\
$\cdots$ & $\cdots$ & 124   & $\cdots$ & $\cdots$ & 350     & $\cdots$ & O8.5\,V+O9.5\,V & Sh22      & 053828.057-690629.05 & 13.79      & 14.09         & 15.03         & S       & (0.25) & (O9\,V) & $\cdots$ & $\cdots$ \\
$\cdots$ & $\cdots$ & 83     & $\cdots$ & $\cdots$ & 339    & $\cdots$ & O9.5\,IV                & Wa14     & 053826.218-690501.84 & 13.83      & 14.15        & 15.52          & S       & (0.24) & (O9\,V) & $\cdots$ & $\cdots$ \\[2pt]
$\cdots$ & $\cdots$ & 169    & $\cdots$ & 107       & 360     & $\cdots$ & O9.7                     & Wa14     & 053829.421-690521.20 & 13.84      & 13.85        & 14.85          & S       & (0.24) &  (O9.5\,III) & $\cdots$ & $\cdots$ \\
$\cdots$ & $\cdots$ & 466    & $\cdots$ & 150      & 415      & $\cdots$ & O9.5\,V                & Wa14     & 053835.480-690457.67 & 13.90       & 14.17        &  15.50         & S       & (0.23) & (O9\,V) & $\cdots$ & $\cdots$ \\
$\cdots$ & 60          & 195     & $\cdots$ & 98       & 363      & $\cdots$ & B0.2\,III-II             & Ev15     & 053829.986-690505.18 & 13.61        & 13.68        & 14.86          & S       & (0.23) & (B0\,III) & $\cdots$ & $\cdots$ \\
$\cdots$ & $\cdots$ & 1052  & $\cdots$ &$\cdots$& 546     & $\cdots$ & O8--9\,III              & Wa14    & 053843.373-690446.34 & 14.02        & 14.21        & 15.41         & S        & (0.20) & (O8\,III) & $\cdot$ & $\cdots$ \\
$\cdots$ & $\cdots$ & 1729  & $\cdots$ &$\cdots$& 686     & $\cdots$ & B0.7\,III                & Ev15     & 053855.634-690723.71 & 14.05        & 14.15        & 15.05          & S       & (0.15) & (B1\,III) & $\cdots$ & $\cdots$ \\[2pt]
$\cdots$ & $\cdots$ & 1756  & $\cdots$ &$\cdots$ & 688    & $\cdots$ & O9.7\,III               & Wa14     & 053856.058-690554.04 & 14.45        & 14.70       & 15.64          & S        & (0.14) & (O9.5\,III) & $\cdots$ & $\cdots$ \\
$\cdots$ & $\cdots$ & 955    & $\cdots$ & $\cdots$ & 529   & $\cdots$ & O9.5(n)                & Wa14    & 053842.374-690443.12   & 14.55       & 14.89      & 16.09          & S      & (0.13) & (O9.5\,III) & $\cdots$ & $\cdots$ \\
$\cdots$ & $\cdots$ & 905 & $\cdots$ & $\cdots$ & 521     & $\cdots$ & O9\,V(n)                & Wa14    & 053841.955-690704.97 & $\cdots$    &14.82       & 15.90          & S      & (0.12) & (O9\,V) & $\cdots$ & $\cdots$ \\
$\cdots$ & $\cdots$ & 171  & $\cdots$ & 216       & 361      & $\cdots$ & O8.5\,V                 & Wa14   & 053829.482-690620.55 & 14.63          &14.81       & 15.75          & S      & (0.12) & (O8\,V) & $\cdots$ & $\cdots$ \\   
$\cdots$ & $\cdots$ & 1840  & $\cdots$ & $\cdots$ & 707   & $\cdots$ & B0.5\,V                 & Ev15    & 053858.907-690642.05  & $\cdots$   &14.55         &15.65          & S       & (0.12) & (B0.5\,V) & $\cdots$ & $\cdots$ \\[2pt]
$\cdots$ & $\cdots$ & $\cdots$ & $\cdots$ & $\cdots$ & 486  & $\cdots$ & B1-2ne+               & Ev15    & 053840.604-690456.02  & 14.20       & 14.16        & 15.28          & S      & (0.11) & (B1\,III) & $\cdots$ & $\cdots$ \\
$\cdots$ & $\cdots$ & $\cdots$ & $\cdots$ & 319    & 412    & $\cdots$ & O9.7                    & Wa14    & 053834.901-690453.74 & 14.71      & 15.14         & 16.47         & S       & (0.11) & (O9.5\,III) & $\cdots$ & $\cdots$ \\
$\cdots$ & $\cdots$ & 1209      & $\cdots$ & 205     &$\cdots$ & $\cdots$ & O9--B0\,V        & Bo99     & 053845.055-690446.88 & 14.74       & 14.85       & 15.67         & S       & (0.11) & (O9\,V) & $\cdots$ & $\cdots$ \\
$\cdots$ & $\cdots$ & 1429      & $\cdots$ & $\cdots$ & 621    & $\cdots$ & O2\,V((f))z       & Wa14   & 053848.089-690442.18 & 14.80         & 14.86      & 15.59        & S        & (0.10) & (O2\,V) & $\cdots$ & $\cdots$ \\
\hline
\end{tabular}\par
{\bf Sp Types}: Sh17: \citet{2017A&A...598A..85S}; Sh19: \citet{2019A&A...627A.151S}; Sh22: \citet{2022A&A...665A.148S} \\
{\bf Spectroscopy:} 1. \citet{1984ApJ...279..578F}
\label{tab:iue-list}
\end{table*}

\clearpage

\section{Far-UV calibrations}\label{FUV}

\begin{figure}
\includegraphics[width=0.9\columnwidth, bb = 15 167 515 667]{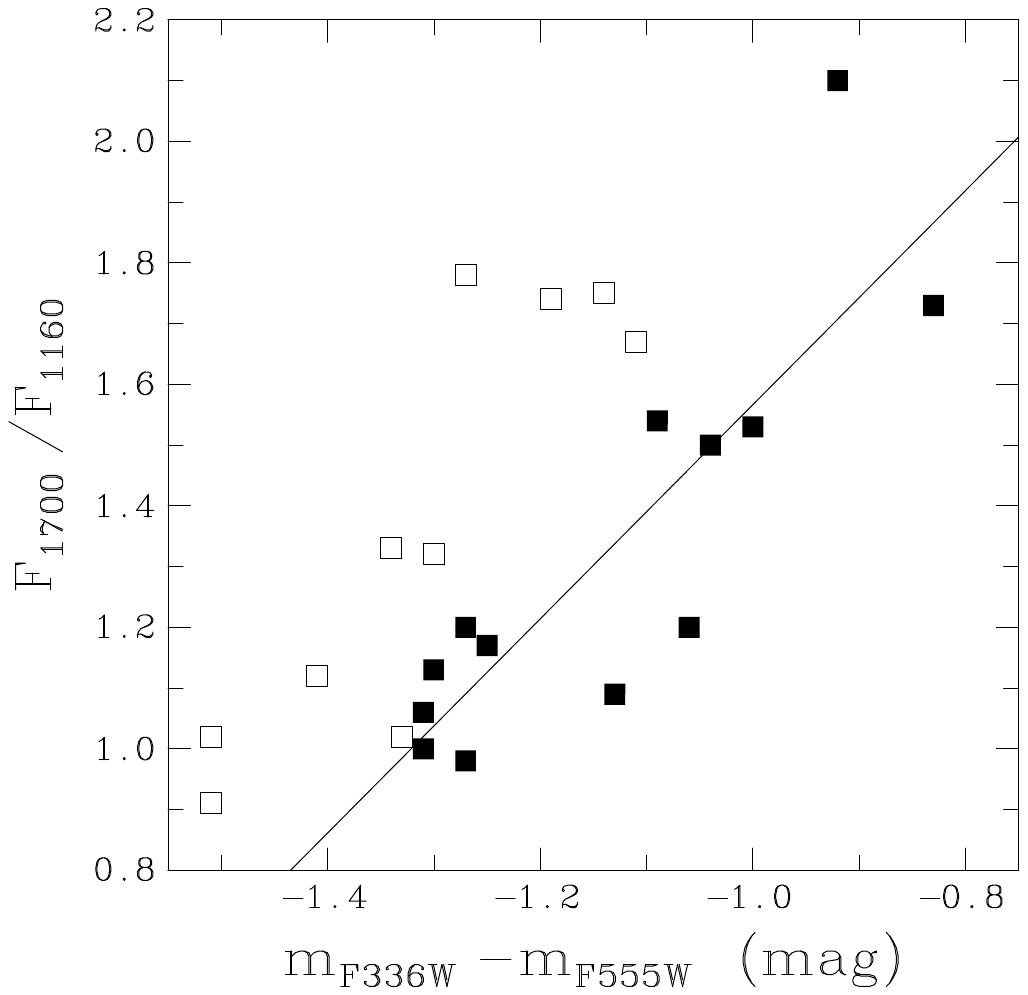}
\caption{Relationship between $(m_{\rm F336W} - m_{\rm F555W})$ colour and far-UV slope, $F_{\rm 1700}/F_{\rm 1160}$, for O stars in 30 Doradus (solid within MUSE field) with far-UV spectroscopy and HST/WFC3 photometry. The solid line is a linear fit to observations within the MUSE field.}
\label{colour}
\end{figure}

We calibrate $F_{\rm 1500}$ fluxes against $m_{\rm F275W}$ or $F_{\rm F336W}$ photometry using reference O stars in 30 Doradus that have been observed in the far-UV. These are listed in Table~\ref{calib}, sorted by $F_{\rm 1500}$ with references to spectral types, photometry and HST datasets identical to Table~\ref{tab:fuv-list}. Column densities, $\log N$(H\,{\sc i}) from fits to Ly$\alpha$ are included and should be reliable to $\pm$0.05 dex (e.g. \citet{2012ApJ...745..173W} obtain $\log N$(H\,{\sc i}/cm$^{2}$) = 21.79 for Mk~42). Comparisons are presented in Fig.~\ref{F275W-F336W}, with calibrations presented in Eqn~\ref{F275W}-\ref{F336W}.
Since we utilise templates spanning a range of extinctions, we also adjust far-UV slopes ($\lambda\lambda$1160-1700) of templates to individual O stars within the MUSE field from their $m_{\rm F336W} - m_{\rm F555W}$ colours. These are presented in Fig.~\ref{colour} with the fit obtained as follows
\begin{equation}
F_{1700}/F_{1160} \simeq 1.76 (m_{\rm F336W} - m_{\rm F555W}) + 3.33.
\end{equation}
30 Doradus O stars external to the MUSE field reveal a somewhat different behaviour, likely as a result of lower dust extinction.

\begin{table*}
\centering
\caption{Reference O stars within 30 Doradus used to calibrate far-UV fluxes (units are erg\,s$^{-1}$\,cm$^{-2}$\AA$^{-1}$), $\lambda\lambda$1160-1700 slopes and $\log N$(H\,{\sc i}) from fits to Ly$\alpha$. Sources within MUSE footprint are indicated. References are as for Table~\ref{tab:fuv-list}.}
\begin{tabular}{
r @{\hspace{2mm}} r @{\hspace{2mm}} r @{\hspace{2mm}}  c @{\hspace{2mm}} c @{\hspace{2mm}} c @{\hspace{2mm}} c @{\hspace{2mm}} r @{\hspace{2mm}} c @{\hspace{2mm}}  l @{\hspace{2mm}}
r @{\hspace{2mm}} r @{\hspace{2mm}} c}
\hline 
Star & Sp Type & Ref & $m_{\rm F275W}$ & $m_{\rm F336W}$ & $m_{\rm F555W}$ & Ref & $F_{\rm 1500}$ & $F_{\rm 1700}/F_{\rm 1160}$ & $\log N$(H\,{\sc i}) & HST dataset & Ref & MUSE \\
       &              &       &           (mag)           &      (mag)              & (mag)                     &             & $10^{-13}$    &                &   cm$^{-2}$      & \\
\hline
Mk~42      & O2\,If             & CW11  & 11.08         & 11.51                   & 12.82                     & S & 3.01                        & 1.00         &  21.75     & STIS/E140M & 3 & $\checkmark$ \\
Mk~39      & O2.5\,If/WN6+ & Cr22 & 11.38         & 11.66                    & 12.95                     & S & 2.74                       & 1.13          & 21.75      & COS/G130M+G160M & 3 & $\checkmark$ \\
VFTS~506 & ON2\,V        & Wa14   & 11.74        & 12.01                    & 13.32                    & S & 2.04                        & 1.06           & 21.8       & COS/G130M+G160M & 3 & $\checkmark$ \\
Mk~30      & O2\,If/WN5    & CW11  & 11.91        & 12.22                   & 13.48                     & S & 1.85                        & 0.98           & 21.75     & COS/G130M+G160M & 3 & $\checkmark$ \\
VFTS~180 & O3\,If*          & Wa14  & $\dots$      & 12.11                  & 13.52                      & S & 1.57                        & 1.12           & 21.6  & COS/G130M+G160M & 3 & $\cdots$ \\
VFTS~87  & O9.7\,Ib-II      & Wa14 & 11.97          & 12.09               & 13.60                       & S & 1.52                         & 1.02           & 21.6   & STIS/E140M & 3 & $\cdots$ \\
VFTS~267 & O3\,III-I        & Wa14  & $\cdots$     & 12.15                 & 13.49                      & S & 1.39                        & 1.33           & 21.75   & COS/G130M+G160M & 3 & $\cdots$ \\
VFTS~440 & O6--6.5\,III & Wa14      & 12.15        & 12.44                   & 13.69                    & S & 0.96                        & 1.17           & 21.75     & STIS/E140M & 2 & $\checkmark$ \\
Mk~37a    & O3.5\,If/WN7 & CW11 & 12.33          & 12.48                   & 13.52                     & S & 0.82                        & 1.50           & $\cdots$    & STIS/G140L & 6 & $\checkmark$ \\
Mk~33S & O3\,III             & Ma15 & 12.48          & 12.76                   & 13.81                     & S & 0.79                        & 1.20           & $\cdots$    & STIS/G140L &6 & $\checkmark$ \\ %
VFTS~586 & O4\,Vz          & Wa14 & 13.06         & 13.50                   & 15.02                     & S & 0.76                        & 0.91            & 21.75    & COS/G130M+G160M & 3 & $\cdots$ \\
VFTS~352 & O4.5\,Vz+O5.5\,Vz & Wa14 & 12.75 & 13.13                 & 14.46                     & S & 0.70                        & 1.02            & 21.8    & COS/G130M+G160M & 3 & $\cdots$ \\ 
Mk~55      & O6\,Vnn         & Wa14 & 12.79         & 13.04                  & 14.34                        & S & 0.61                      & 1.32             & 21.75   & COS/G130M+G160M & 3 & $\cdots$ \\ 
VFTS~385 & O4--5\,V        & Wa14 & 13.22         & 13.38                  & 14.65                      & S & 0.51                       & 1.20              & 21.8:  & COS/G140L & 2 & $\checkmark$ \\
VFTS~404 & O3.5\,V          & Wa14 & 12.76         & 12.95                 & 14.13                    & S & 0.49                         & 1.74               & 21.9  & COS/G130M+G160M & 3 & $\cdots$ \\
Mk~51     & O3.5\,If/WN7  & CW11 & $\cdots$     & 12.98                   & 13.81                     & S & 0.45                       & 1.73              & 21.9: & COS/G140L & 2 & $\checkmark$ \\
Mk14S     & O4\,III              & Wa14  & 13.01         & 13.19                 & 14.29                      & S & 0.43                      & 1.54               & 21.9:  & COS/G140L & 2 & $\checkmark$ \\ 
Mk~4        & O7\,II               & Wa14  & $\cdots$   & 13.37                & 14.38                     & S & 0.41                        & 1.53               & 22.0:  &  COS/G140L & 2 & $\checkmark$ \\  
VFTS~667 & O6\,V            & Wa14   & $\cdots$   & 13.95                 & 15.08                      & S & 0.33                      & 1.09                & 21.9: & COS/G140L & 2 & $\checkmark$ \\
VFTS~169 & O2.5\,V          & Wa14  & 13.64       & 13.46                 & 14.60                      & S & 0.32                     & 1.75                  & 22.0 & COS/G130M+G160M & 3 & $\cdots$ \\
VFTS~190 & O7\,Vnnp       & Wa14  & 13.70       & 13.43                  & 14.70                     & S & 0.28                     & 1.78                  & 21.7 & COS/G130M+G160M & 3 & $\cdots$ \\
VFTS~66   & O9.5\,III        & Wa14   & 14.40        & 14.49                & 15.61                      & S & 0.14                       & 1.67                & 22.0  & COS/G130M+G160M & 3 & $\cdots$ \\
VFTS~422 & O4III(f)          & Wa14 & 14.15         & 14.23                   & 15.15                    & S & 0.13                        & 2.10               & 21.9: & COS/G140L & 2 & $\checkmark$ \\
\hline
\end{tabular}\par
\label{calib} 
\end{table*}


\bsp	
\label{lastpage}
\end{document}